\documentclass[%
 reprint,
 showpacs,
nofootinbib,
 amsmath,amssymb,
 aps,
 prc,
]{revtex4-1}
\usepackage{graphicx}% Include figure files

\usepackage[caption=false]{subfig}
\usepackage{amsmath}
\usepackage{mathtools}
\usepackage{float}
\usepackage{dsfont}
\usepackage{dcolumn}% Align table columns on decimal point
\usepackage{bm}% bold math
\usepackage{lipsum}
\usepackage{csquotes}
\usepackage{textcomp}
\usepackage{nicefrac} 
\usepackage{feynmp}
\usepackage{color}
\usepackage{ifpdf}
\ifpdf
\fi

\makeatletter
\def\endfmffile{%
  \fmfcmd{\p@rcent\space the end.^^J%
          end.^^J%
          endinput;}%
  \if@fmfio
    \immediate\closeout\@outfmf
  \fi
  \IfFileExists{\thefmffile.mp}{\immediate\write18{mpost \thefmffile}}{}
  \let\thefmffile\relax
}
\makeatother

\newcommand{\gbs}[1]{{\widetilde{#1}}}
\newcommand{\hfb}[1]{{\rm #1}}
\newcommand{\hfe}[1]{{\epsilon^{(0)}_{#1}}}
\newcommand{\CG}[6]{ {\rm C}^{#5 #6}_{#1 #3 \, #2 #4} }
\newcommand{\CSJ}[6]{ \left\{ \begin{array}{@{\!~}c@{\!~}c@{\!~}c@{\!~}}  #1 & #2 & #3 \\[2mm]  #4 & #5  &  #6  \end{array}\right\} }

\newcommand{\dpjq}[1]{ \delta^{(\pi j q)}_{#1} }

\def\orcid#1{\kern .08em\href{https://orcid.org/#1}{\includegraphics[keepaspectratio,width=0.7em]{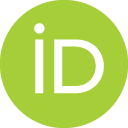}}}

\usepackage{hyperref}% add hypertext capabilities
\allowdisplaybreaks

\begin{document}
\captionsetup[subfigure]{labelformat=empty}

\preprint{APS/123-QED}

%\title{Gorkov algebraic diagrammatic construction formalism at second order with three-body interactions}
\title{Gorkov algebraic diagrammatic construction formalism at third order  }% Force line breaks with \\

%\author{Francesco Raimondi$^{1}$ and Carlo Barbieri$^{1}$}
\author{Carlo Barbieri \!\!\orcid{0000-0001-8658-6927} }
\affiliation{ Department of Physics, Via Celoria 16, 20133, Milano, Italy}
\affiliation{ INFN, Via Celoria 16, 20133, Milano, Italy}

\author{Thomas Duguet}
\affiliation{IRFU, CEA, Universit\'e Paris-Saclay, 91191 Gif-sur-Yvette, France}
\affiliation{KU Leuven, Instituut voor Kern- en Stralingsfysica, 3001 Leuven, Belgium}

\author{Vittorio Som\`a \!\!\orcid{0000-0001-9386-4104} }
\affiliation{ IRFU, CEA, Universit\'e Paris-Saclay, 91191 Gif-sur-Yvette, France}

\date{\today}% It is always \today, today,

\begin{abstract}
\begin{description}

\item[Background] The Gorkov approach to self-consistent Green's function theory has been formulated in [V. Som\`a, T. Duguet, C. Barbieri, {\color{blue}Phys. Rev. C \textbf{84}, 064317 (2011)}]. Over  the past decade, it has become a method of reference for first-principle computations of semi-magic nuclear isotopes. The currently available implementation is limited to a second-order self-energy and neglects particle-number non-conserving terms arising from contracting three-particle forces with anomalous propagators. For nuclear physics applications, this is sufficient to address first-order energy differences (i.e. two neutron separation energies, excitation energies of states dominating the one-nucleon spectral function), ground-state radii and moments on an accurate enough basis. However, addressing absolute binding energies, fine spectroscopic details of $N\pm1$ particle systems or delicate quantities such as second-order energy differences associated to pairing gaps, requires to go to higher truncation orders.

\item[Purpose]  The formalism is extended to third order in the algebraic diagrammatic construction (ADC) expansion with two-body Hamiltonians.

\item[Methods]  The expansion of Gorkov propagators in Feynman diagrams is combined with the algebraic diagrammatic construction up to the third order as an organization scheme to generate the Gorkov self-energy.

\item[Results]  Algebraic expressions for the static and dynamic contributions to the self-energy, along with equations for the matrix elements of the Gorkov eigenvalue problem, are derived. It is first done for a general basis before specifying the set of equations to the case of spherical systems displaying rotational symmetry. Workable approximations to the full self-consistency problem are also elaborated on. The formalism at third order it thus complete for a general two-body Hamiltonian.

\item[Conclusion] Working equations for the full Gorkov-ADC(3) are now available for numerical implementation.  

\end{description}
\end{abstract}

% PACS, the Physics and Astronomy Classification Scheme.


\maketitle

%\tableofcontents

\section{\label{Sec:Intro}Introduction}

%Quantum many-body problems are key to understanding a variety of complex systems in Nature. For example, they are at the basis of a vast range of technological advances in material sciences and chemistry and inform us on the structure of atomic nuclei and their role in astrophysics scenarios like nucleosynthesis and even compact stars.
%Naturally, ab initio many-body methods have become a crucial tools to enable high precision investigation in these fields of physics.
%In the form of finite size fermionic systems, most ab initio applications concern Nuclear Physics and Quantum Chemistry and cross fertilization among the two disciplines has lead to advancements of ab initio method over the years.

Ab initio quantum many-body computations are crucial to high precision investigations in several fields of physics. Most applications to finite-size fermion systems concern Nuclear Physics and Quantum Chemistry, to the point that these disciplines often share the same computational techniques and cross fertilization among the two has led to advancements of ab initio theories over the years.
For nuclear physics, the past two decades have witnessed remarkable breakthroughs in first-principle computations of nuclear structure that exploited soft nuclear interactions based on chiral effective field theory~\cite{LongQuestNF2020Front}. The availability of many-body methods that scale favourably with particle number has enabled precision predictions of medium-mass isotopes and the possibility to confront experimental information of exotic isotopes at the limits of stability (see Refs.~\cite{Coraggio2020Editoral,Hergert2020Front} for a review).

Many successful approaches, such as many-body perturbation theory (MBPT)~\cite{Tichai20}, self-consistent Green's function (SCGF)~\cite{Soma2020Front}, coupled cluster (CC)~\cite{Hagen2014CCrev} and in-medium similarity renormalization group~\cite{Hergert2016IMSRGrev} can reach sizable systems by restricting the Fock space to selected excited configurations for which it is possible to resum infinite series of diagrams. However, in their basic formalism, they are limited to closed-shell systems. For open-shell cases, near-degeneracies in the single-particle spectrum often prevent the use of any perturbation expansion. 
The possible ways around this issue are either multi-reference approaches or the use of symmetry-breaking reference states. In the first case, all degenerate configurations are diagonalized explicitly, which however adds a costly step to the calculation that scales exponentially with system's size~\cite{Tichai:2017rqe}, with the notable exception of a recently proposed multi-reference many-body perturbation theory~\cite{Frosini:2021fjf,Frosini:2021sxj,Frosini:2021ddm}. The second path relies on using a reference state that explicitly breaks some symmetries of the Hamiltonian, in exchange for lifting the energy degeneracy. One is left with similar computational requirements as the original approach but needs to worry about projecting the final wave function, when possible~\cite{Duguet2014Jproj,Duguet2016Nproj,qiu2017,qiu2019}, or addressing the uncertainties due to an only partially restored symmetry.

Besides ground-state properties, the SCGF approach is particularly suited to inform on the spectroscopic response to the addition and removal of a nucleon to the system~\cite{Cipollone2015,Atar2018prl}, on the shell structure~\cite{Cipollone2013prl,Raimondi2019EffCh} and on elastic nucleon scattering~\cite{Idini2019PRL}. 
State-of-the-art SCGF calculations exploit the algebraic diagrammatic construction (ADC) truncation scheme that provides a hierarchy for systematic improvements of the method, i.e. of the computation of the self-energy~\cite{Schirmer1982,Schirmer1983,Barb2017LNP}.
The third-order truncation, or ADC(3), resums full Tamm-Dancoff series of ladder and rings diagrams among other terms and it  has become a method of reference for closed-shell nuclei and molecules, providing chemical accuracy predictions for binding energies and ionization potentials~\cite{Danovich2011}. 

In Refs.~\cite{Soma2011GkI,Soma2014GkII}, the standard SCGF formalism was extended to the Gorkov formulation to handle open-shell systems. In this formulation, the reference state is allowed to break $U(1)$ global-gauge symmetry associated with particle-number conservation thus accounting for pairing  correlations and lifting the problematic degeneracy of symmetry-conserving reference states with respect to elementary excitations. Doing so, the reach of SCGF calculations was enlarged from the small set of doubly closed-shell nuclei to the much larger set of semi-magic nuclei. Applications have covered complete isotopic and isotonic chains around  O, Ca and Ni~\cite{Lapoux2016,Soma2020LNL,Mougeot20,Sun20,Soma2021ejpa,Linh21}, eventually stretching to heavier isotopes up to A=140~\cite{Arthuis2020Xe}.
The Gorkov SCGF formalism has so far been been devised only for a second-order self-energy, i.e. at the ADC(2) truncation level, which can grasp around 90\% of correlation energy and predict accurate trends of nuclear binding energies and radii with varying proton-neutron asymmetry. Nevertheless, confronting the predictive power of chiral Hamiltonians on absolute nuclear masses and spectroscopic data requires more accurate computations by going to ADC(3) or higher orders. At the same time, simple truncations such as ADC(2) remain significantly affected by the violation of global-gauge symmetry. While this issue is expected to resolve when going to higher truncation levels, it is not possible to assess the rate at which good particle number is restored without computing a proper sequence of ADC($n$) results with increasing orders $n=1,2,3$ and so on. In this work we pave the way to addressing these open questions by deriving the Gorkov ADC(3) approximation in full for a Hamiltonian with up to two-body forces.

Section~\ref{Sec:GPros} reviews key concepts of the Gorkov Green's function formalism and sets out the details needed for its implementation. In particular, the analytical form of the self-energy and the Gorkov eigenmatrix problem, which are central to the following developments, are discussed.  Working ADC(2) and ADC(3) equations are presented in full in Sec.~\ref{Sec:ADCn}. While the focus is eventually on a self-consistent implementation, Sec.~\ref{Sec:Comb_diags} provides a detailed overview of the composite diagrams required in a standard (non self-consistent) theory. Sec.~\ref{Sec:scgf_approx} deals with practical limitations in the application of SCGF theory and sets out a systematic way to implement partial self-consistency without introducing uncontrolled uncertainties.
The main results of this work are then collected in Sec.~\ref{Sec:Implem}, which summarises the specific terms and working equations at each level of ADC($n$) truncation.
A number of further technical details are relegated to the appendices, in decreasing order of importance. Appendix~\ref{App:Jcoupl} discusses the angular-momentum coupling of the ADC($n$) equations applicable to spherical bases. While these are conceptually the same formulae as the ones discussed in the main text, they provide implementation-ready working equations for applications to semi-magic nuclei. The contributions of composite diagrams to the static self-energy are not needed in most application but are derived in App.~\ref{App:SigInf_3rd_ord} for completeness. App.~\ref{App:Freq_ints} demonstrates some (somewhat pedagogical) details regarding how the final ADC(3) equations discussed in this paper are derived. Conclusions are drawn in Sec.~\ref{Sec:Concl}.

\section{\label{GkvIntro} Gorkov Green's function formalism}

We are interested in solving for a general many-fermion system described by an energy-independent Hamiltonian with up to two-body interactions
%three-body interactions:
\begin{align}
\nonumber
 H ={}& T ~+~ V \\ %~+~\hat{W}   \\
    ={}& \sum_{\alpha \beta}  \; t_{\alpha \beta}  \; c^\dagger_{\alpha} c_\beta  
     + \frac{1}{4} \sum_{\alpha \beta \gamma \delta}  \; v_{\alpha \beta ,  \gamma \delta} \; c^\dagger_{\alpha} c^\dagger_{\beta} c_\delta c_\gamma \;,    \label{eq:H} 
%    &+ \frac{1}{6} \sum_{\alpha \beta \gamma \mu \nu \lambda} \;  w_{\alpha \beta \gamma, \mu \nu \lambda}  \; c^\dagger_{\alpha} c^\dagger_{\beta} c^\dagger_{\gamma} c_\lambda  c_\nu c_\mu \; , 
\end{align}
where   $\alpha, \beta, \gamma\ldots$ label a complete orthonormal one-body basis whereas $c^\dagger_{\alpha}$~($c_{\alpha}$) denote associate creation~(annihilation) operators\footnote{Note that greek letters refer to a general single-particle basis throughout this work, while latin letters are reserved for \hbox{j-coupled} bases
as discussed in App.~\ref{App:Jcoupl}. This choice differs from our initial Gorkov work of Ref.~\cite{Soma2011GkI} but maintains a continuity of notation with our other SCGF developments~\cite{Barbieri2001,Barbieri2007,Barbieri2003DRPA,Barbieri2009Ni56,Carbone2013TNF,Cipollone2015,Barb2017LNP,Raimondi2018adc3,Drissi2021ncpt,Drissi2021ncgf}}.
In Eq.~\eqref{eq:H},  $T$ captures the complete one-body sector of the Hamiltonian: it typically reduces to the kinetic energy  for self-bound systems, such as atomic nuclei, but it may include an external potential in the general case. 
The two-body matrix elements, $v_{\alpha \beta ,  \gamma \delta}$,  are intended as being properly antisymmetrized.
%The two- and three-body matrix elements, $v_{\alpha \beta ,  \gamma \delta}$ and $w_{\alpha \beta \gamma, \mu \nu \lambda}$, are intended as being properly antisymmetrized.

In this work we follow Ref.~\cite{Soma2011GkI} and associate to a given basis, $\{|\alpha\rangle\}$, of the one-body Hilbert-space $\mathcal{H}_1$ a dual basis 
 $\{|\bar{\alpha}\rangle\}$ that is related to the former through an antiunitary transformation $\mathcal{T}$.
 Specifically, one starts with the set of quantum numbers $\alpha$ specifying a state of the original basis and associate a new set $\widetilde\alpha$ that is in a self-inverse \hbox{one-to-one} correspondence with the former, i.e.
 $\widetilde{\widetilde\alpha}=\alpha$.  The dual basis state is defined by adding the antiunitary \emph{real} phase $\eta_\alpha$
 ($\eta_\alpha \eta_{\widetilde\alpha}=-1$)
\begin{align}
\label{eq:dualA}
  |\bar{\alpha}\rangle ~=& ~\mathcal{T} \, |\alpha\rangle ~\equiv~ \eta_\alpha \, |\widetilde{\alpha}\rangle \, ,
 \end{align}
such that dual creation and annihilation operators are related to the original ones through
\begin{subequations}
\label{eq:ccdag_bar}
 \begin{align}
    \bar{c}^\dagger_{\alpha}  \equiv{}& \eta_\alpha \, c^\dagger_{\widetilde{\alpha}} \; ,  \\
    \bar{c}_{\alpha}                \equiv{}& \eta_\alpha \, c_{\widetilde{\alpha}} \; , 
 \end{align}
\end{subequations}
while the matrix elements of the operators entering the Hamiltonian can be expressed completely in the dual basis or in a mixed representation
\begin{subequations}
\label{eq:TV_bar} 
 \begin{align}
    t_{\bar{\alpha} \bar{\beta}}  ={}&   \eta_\alpha \,  \eta_\beta \,  t_{\widetilde{\alpha} \widetilde{\beta}}  \; ,   \\
    v_{\bar{\alpha} \bar{\beta} ,  \bar{\gamma} \bar{\delta}} ={}& \eta_\alpha \,  \eta_\beta \, \eta_\gamma \,  \eta_\delta \,
        v_{\widetilde{\alpha} \widetilde{\beta} ,  \widetilde{\gamma} \widetilde{\delta}}  \; ,  
        \\
    v_{\bar{\alpha} \beta  ,  \gamma  \bar{\delta}} ={}&  \eta_\alpha \,  \eta_\delta \, v_{\widetilde{\alpha} \beta ,  \gamma \widetilde{\delta}} \, ,
 \end{align}
\end{subequations}
and so on.
 The advantage of the above relations is that many-body operators are invariant with respect to (partial) changes of the single-particle basis as long as barred quantities are transformed consistently for each  separate index. For example,
 \begin{align}
 \sum_{\alpha \beta}  \; t_{\alpha \beta}  \; c^\dagger_{\alpha} c_\beta ={}&
 \sum_{\bar{\alpha} \beta}  \; t_{\bar{\alpha} \beta}  \; \bar{c}^\dagger_{\alpha} c_\beta =
 \sum_{\bar{\alpha} \bar{\beta}}  \; t_{\bar{\alpha} \bar{\beta}}  \; \bar{c}^\dagger_{\alpha} \bar{c}_\beta
 \end{align}
and similarly for all other components of Eq.~\eqref{eq:H}. This property facilitates the definition of the Gorkov propagators in Sec.~\ref{Sec:GPros} and propagates to all tensor products of propagators and operators arising in the diagrammatic expansion of perturbation and SCGF theories.

The introduction of the dual basis is not strictly mandatory such that the Gorkov formalism presented in this work could be derived without making use of barred indices. However, definition~\eqref{eq:dualA} makes it easier to elegantly handle Nambu indices for normal and anomalous propagators and accounts automatically for the phases that are related to broken symmetries in the formalism.  Only in the last step of deriving working Gorkov-ADC(3) equations the transformation $\mathcal{T}$ is identified with the time reversal operator and the phases $\eta_\alpha$ explicitly stated (see also App.~\ref{App:Jcoupl}).
More importantly, the combined use of Nambu indices and an appropriate dual basis can be extended into a generalised Nambu-covariant formalism as discussed in Refs.~\cite{Drissi2021ncpt,Drissi2021ncgf}. In Nambu-covariant Green's function theory, all normal and anomalous propagators appear as specific
elements of a unique propagator carrying the common features in their spectral representations.

\subsection{\label{Sec:GPros}Gorkov propagators}

The Gorkov-SCGF approach builds on relaxing the requirement that the unperturbed state is an eigenstate of the particle-number operator and seeking for the solution of the grand-canonical-like Hamiltonian\footnote{Being presently interested in a zero-temperature formalism the T-dependent term of the grand-canonical potential drops out. Moreover, it is understood that a separate chemical potential for each different fermion is to be considered when the system consists of more than one type of particle.}
\begin{align}
\Omega \equiv{}& H - \mu N \, ,
 \label{eq:Om_def0}
\end{align}
where $\mu$ denotes the chemical potential and $N$ the particle number operator. The Hamiltonian is partitioned into a unperturbed term $\Omega_U$ containing only one-body vertices and an interacting part as follows
\begin{align}
\Omega \equiv{}& \Omega_U + \Omega_I \nonumber \\
           ={}&  \left(T ~+~ U  - \mu N\right)  +  
             \left(-U ~+~ V  \right) \, , %~+~\hat{W} \right) \, ,
 \label{eq:Om_def}
\end{align}
where $U$ denotes an external mean-field like potential.

We consider eigenstates of the Hamiltonian conserving even- ($e$) or odd- ($o$) number parity
\begin{align}
  \Omega \, |\Psi^{e(o)}_k \rangle ={}&  \Omega_k \, |\Psi^{e(o)}_k \rangle  \, ,
\label{eq:Psi_k}
\end{align}
where
\begin{align}
  |\Psi^{e(o)} \rangle ={}& \sum_{n=0}^\infty   \;  c_{2n(2n+1)} |\psi^{2n (2n+1)} \rangle  
\label{eq:Psi_eo}
\end{align}
is a superposition of states $|\psi^{l} \rangle$ that are eigenstates of $N$ with eigenvalue $l$. %(and the '$+1$' is added only in the odd case.
Rather than the ground state of $H$, Gorkov SCGF formalism targets the state $|\Psi_0 \rangle$ minimizing
\begin{align}
  \Omega_0 ={}&   \min_{|\Psi_0 \rangle} \left\{ \langle\Psi_0 | \Omega |\Psi_0 \rangle \right\}
\label{eq:DefPsi0}
\end{align}
under the constraint
\begin{align}
  \text{N} ={}& \langle\Psi_0 | N |\Psi_0 \rangle \, ,
\label{eq:DefPsi0N}
\end{align}
%where $\text{N}$ denotes the number of particles for the system under consideration. While the exact $|\Psi_0 \rangle$ associated with a finite system is indeed an eigenstate of $N$, it is not enforced to do so when being approximated, i.e. it is only constrained to carry the particle number $\text{N}$  {\it on average}.
%
%Clearly, whenever the approximate treatment approaches the exact solution, the exact particle number shall be restored. An obvious question of present interest is to understand to which extent going from the ADC(2) to the ADC(3) truncation level can do so to a good enough accuracy. Alternatively, an exact symmetry restoration can hopefully be formulated in a similar way to what has recently been done within the frame of MBPT and CC formalisms~\cite{Duguet2014Jproj,Duguet2016Nproj,qiu2017,qiu2019}.
%
where $\text{N}$ denotes the number of particles for the system under consideration. 
While the exact $|\Psi_0 \rangle$ associated with a finite system is indeed an eigenstate of $N$, it is not enforced to do so in the thermodynamic limit or when being approximated.
In such cases, it is only constrained to carry the particle number $N$~\emph{on average}.

For a typical superfluid system approaching the thermodynamic limit, the ground state  energies of Eq.~\eqref{eq:H} associated with $N$ particles, $H |\psi^N_0\rangle = E^N_0  |\psi^N_0\rangle$,
will differ from each other only by multiples of the chemical potential
\begin{align}
 E^{N \pm 2n}_0 \approx{}& E^N_0 ~\pm~ 2n\mu \pm n \Delta\varepsilon_P \, , \qquad \hbox{for } n = 1, 2, 3 \ldots ,
\label{eq:Emu}
\end{align}
since $\mu$ is substantially independent of $N$ at large particle number and, likewise, the average cost for the possible creation of Cooper pairs, $\Delta\varepsilon_P$, will be the same every time two particles are added. 
Eqs.~\eqref{eq:DefPsi0} and~\eqref{eq:DefPsi0N} naturally allow to interpret state $|\Psi_0\rangle$ as the fermionic part of a ground-state wave function in equilibrium with a reservoir of Cooper pairs. Hence, defining Gorkov propagators with respect to $|\Psi_0\rangle$ directly provides a theory for superconductivity and superfluidity.
For finite-size systems, such as atomic nuclei or molecules, Eq.~\eqref{eq:Emu} may hold only in a very approximate way.  Because both Hamiltonians $H$ and $\Omega$ preserve particle number, the requirements~\eqref{eq:DefPsi0} and~\eqref{eq:DefPsi0N}  will force $|\Psi_0\rangle$ to be the true ground state  $|\psi^N_0\rangle$, with an exact number of particles.

The breaking of particle-number symmetry arises naturally, in most cases, whenever approximations have to be made, typically in computing the self-energy. This is true for \emph{both} Dyson and Gorkov formulations of Green's function theory since they  equally rely on an open Fock space, where mixing of particle number as in Eq.~\eqref{eq:Psi_eo} is fully allowed. In fact both, formulations can be seen as just one theory where in the first case the reference state preserves the symmetries of the Hamiltonian from the start, whereas in the second case one begins with a symmetry-broken reference but with the advantage of a better radius of convergence for the perturbative expansion.
Clearly, whenever the approximate treatment approaches the exact solution, the exact particle number shall be restored.

For Gorkov theory, the symmetry breaking is more substantial because it is imposed into the formalism form the start through~$\Omega_U$.  Hence, one may wish to eventually restore the exact symmetries of the Hamiltonian.
Several works for the standard, Dyson, theory have investigated how approximations based on the self-consistency principle can guarantee the conservation of particle number and other symmetries associated to $H$~\cite{Baym1961b,Baym1962,VanNeck2001JCP2nd,Rios2006Entrop,Rios2020Front}.
Two obvious questions of present interest are whether a truly self-consistency computation could have implications on particle-number conservation also for Gorkov and, otherwise, understanding to which extent going from the ADC(2) to the ADC(3) truncation level can do so to a good enough accuracy. Alternatively, one will need to seek formulations for exact symmetry restoration at the final stage of each computation, in a similar way to what has recently been done within the frame of MBPT and CC formalisms~\cite{Duguet2014Jproj,Duguet2016Nproj,qiu2017,qiu2019}.

The crucial feature of the Gorkov SCGF formulation is that the unperturbed Hamiltonian $\Omega_U$  breaks particle number explicitly. %, through the choice of the auxiliary interaction $\hat U$. 
Open-shell systems are characterised by partially filled orbitals at the level of mean filed theory. This causes degeneracies between the energies of particle and hole states which are sufficient to invalidate any perturbation expansion (or even partial resummations of it) if not dealt with in advance, as it is the case of the Dyson SCGF formulation. In this context, a superposition of states of the type~\eqref{eq:Psi_eo} provides a better approximation to the ground state wave function.  Following this consideration, we exploit an auxiliary interaction $U$ to break particle number symmetry explicitly:
\begin{align}
 U ={}& \sum_{\alpha \beta}  \left[
       u_{\alpha \beta}  \, c^\dagger_{\alpha} c_\beta  
      + \frac 1 2    u^{an.}_{\alpha \beta}  \, c^\dagger_{\alpha} c^\dagger_\beta  +  \frac 1 2    (u^{an.}_{\alpha \beta})^{*}  \, c_{\alpha} c_\beta   \right] \, ,
\label{eq:def_U}
\end{align}
where $u_{\alpha \beta} = u_{\beta \alpha}^*$ and $u^{an.}_{\alpha \beta}=-u^{an.}_{\beta \alpha}$ without loss of generality.

Following the above considerations we formulate Gorkov SCGF theory with respect to the ground state of $\Omega$ as defined by Eqs.~\eqref{eq:DefPsi0} and~\eqref{eq:DefPsi0N}. Since $|\Psi_0\rangle$ breaks particle number symmetry, it is possible to define four Gorkov propagators,
\begin{subequations}
\label{eq:Gkv_props_time}
\begin{align}
  G^{11}_{\alpha \beta}(t-t') \equiv{}& -i \langle \Psi_0 |  \; T [            c_{\alpha}(t)                c^\dagger_\beta(t') ] \; |\Psi_0\rangle \, ,  \\
  G^{12}_{\alpha \beta}(t-t') \equiv{}& -i \langle \Psi_0 |  \; T [            c_{\alpha}(t) \bar{c}_                    \beta(t') ] \; |\Psi_0\rangle \, ,   \\
  G^{21}_{\alpha \beta}(t-t') \equiv{}& -i \langle \Psi_0 |  \; T [   \bar{c}^\dagger_{\alpha}(t)   c^\dagger_\beta(t') ] \; |\Psi_0\rangle \, ,   \\
  G^{22}_{\alpha \beta}(t-t') \equiv{}& -i \langle \Psi_0 |  \; T [   \bar{c}^\dagger_{\alpha}(t)        \bar{c}_\beta(t') ] \; |\Psi_0\rangle \, ,
\end{align}
\end{subequations}
where $T[\cdots]$ denotes the usual time ordering operator and the superscripts `$1$'~(`$2$') are the Nambu indices referring to creation and annihilation of normal~(anomalous) single-particle excitations.  In Eqs.~\eqref{eq:Gkv_props_time}, creation and annihilation operators depend on time\footnote{We use natural units with dimensionless $\hbar$=1 throughout this work.} according to the Heisenberg picture with respect to $\Omega$ 
\begin{align}
     c^{(\dagger)}_{\alpha}(t) = e^{i\Omega t} \, c^{(\dagger)}_{\alpha} \, e^{-i\Omega t} \, .
\label{eq:Heis_pic}
\end{align}
It is useful to collect the four propagators in a $2\times2$ matrix according to their Nambu indices and whose elements are tensors in the single-particle basis indices
\begin{align}
 \mathbf{G}_{\alpha \beta}(t-t') \equiv  \left(
  \begin{array}{ccc}
    G^{11}_{\alpha \beta}(t-t') &&   G^{12}_{\alpha \beta}(t-t')  \\ ~ \\
  G^{21}_{\alpha \beta}(t-t')  & &  G^{22}_{\alpha \beta}(t-t') 
\end{array} \right)  \, .
\label{eq:Gnmb}
\end{align}

Rather that the time representation, the frequency representation is presently used and is obtained via Fourier transformation
\begin{align}
  \mathbf{G}_{\alpha \beta}(\omega) = \int_{-\infty}^{\infty} e^{i \omega \tau} \,  \mathbf{G}_{\alpha \beta}(\tau) \; d\tau \, .
\label{eq:Fourier}
\end{align}
Exploiting the completeness of Eq.~\eqref{eq:Psi_k}, the spectral representation of the propagators can be obtained as
\begin{subequations}
\label{eq:Gkv_props}
\begin{align}
  G^{11}_{\alpha \beta}(\omega) ={}& \sum_k \left\{   \frac{        {\mathcal U}^k_\alpha    \; \;        {\mathcal U}^k_\beta{}^*     }{\omega - \omega_k + i\eta} +    \frac{  \bar{\mathcal V}^k_\alpha{}^*        \;\,  \bar{\mathcal V}^k_\beta }{\omega + \omega_k - i\eta}     \right\}  \, ,  \\
  G^{12}_{\alpha \beta}(\omega) ={}& \sum_k \left\{   \frac{        {\mathcal U}^k_\alpha    \; \;        {\mathcal V}^k_\beta{}^*     }{\omega - \omega_k + i\eta} +    \frac{  \bar{\mathcal V}^k_\alpha{}^*        \;\,  \bar{\mathcal U}^k_\beta }{\omega + \omega_k - i\eta}     \right\}  \, ,  \\
  G^{21}_{\alpha \beta}(\omega) ={}& \sum_k \left\{   \frac{        {\mathcal V}^k_\alpha    \; \;        {\mathcal U}^k_\beta{}^*     }{\omega - \omega_k + i\eta} +    \frac{  \bar{\mathcal U}^k_\alpha{}^*        \;  \bar{\mathcal V}^k_\beta }{\omega + \omega_k - i\eta}     \right\}  \, ,  \\
  G^{22}_{\alpha \beta}(\omega) ={}& \sum_k \left\{   \frac{        {\mathcal V}^k_\alpha    \; \;        {\mathcal V}^k_\beta{}^*     }{\omega - \omega_k + i\eta} +     \frac{  \bar{\mathcal U}^k_\alpha{}^*        \; \,  \bar{\mathcal U}^k_\beta }{\omega + \omega_k - i\eta}     \right\}  \, , 
\end{align}
\end{subequations}
where the spectroscopic amplitudes for the addition and removal of a particle are defined as
\begin{subequations}
\label{eq:def_UV}
\begin{align}
    {\mathcal U}^k_\alpha \equiv{}&  \langle \Psi_0 |              c_\alpha  |\Psi_k \rangle \, ,  \\
    {\mathcal V}^k_\alpha \equiv{}&  \langle \Psi_0 |     \bar{c}^\dagger_\alpha  |\Psi_k\rangle \, ,  
\end{align}
\end{subequations}
so that
\begin{subequations}
\label{eq:def_UVbar}
\begin{align}
    \bar{\mathcal U}^k_\alpha \equiv{}&    \langle \Psi_0 |    \bar{c}_\alpha  |\Psi_k \rangle  = ~~ \eta_\alpha {\mathcal U}^k_{\widetilde \alpha}   \, ,  \\
   \bar {\mathcal V}^k_\alpha \equiv{}&    \langle \Psi_0 |     c^\dagger_\alpha  |\Psi_k \rangle  = -  \eta_\alpha {\mathcal V}^k_{\widetilde \alpha}  \, .   
\end{align}
\end{subequations}
The index $k$ in Eqs.~\eqref{eq:Gkv_props} labels all possible excitations from Eq.~\eqref{eq:Psi_k}, which combine both the Landau and Bogoliubov meanings of \emph{quasiparticle}. The respective poles are given by
\begin{align}
 \omega_k \equiv \Omega_k - \Omega_0 \, .
 \label{eq:wk}
\end{align}

Whenever the targeted ground state $|\Psi^e_0\rangle$ belongs to a system carrying an even particle number, the completeness set $\{|\Psi^o_k\rangle\}$  runs over odd particle numbers states only. And vice versa for an odd-N ground state. While the ADC($n$) equations derived in Secs.~\ref{Sec:adc_1_2} and~\ref{Sec:adc_3} are general and apply to both cases indistinctly, App.~\ref{App:Jcoupl} discusses their j-coupling reduction for $J=0$ ground states, i.e. for even-N systems. 

Once the spectral representation~\eqref{eq:Gkv_props} is known, it is possible to extract normal and anomalous one-body density matrices according to
\begin{subequations}
\label{eq:rho_ab}
\begin{align}
  \rho_{\alpha \beta} \equiv{}&  \langle \Psi_0 |   c^\dagger_\beta   c_\alpha |\Psi_0\rangle   \nonumber \\        
  =& \frac{1}{\pi} \int_{-\infty}^0   \hbox{Im} \, G^{11}_{\alpha \beta} (\omega) \; d\omega \nonumber \\        
  =& \sum_k   \,    \bar {\mathcal V}^k_\alpha{}^* \,  \bar {\mathcal V}^k_\beta \, , \\
  \tilde{\rho}_{\alpha \beta} \equiv{}&  \langle \Psi_0 |   \bar{c}_\beta   c_\alpha |\Psi_0\rangle   \nonumber \\        
  =&  \frac{1}{\pi} \int_{-\infty}^0   \hbox{Im} \, G^{12}_{\alpha \beta} (\omega) \; d\omega \nonumber \\        
  =& \sum_k  \,     \bar {\mathcal V}^k_\alpha{}^* \, \bar {\mathcal U}^k_\beta    \, .
\end{align}
\end{subequations}
The expectation value of any one-body operator $O$ is given by
\begin{align}
   \langle \Psi_0 |  O |\Psi_0\rangle  ={}& \sum_{\alpha \beta}  o_{\alpha \beta} \rho_{\beta \alpha}  \, ,
\label{eq:O_exp_vl}
\end{align}
whereas the Migdal-Galitski-Koltun energy sum rule delivering the ground-state energy
\begin{align}
  \Omega_0 ={}& \frac{1}{2 \pi} \int_{-\infty}^0  \left[  t_{\alpha \beta} - \mu \delta_{\alpha \beta} + \omega \delta_{\alpha \beta}  \right] \hbox{Im} \, G^{11}_{\beta \alpha} (\omega) \; d\omega
%CC%\nonumber \\
%CC%  ={}& \frac{1}{4 \pi i} \int_{C\uparrow}  \left[  t_{\alpha \beta}  - \mu \delta_{\alpha \beta} + \omega \delta_{\alpha \beta}  \right]  \, G^{11}_{\beta \alpha} (\omega) \; d\omega
\label{eq:Koltun}
\end{align}
is exact for a Hamiltonian with up to two-particle interactions.

\subsection{\label{Sec:GkvEqs}Gorkov equations}

The perturbative expansion of Gorkov propagators is devised following the standard approach of defining an unperturbed propagator, 
$\mathbf{G}^{(0)}(t-t')$, according to definitions~\eqref{eq:Gkv_props_time} and~\eqref{eq:Heis_pic} but with $\Omega$ replaced by the one-body 
grand potential $\Omega_U$.  After Fourier transform to frequency domain, one finds
\begin{align}
     \mathbf{G}^{(0)}(\omega) = [\omega \mathbb{I}  - \mathbf{\Omega}_U]^{-1} \, ,
\label{eq:G0}
\end{align}
where model space and Nambu indices are implicit and the matrix inversion is performed with respect to both.
One then exploits the interaction picture to devise a perturbative expansion of the full propagator of Eq.~\eqref{eq:Gnmb}
that can represented as a series of Feynman diagrams in powers of the perturbation~$\Omega_I$~\cite{Soma2011GkI}.

Doing so, the standard Dyson equation for the interacting propagator $\mathbf{G}(\omega)$ is generalised to the set of coupled Gorkov equations  for the four propagators~\eqref{eq:Gkv_props}. Using Nambu's matrix notation, they read as
\begin{align}
   \mathbf{G}_{\alpha \beta}(\omega)  ={}&  \mathbf{G}^{(0)}_{\alpha \beta}(\omega)  \, + \,   \sum_{\gamma \, \delta} \mathbf{G}^{(0)}_{\alpha \gamma}(\omega) \, \mathbf{\Sigma}^\star_{\gamma \delta}(\omega) \, \mathbf{G}_{\delta \beta}(\omega) \, ,
\label{eq:Gkv_eqs}
\end{align}
where the four self-energies 
\begin{align}
\mathbf{\Sigma}^\star_{\alpha \beta}(\omega)  \equiv  \left(
  \begin{array}{ccc}
    \Sigma^{\star \,11}_{\alpha \beta}(\omega) &&   \Sigma^{\star \,12}_{\alpha \beta}(\omega)  \\ ~ \\
  \Sigma^{\star \,21}_{\alpha \beta}(\omega)  & &  \Sigma^{\star \,22}_{\alpha \beta}(\omega) 
\end{array} \right)
\label{eq:Slfen_mtx}
\end{align}
include  all possible \emph{one-particle irreducible } diagrams stripped of their external legs. The remaining reducible diagrams are then
generated in a non-perturbative way through the all-orders resummation generated by Eq.~\eqref{eq:Gkv_eqs}.
In standard perturbation theory, a given approximation to $\mathbf{\Sigma}^\star(\omega)$ is a functional of the
unperturbed propagators $\mathbf{G}^{(0)}(\omega)$ and hence depends directly on the choice of the reference state 
associated with $\Omega_U$.
In SCGF theory, the series of diagrams to be resummed is further restricted to \emph{skeleton} diagrams displaying no self-energy insertion, provided that all propagator lines are replaced by the interacting propagator~$\mathbf{G}(\omega)$.
Since the full Dyson-Gorkov series is included in such a propagator, the SCGF procedure not only reduces the number of Feynman diagrams that need to be dealt with but it implicitly accounts for higher-order terms that are beyond the perturbative truncation chosen for the self-energy. The self-energy becomes a functional of the interacting propagator, $\mathbf{\Sigma}^\star[\mathbf{G};T, V]$ and is no longer affected by the choice of the unperturbed state. 
The price to pay for such improvements is that diagrams expressed in terms of $\mathbf{G}(\omega)$ are more demanding to deal with, due to the rich pole structure of Eqs.~\eqref{eq:Gkv_props}. Furthermore, $\mathbf{\Sigma}^\star(\omega)$ and the Gorkov equations~\eqref{eq:Gkv_eqs} have respectively to be computed and solved repeatedly through an iterative procedure.

The most general structure of the Gorkov self-energy can be written as
\begin{align}
   \mathbf{\Sigma}^\star_{\alpha \beta}(\omega)  = -\mathbf{U} +  \mathbf{\Sigma}^{(\infty)}_{\alpha \beta}  +  \widetilde{\mathbf{\Sigma}}_{\alpha \beta}(\omega)  \, ,
\label{eq:SE_split}
\end{align}
where the auxiliary potential term $U$ arising from $\Omega_I$ at first order is separated from the proper part of the self-energy. The
term $\mathbf{\Sigma}^{(\infty)}$ embodies the limit of the proper self-energy to $\omega\rightarrow\pm\infty$ and represents the mean field experienced by a particle in the correlated medium. It reduces to the Hartree-Fock-Bogoliubov (HFB) potential for a self-consistent first-order truncation of $\mathbf{\Sigma}^\star(\omega)$ but otherwise it includes additional in-medium corrections at higher orders. Hence, it is referred to as the \emph{correlated} HFB (cHFB) potential.

The components of the dynamic self-energy $\widetilde{\mathbf{\Sigma}}(\omega)$ also have a spectral representation analogous to Eqs.~\eqref{eq:Gkv_props}.
In this case, the poles of the Lehmann representation are associated to  intermediate state configurations (ISCs) combining different quasiparticle excitations $\{|\Psi_k\rangle;\omega_k\}$.  In order to write the most general form of the dynamic self-energy, a generic index $r$ is employed to label all possible ISCs that are eventually made explicit in Sec.~\ref{Sec:ADCn}.
Thus, the general form writes
\begin{widetext}
\begin{subequations}\label{eq:Sig_tild}
\begin{align}
  \widetilde{\Sigma}^{11}_{\alpha \beta} (\omega)={}& \sum_{r \, r'} \left\{         \mathcal{C}_{\alpha, r}                \left[ \frac 1{\omega\mathbb{I} - \mathcal{E} + i\eta} \right]_{r, r'}       \mathcal{C}^\dagger_{r', \beta}  
                                                                                                 ~+~ \bar{\mathcal{D}}^\dagger_{\alpha, r}  \left[ \frac 1{\omega\mathbb{I} + \mathcal{E}^T - i\eta} \right]_{r, r'}   \bar{\mathcal{D}}_{r', \beta} \right\} \, ,  \label{eq:Sig_tild11} \\
                                                             ~
  \widetilde{\Sigma}^{12}_{\alpha \beta} (\omega)={}& \sum_{r \, r'} \left\{         \mathcal{C}_{\alpha, r}                \left[ \frac 1{\omega\mathbb{I} - \mathcal{E} + i\eta} \right]_{r, r'}       \mathcal{D}^*_{r', \beta}  
                                                                                                 ~+~ \bar{\mathcal{D}}^\dagger_{\alpha, r}  \left[ \frac 1{\omega\mathbb{I} + \mathcal{E}^T - i\eta} \right]_{r, r'}   \bar{\mathcal{C}}^T_{r', \beta} \right\} \, ,  \label{eq:Sig_tild12} \\
                                                               ~                                  
  \widetilde{\Sigma}^{21}_{\alpha \beta} (\omega)={}& \sum_{r \, r'} \left\{         \mathcal{D}^T_{\alpha, r}                \left[ \frac 1{\omega\mathbb{I} - \mathcal{E} + i\eta} \right]_{r, r'}       \mathcal{C}^\dagger_{r', \beta}  
                                                                                                 ~+~ \bar{\mathcal{C}}^*_{\alpha, r}  \left[ \frac 1{\omega\mathbb{I} + \mathcal{E}^T - i\eta} \right]_{r, r'}   \bar{\mathcal{D}}_{r', \beta} \right\} \, ,  \label{eq:Sig_tild21} \\
                                                           ~
  \widetilde{\Sigma}^{22}_{\alpha \beta} (\omega)={}& \sum_{r \, `r'} \left\{         \mathcal{D}^T_{\alpha, r}                \left[ \frac 1{\omega\mathbb{I} - \mathcal{E} + i\eta} \right]_{r, r'}       \mathcal{D}^*_{r', \beta}  
                                                                                                 ~+~ \bar{\mathcal{C}}^*_{\alpha, r}  \left[ \frac 1{\omega\mathbb{I} + \mathcal{E}^T - i\eta} \right]_{r, r'}   \bar{\mathcal{C}}^T_{r', \beta} \right\} \, ,   \label{eq:Sig_tild22}
\end{align}
\end{subequations}
%CC%\begin{subequations}\label{eq:Sig_tild2}
%CC%\begin{align}
%CC%  \widetilde{\Sigma}^{11}_{\alpha \beta} (\omega)={}&  ~~\, \mathcal{C}  \left[ \frac 1{\omega\mathbb{I} - \mathcal{E} + i\eta} \right]  \mathcal{C}^\dagger ~+ \bar{\mathcal{D}}^\dagger  \left[ \frac 1{\omega\mathbb{I} + \mathcal{E}^T - i\eta} \right] \bar{\mathcal{D}} \, ,      \label{eq:Sig_tild2_11} \\
%CC%  \widetilde{\Sigma}^{11}_{\alpha \beta} (\omega)={}&  ~~\, \mathcal{C}  \left[ \frac 1{\omega\mathbb{I} - \mathcal{E} + i\eta} \right]  \mathcal{D}^*             + \bar{\mathcal{D}}^\dagger  \left[ \frac 1{\omega\mathbb{I} + \mathcal{E}^T - i\eta} \right] \bar{\mathcal{C}}^T \, ,  \label{eq:Sig_tild2_12} \\
%CC%  \widetilde{\Sigma}^{11}_{\alpha \beta} (\omega)={}&    \mathcal{D}^T   \left[ \frac 1{\omega\mathbb{I} - \mathcal{E} + i\eta} \right]  \mathcal{C}^\dagger ~+ \bar{\mathcal{C}}^*             \left[ \frac 1{\omega\mathbb{I} + \mathcal{E}^T - i\eta} \right] \bar{\mathcal{D}} \, ,     \label{eq:Sig_tild2_21} \\
%CC%  \widetilde{\Sigma}^{11}_{\alpha \beta} (\omega)={}&    \mathcal{D}^T   \left[ \frac 1{\omega\mathbb{I} - \mathcal{E} + i\eta} \right]  \mathcal{D}^*            + \bar{\mathcal{C}}^*              \left[ \frac 1{\omega\mathbb{I} + \mathcal{E}^T - i\eta} \right] \bar{\mathcal{C}}^T \, ,  \label{eq:Sig_tild2_22}
%CC%\end{align}
%CC%\end{subequations}
\end{widetext}
where $\mathcal{E}_{r, r'}$ denotes the elements of an energy matrix associated with an interaction among ISCs $r$ and $r'$. Matrix $\mathcal{E}$ is hermitian, so that $\mathcal{E}^T=\mathcal{E}^*$. 
The coupling matrices $\mathcal{C}$ and~$\mathcal{D}$ couple single-particle and ISC spaces, with the elements of the barred 
matrices defined as 
\begin{subequations}
\label{eq:def_CDbar}
\begin{align}
  \bar{\mathcal C}_{\alpha, r} ={}& ~~ \eta_\alpha  \mathcal{C}_{\widetilde{\alpha}, r} \, , \\
    \bar{\mathcal D}_{r, \alpha} ={}&   -  \eta_\alpha  \mathcal{D}_{r, \widetilde{\alpha}} \, .  
%%  \bar{\mathcal D}_{\alpha, r} ={}& -  \eta_\alpha  \mathcal{D}_{\widetilde{\alpha}, r} \, .
\end{align}
\end{subequations}

By exploiting the spectral representation of $\mathbf{G}(\omega)$ in Eq.~\eqref{eq:Gkv_eqs} and extracting each pole $\omega_k$ separately, Gorkov's equations can be transformed into
a set of energy-dependent eigenvalue equations for vectors ($ \mathcal{U}^k$,$\mathcal{V}^k$) of the form
\begin{widetext}
\begin{align}
   \omega_k  \left( \begin{array}{c} \mathcal{U}^k_\alpha  \\ ~ \\ \mathcal{V}^k_\alpha \end{array} \right) 
  ={}& \sum_\beta
 \left( \begin{array}{ccc}
              t_{\alpha \beta} - \mu \delta_{\alpha \beta} + \Sigma^{(\infty)\,11}_{\alpha \beta} + \widetilde{\Sigma}^{11}_{\alpha \beta}(\omega)        &&         \Sigma^{(\infty)\,12}_{\alpha \beta} + \widetilde{\Sigma}^{12}_{\alpha \beta}(\omega)  \\ ~\\
                          \Sigma^{(\infty)\,21}_{\alpha \beta} + \widetilde{\Sigma}^{21}_{\alpha \beta}(\omega)                             & & - t_{\alpha \beta} + \mu \delta_{\alpha \beta} + \Sigma^{(\infty)\,22}_{\alpha \beta} + \widetilde{\Sigma}^{22}_{\alpha \beta}(\omega)
 \end{array} \right) \, 
  \left( \begin{array}{c} \mathcal{U}^k_\beta  \\  ~ \\   \mathcal{V}^k_\beta \end{array} \right) \, .
\label{eq:Gkv_wdep}
\end{align}
%\end{widetext}

The eigenvalue problem of Eq.~\eqref{eq:Gkv_wdep} can be further optimised by introducing two new vectors $\mathcal{W}^k$ and $\bar{\mathcal{Z}}^k$ that belong to ISC space:
%\begin{widetext}
\begin{subequations}
\label{eq:def_WZ}
\begin{align}
   \mathcal{W}^k_r  \equiv {}& \sum_{r'} \left[\frac 1{\omega \mathbb{I} -  \mathcal{E}} \right]_{r, r'}   \left[  \sum_\alpha \left(       \mathcal{C}^*_{\alpha, r'} \, \mathcal{U}^k_\alpha +          \mathcal{D}^*_{r', \alpha}  \, \mathcal{V}^k_\alpha  \right) \right] \, , \\
   \bar{\mathcal{Z}}^k_r  \equiv {}& \sum_{r'} \left[\frac 1{\omega \mathbb{I} +  \mathcal{E}^T} \right]_{r, r'}   \left[  \sum_\alpha \left(  \bar{\mathcal{D}}_{r', \alpha}    \, \mathcal{U}^k_\alpha +  \bar{\mathcal{C}}_{\alpha, r'} \, \mathcal{V}^k_\alpha  \right) \right] \, .
\end{align}
\end{subequations}
Note that there will be a pair of such vectors for any quasiparticle solution $\omega_k$ from Eq.~\eqref{eq:Gkv_wdep}. In fact they can be interpreted as projections of a quasiparticle wave function  onto the subspace of ISCs.
Definitions~\eqref{eq:Sig_tild} and~\eqref{eq:def_WZ} allow to eventually re-expressed Gorkov's equations as a single (energy-independent) matrix diagonalization
\begin{align}
   \omega_k  \left( \begin{array}{c} \mathcal{U}^k \\ ~  \\  \mathcal{V}^k \\ ~ \\ \mathcal{W}^k \\ ~  \\  \bar{\mathcal{Z}}^k  \end{array} \right) 
  ={}&
   \left( \begin{array}{ccccccc}
              T - \mu \mathbb{I} + \Sigma^{(\infty)\,11}       &&         \Sigma^{(\infty)\,12}       &&          \mathcal{C}   &&    \bar{\mathcal{D}}^\dagger   \\ ~ \\
             \Sigma^{(\infty)\,21}            & & - T + \mu \mathbb{I} + \Sigma^{(\infty)\,22}      &&          \mathcal{D}^T   & &   \bar{\mathcal{C}}^*    \\ ~ \\
                       \mathcal{C}^\dagger         &&    \mathcal{D}^*      &&     \mathcal{E}     &   \\ ~ \\
                        \bar{\mathcal{D}}   &&     \bar{\mathcal{C}}^T  & &    &&   - \mathcal{E}^T \\
 \end{array} \right) \, 
  \left( \begin{array}{c}  \mathcal{U}^k  \\ ~ \\ \mathcal{V}^k  \\ ~ \\ \mathcal{W}^k  \\ ~ \\  \bar{\mathcal{Z}}^k \end{array}  \right) \, ,
\label{eq:ADC_mtx}
\end{align}
\end{widetext}
with normalisation condition
\begin{align}
  \sum_\alpha  | \mathcal{U}^k_\alpha  |^2  +  \sum_\alpha | \mathcal{V}^k_\alpha  |^2  +    \sum_r | \mathcal{W}^k_r  |^2  +   \sum_r  | \bar{\mathcal{Z}}^k_r  |^2  = 1 \, .
\label{eq:ADC_norm}
\end{align}
Although the  Gorkov matrix in Eq.~\eqref{eq:ADC_mtx} can have large dimensionality, the latter approach provides the entire quasiparticle spectrum in one single diagonalization. Often, this is by far the most efficient way to solve Dyson or Gorkov equations of many-body Green's function theory.  In practical applications, the computing time can be highly reduced by performing Krylov subspace projections without loosing details of the full spectral distribution~\cite{Barb2017LNP,Soma2014GkII}.

\section{\label{Sec:ADCn} Gorkov ADC($n$) }

The algebraic diagrammatic construction method is a systematic approach to construct accurate approximations to the self-energy. It is based on two fundamental requirements. 

First, the correct analytic form of the self-energy given by Eqs.~\eqref{eq:SE_split} and~\eqref{eq:Sig_tild} must be preserved. In particular, the Lehmann representation is pivotal as it follows from the causality principle, while the relations between poles and residues among different Nambu components are dictated by the Gorkov superfluid assumptions. Thus, the ADC($n$) scheme aims at directly formulating approximations to the energy-independent self-energy $\mathbf{\Sigma}^{(\infty)}$, as well as to the interaction and coupling matrices $\mathcal{E}$, $\mathcal{C}$ and $\mathcal{D}$.

Second, given the chosen truncation order $n$, all Feynman diagrams up to $n$-th order in the interaction $\Omega_I$ are required to be included. 

Thus, the general procedure to work out the Gorkov ADC($n$) approximation to the self-energy results from the following steps
\begin{enumerate}
\item Formally expand coupling and interaction matrices in powers of $\Omega_{I}$ and generate the associated expansion of the self-energy starting from Eq.~\ref{eq:Sig_tild},
\item Produce and evaluate algebraically all Feynman self-energy diagrams up to order $n$ in powers of $\Omega_{I}$,
\item Extract the algebraic expressions of the perturbative contributions to coupling and interaction matrices up to order $n$ by matching the form obtained in step 1. onto the explicit expressions of the diagrams generated in step 2.
\end{enumerate}
Such a procedure is explicated in Ref.~\cite{Barb2017LNP} and it has been applied to the standard SCGF theory up to the ADC(3) level~\cite{Raimondi2018adc3}. The very same steps are followed here to develop the ADC(3) approximation within the generalized frame of Gorkov SCGF theory. Below, the results obtained in step 3. are directly provided without making steps 1. and 2. explicit. While step 1. is easily adapted from Ref.~\cite{Barb2017LNP}, an example of step 2. is outlined in App.~\ref{App:Freq_ints} and results from a direct application of the topological and algebraic Feynman diagrammatic rules laid out in Ref.~\cite{Soma2011GkI}.

It is interesting to note that the straight summation of Feynman diagrams up to $n$-th order in $\Omega_I$--used to match the perturbative corrections to coupling and interaction matrices--does not satisfy the spectral representation~\eqref{eq:Sig_tild} in general. 
The ADC approach corrects this defect by implicitly including additional terms beyond order $n$. Moreover, it generates all-order resummations of infinite subsets of diagrams, starting with ladders and rings topologies at orders $n\geq3$. 
All together, the ADC($n$) expansion scheme provides a sequence of systematically improvable many-body approximations that is expected to converge toward the exact resummation of the full diagrammatic series.

In this work, the ADC($3$) truncation scheme with at most two-particle interactions is investigated within the Gorkov framework. Since Feynman diagrams of interest involve poles including at most three lines, the ISC space at play spans all quasiparticle triplets. Thus, the relevant collective ISC indices $r$ and $r'$ denote in the present case 
\begin{subequations}
\label{eq:r_indx}
\begin{align}
   r\; \equiv{}& (k_1, k_2, k_3)  \, , \\
   r' \equiv{}& (k_4, k_5, k_6)  \, .
\end{align}
\end{subequations}
The interaction matrix $\mathcal{E}$ splits into the energy $\mathcal{E}^{(\rm{0})}$ of uncorrelated excitations plus correction terms at first order in $\Omega_I$. Contrarily, $\mathcal{C}$ and $\mathcal{D}$ are interaction matrices between single quasiparticles and ISCs and do not carry zeroth-order contributions. Only contributions at first (I) and second (II) order are needed to build all third-order self-energy diagrams. Hence, for $n\leq3$, the expansion reads as
\begin{subequations}
\label{eq:CDE_exp}
\begin{align}
  \mathcal{E} \equiv {}& \mathcal{E}^{(\rm{0})} + \mathcal{E}^{(\rm{I})} \, ,  \\
  \mathcal{C} \equiv {}& \mathcal{C}^{(\rm{I})} + \mathcal{D}^{(\rm{II})} \, , \\
  \mathcal{D} \equiv {}& \mathcal{D}^{(\rm{I})} + \mathcal{D}^{(\rm{II})} \, .
\end{align}
\end{subequations}
The corresponding barred quantities are obtained from these amplitudes through Eqs.~\eqref{eq:def_CDbar}. While the Gorkov approach can be equivalently formulated without the aid of the dual basis  $\{|\bar{\alpha}\rangle\}$~\eqref{eq:dualA},  our formalism takes explicit advantage of it to handle time-reversal phases automatically. 

Because Gorkov objects with \emph{all} normal Nambu indices, e.g $G^{11}_{\alpha \beta}(\omega)$ or $\widetilde{\Sigma}^{11}_{\alpha \beta}(\omega)$, are not affected by time-reversal, they remain unchanged in the two approaches. Contrarily, the presence of
\emph{any} anomalous index as in $\Sigma^{\star\,12}_{\alpha \beta}\!(\omega)$ or $\Sigma^{\star\, 22}_{\alpha \beta}\!(\omega)$ implies differences due to the phases entering Eqs.~\eqref{eq:ccdag_bar}.
In the following, results are presented in terms of the amplitudes $\mathcal{U}$, $\bar{\mathcal V}$, $\mathcal{C}$ and $\bar{\mathcal D}$ entering the normal self-energy and propagators of the first group. Thus, the displayed equations can be used unchanged in a formulation that does not make distinction between barred and non-barred basis states.

\subsection{\label{Sec:adc_1_2}First- and second-order diagrams }

\begin{figure}[t]
  \centering
  \subfloat[(a)]{\label{Fig_SigInf_11}\includegraphics[scale=0.45]{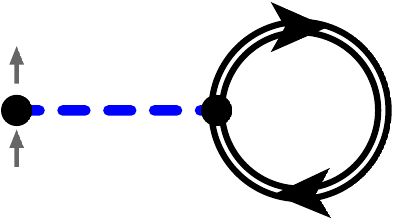}}
  \hspace{1.2cm}
  \subfloat[(b)]{\label{Fig_SigInf_12}\includegraphics[scale=0.45]{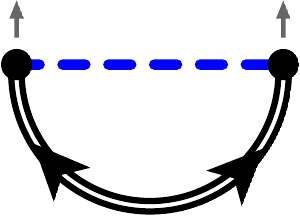}}
  \newline   \vskip .4cm
  \hspace{.5cm}
  \subfloat[(c)]{\label{Fig_SigInf_21}\includegraphics[scale=0.45]{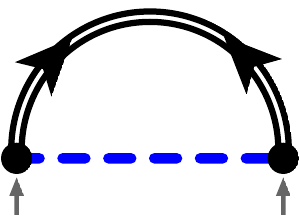}}
  \hspace{1.2cm}
  \subfloat[(d)]{\label{Fig_SigInf_22}\includegraphics[scale=0.45]{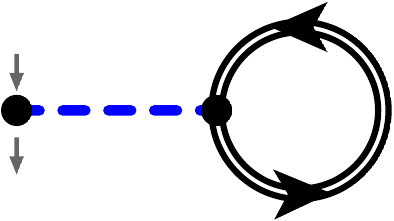}}
%  \newline
%  \includegraphics[width=\columnwidth]{Fig1_tmp2.pdf}
  \caption{Energy-independent {skeleton} diagrams defining the self-consistent contributions to the static self-energy $\mathbf{\Sigma}^{(\infty)}$, Eqs.~\eqref{eq:SigI_expr}. Dashed lines represent matrix elements of the two-particle interaction $V$, while double lines are correlated propagators from Eqs.~\eqref{eq:Gkv_props}.}
%  \label{M_N_3order_2N_2N}
\label{fig:SigI}
\end{figure} 

The Gorkov SCGF expressions for ADC(1) and ADC(2) provided in Ref.~\cite{Soma2011GkI} are presently recalled for completeness.

Only the four self-energy diagrams  depicted in Figure~\ref{fig:SigI} contribute at first order. They are depicted in terms of self-consistent interacting propagators (double lines) and contribute each to one of the four Nambu components of $\mathbf{\Sigma}^{(\infty)}$. The corresponding algebraic expressions are
\begin{subequations}\label{eq:SigI_expr}
\begin{align}
  \Sigma^{(\infty)\,11}_{\alpha \beta} ={}& \sum_{k \, \gamma \, \delta}    v_{\alpha \gamma, \beta \delta}\,  \bar{\mathcal V}^k_\delta{}^* \bar{\mathcal V}^k_\gamma = \sum_{ \gamma \, \delta}  v_{\alpha \gamma, \beta \delta} \,  \rho_{\delta \gamma}  \equiv  \Lambda_{\alpha \beta} \, , \label{eq:SigI_expr_11} \\
  \Sigma^{(\infty)\,12}_{\alpha \beta} ={}& \frac 1 2 \sum_{k \, \gamma \, \delta}    v_{\alpha \bar{\beta}, \gamma \bar{\delta}} \,  \bar{\mathcal V}^k_\gamma{}^* \bar{\mathcal U}^k_\delta = \frac 1 2 \sum_{ \gamma \, \delta}  v_{\alpha \bar{\beta}, \gamma \bar{\delta}} \, \tilde{\rho}_{ \gamma \delta}  \equiv  \tilde{h}_{\alpha \beta} \, , \label{eq:SigI_expr_12} \\
  \Sigma^{(\infty)\,21}_{\alpha \beta} ={}& \frac 1 2 \sum_{ \gamma \, \delta} \tilde{\rho}^*_{ \gamma \delta} \,  v_{\gamma \bar{\delta},  \beta \bar{\alpha}}  = \tilde{h}^*_{\beta \alpha }  \, , \label{eq:SigI_expr_21} \\
  \Sigma^{(\infty)\,22}_{\alpha \beta} ={}& - \sum_{ \gamma \, \delta}    v_{\bar{\beta}  \gamma, \bar{\alpha} \delta} \,  \rho_{\delta \gamma} = - \Sigma^{(\infty),11}_{\bar{\beta}  \bar{\alpha}} \,  =  -  \Lambda_{\bar{\beta}  \bar{\alpha}} \, , \label{eq:SigI_expr_22}
\end{align}
\end{subequations}
where matrices $\Lambda$ and $\tilde{h}$ denote normal and anomalous cHFB potentials.
Additional first-order diagrams arising from the $-U$ term in $\Omega_I$ cancel in the Gorkov Eqs.~\eqref{eq:Gkv_wdep} or~\eqref{eq:ADC_mtx}, as already discussed above, and do not need to be considered at any level. Any higher-order contribution to $\mathbf{\Sigma}^{(\infty)}$ relates to a \emph{composite}, i.e. non-skeleton, diagram. Thus, Eqs.~\eqref{eq:SigI_expr} completely defines the energy-independent self-energy of Gorkov SCGF theory.

\begin{figure}[t]
  \centering
  \subfloat[(a)]{\label{Fig_SigADC2_11a}\includegraphics[scale=0.45]{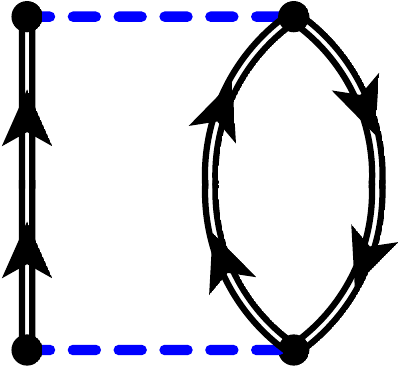}}
  \hspace{1.cm}
  \subfloat[(b)]{\label{Fig_SigADC2_11b}\includegraphics[scale=0.45]{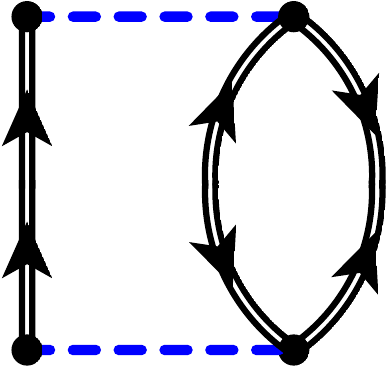}}
  \newline   \vskip .4cm
  \hspace{.5cm}
  \subfloat[(c)]{\label{Fig_SigADC2_21a}\includegraphics[scale=0.45]{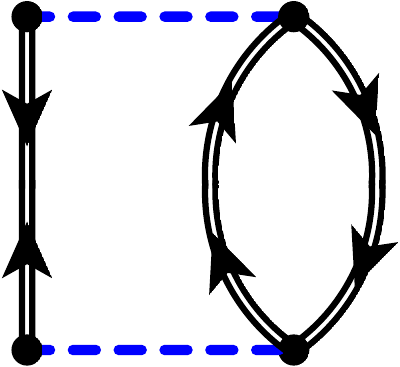}}
  \hspace{1.cm}
  \subfloat[(d)]{\label{Fig_SigADC2_21b}\includegraphics[scale=0.45]{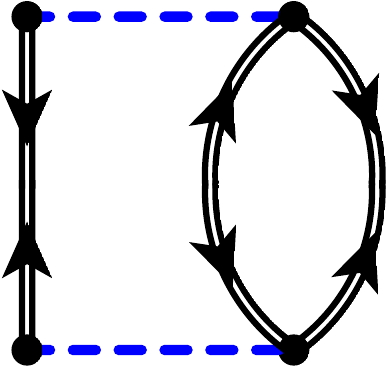}}
%  \newline
%  \includegraphics[width=\columnwidth]{Fig2_tmp2.pdf}
  \caption{Second-order skeleton diagrams contributing to the normal $\widetilde{\Sigma}^{11}(\omega)$ (a and b) and anomalous $\widetilde{\Sigma}^{21}(\omega)$ (c and d) self-energies. Similar diagrams apply to the remaining Nambu components, $\widetilde{\Sigma}^{12}(\omega)$ and $\widetilde{\Sigma}^{22}(\omega)$.}
\label{fig:SigIIab}
\end{figure}

Figure~\ref{fig:SigIIab} displays all second-order diagrams associated with $\widetilde\Sigma^{11}(\omega)$ and $\widetilde\Sigma^{12}(\omega)$. The corresponding diagrams for $\widetilde\Sigma^{21}(\omega)$ and $\widetilde\Sigma^{22}(\omega)$ are analogous. 
The algebraic derivation of these diagrams was performed in~Ref.~\cite{Soma2011GkI}. The sum of diagrams~\ref{Fig_SigADC2_11a} and~\ref{Fig_SigADC2_11b} reads as
\begin{align}
   &\widetilde{\Sigma}^{11\, \rm{\ref{Fig_SigADC2_11a}}}_{\alpha \beta}(\omega) + 
   \widetilde{\Sigma}^{11\, \rm{\ref{Fig_SigADC2_11b}}}_{\alpha \beta}(\omega) = \nonumber \\
%   \widetilde{\Sigma}^{11\, ( \ref{fFig_SigADC2_11a} + \ref{Fig_SigADC2_11b} ) }_{\alpha \beta}(\omega) \nonumber \\  %={}& 
                  &  \!\!\!\! \sum_{k_1 k_2 k_3}  \!\! \left\{   \frac { \mathcal{C}_{\alpha, k_1 k_2 k_3 }   \; \mathcal{C}_{\beta, k_1 k_2 k_3}^* }{\omega \!-\! \omega_{k_1} \!-\! \omega_{k_2}\!-\! \omega_{k_3}  \!+\! i\eta} 
                         \! + \! \frac{ \bar{\mathcal D}^*_{k_1 k_2 k_3, \alpha}  \; \bar{\mathcal D}_{k_1 k_2 k_3, \beta}}{\omega \!+ \! \omega_{k_1} \!+\! \omega_{k_2} \!+\! \omega_{k_3} \!- \! i\eta}
                                                                                                 \right\}  .
\label{eq:Sig11ab}
\end{align}

%\begin{figure}[t]
\begin{figure*}[th]
  \centering
  \includegraphics[scale=0.35]{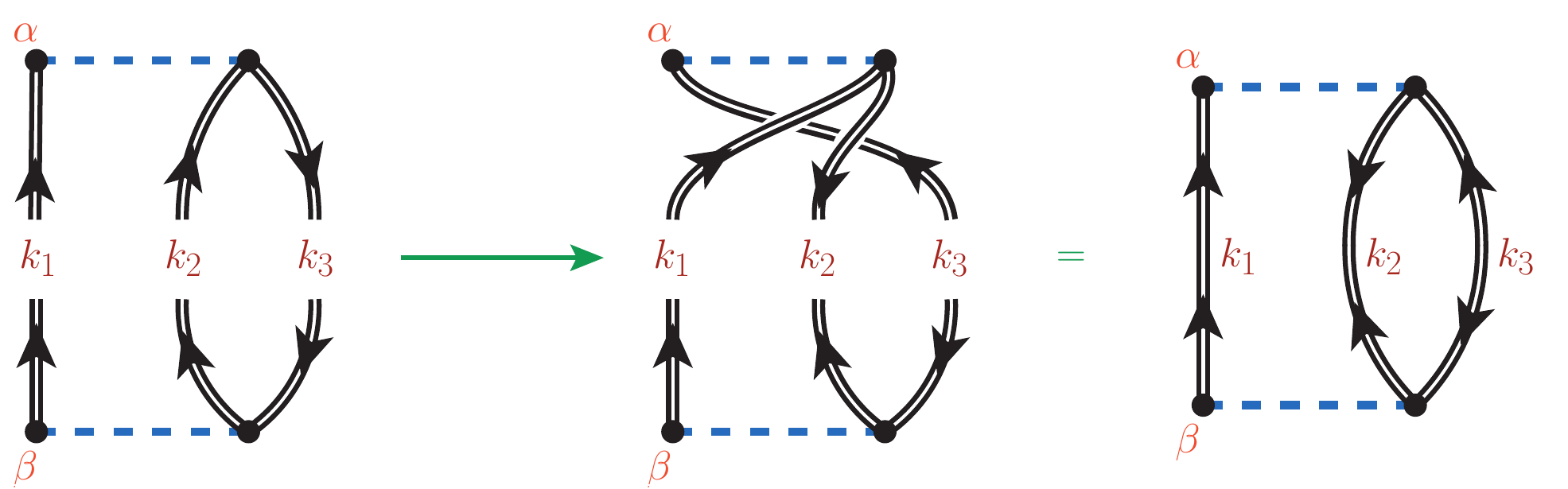}
  \caption{\emph{Left:}  The second-order diagram from Fig.~\ref{Fig_SigADC2_11a} has been separated to highlight the propagation of the ISCs with $r=(k_1,k_2,k_3)$. \emph{Center and right:} Performing a cyclic permutation of the intermediate quasiparticle states changes the topological structure of the diagram and transforms it into the contribution of Fig.~\ref{Fig_SigADC2_11b}.}
\label{fig:SigII_perm}
%\end{figure} 
\end{figure*}

By comparison with Eq.~\eqref{eq:Sig_tild11} and exploiting the expressions derived in~Ref.~\cite{Soma2011GkI}, the expressions of the interaction and coupling matrices at play in Gorkov ADC(2) are easily obtained as
\begin{subequations}\label{eq:ADC2}
\begin{align}
  \mathcal{C}^{[ADC(2)]}_{\alpha, r} \equiv{}&  \mathcal{C}^{(\rm{I})}_{\alpha, r}   = \!\frac 1{\sqrt{6}} \mathcal{P}_{123}  \! \sum_{\mu \nu \lambda} \! v_{\alpha \lambda, \mu \nu} \, \mathcal{U}^{k_1}_\mu \mathcal{U}^{k_2}_\nu \bar{\mathcal V}^{k_3}_{\lambda} , \label{eq:ADC2_C} \\
  \bar{\mathcal D}^{[ADC(2)]}_{r, \alpha} \equiv{}&   \bar{\mathcal D}^{(\rm{I})}_{r, \alpha} \!= \!\frac 1{\sqrt{6}} \mathcal{P}_{123} \! \sum_{\mu \nu \lambda}  \! \bar{\mathcal V}^{k_1}_\mu \bar{\mathcal V}^{k_2}_\nu \mathcal{U}^{k_3}_{\lambda}  v_{ \mu \nu, \alpha \lambda} \, , \label{eq:ADC2_D}  \\
  \mathcal{E}^{[ADC(2)]}_{r, r'} \equiv{}& \delta_{r,r'} \mathcal{E}^{(\rm{0})}_r = \rm{diag} \{ \omega_{k_1} + \omega_{k_2} + \omega_{k_3} \}  \, ,  \label{eq:ADC2_E}
\end{align}
\end{subequations}
where the cyclic permutation operator acting on a generic function of quasiparticle indices~$k_1$, $k_2$,\ldots reads
\begin{align}
 &\mathcal{P}_{ij\ell} \,  f(k_i, k_j, k_\ell)  \nonumber \\
    & \qquad  \equiv  f(k_i, k_j, k_\ell)+  f(k_j, k_\ell, k_i ) + f(k_\ell, k_i, k_j) \, .
\label{eq:Pijk_def}
\end{align}

Before moving to higher-order contributions, it is worth to make a few qualitative considerations on how different diagrams combine to yield the correct coupling matrices~$\mathcal{C}$ and~$\mathcal{D}$ and to impose Pauli antisymmetry.
Fig.~\ref{fig:SigII_perm} shows the diagram from Fig.~\ref{Fig_SigADC2_11a} split across the lines propagating the energy denominator $\mathcal{E}^{(0)}_{r=k_1,k_2,k_3}$. The upper half of the diagram gives the contribution 
\begin{equation}
 \sum_{\mu \nu \lambda} \! v_{\alpha \lambda, \mu \nu} \, \mathcal{U}^{k_1}_\mu \mathcal{U}^{k_2}_\nu \bar{\mathcal V}^{k_3}_{\lambda}
\end{equation}
to  matrix~$\mathcal{C}$. It is antisymmetric under the exchange of quasiparticles $k_1$ and $k_2$ by construction, due to the two-body matrix element of $V$, but not for other permutations.
The second diagram in Fig.~\ref{fig:SigII_perm} illustrates one of the cyclic permutations needed to achieve complete antisymmetry: rejoining the propagator lines after such a transformation delivers nothing but the other second-order diagram, 
Fig.~\ref{Fig_SigADC2_11b}. This demonstrates how the sum of both diagrams does provide Eq.~\eqref{eq:Sig11ab} with~$\mathcal{C}$ and~$\mathcal{D}$ satisfying the correct antisymmetry as in Eqs.~\eqref{eq:ADC2}.
Similar combinations of diagrams also appear in standard Dyson Green's function theory. However, these are less frequent due to topological constraints and to
the need to antisymmetrize the quasiparticles and quasiholes separately%
\footnote{To be specific, the first instances of the Dyson expansion in which different Feynman diagrams have to be grouped to satisfy Pauli antisymmetry are at third order if three-body forces are present~\cite{Raimondi2018adc3}. With just two-body interaction, this happens only at fourth order in perturbation theory.}.
In Gorkov theory, the presence of anomalous propagators permits the exchange of any pair of propagator lines so that the ISCs corresponding to each intermediate energy denominator are forcefully antisymmetrized with respect to all quasiparticle excitations.
The important consequence in seeking for proper approximations to the self-energy is that one always needs to group together specific sets of Feynman diagrams, related by exchanges of propagator lines.

 \begin{figure}[t]
  \centering
  \includegraphics[width=\columnwidth]{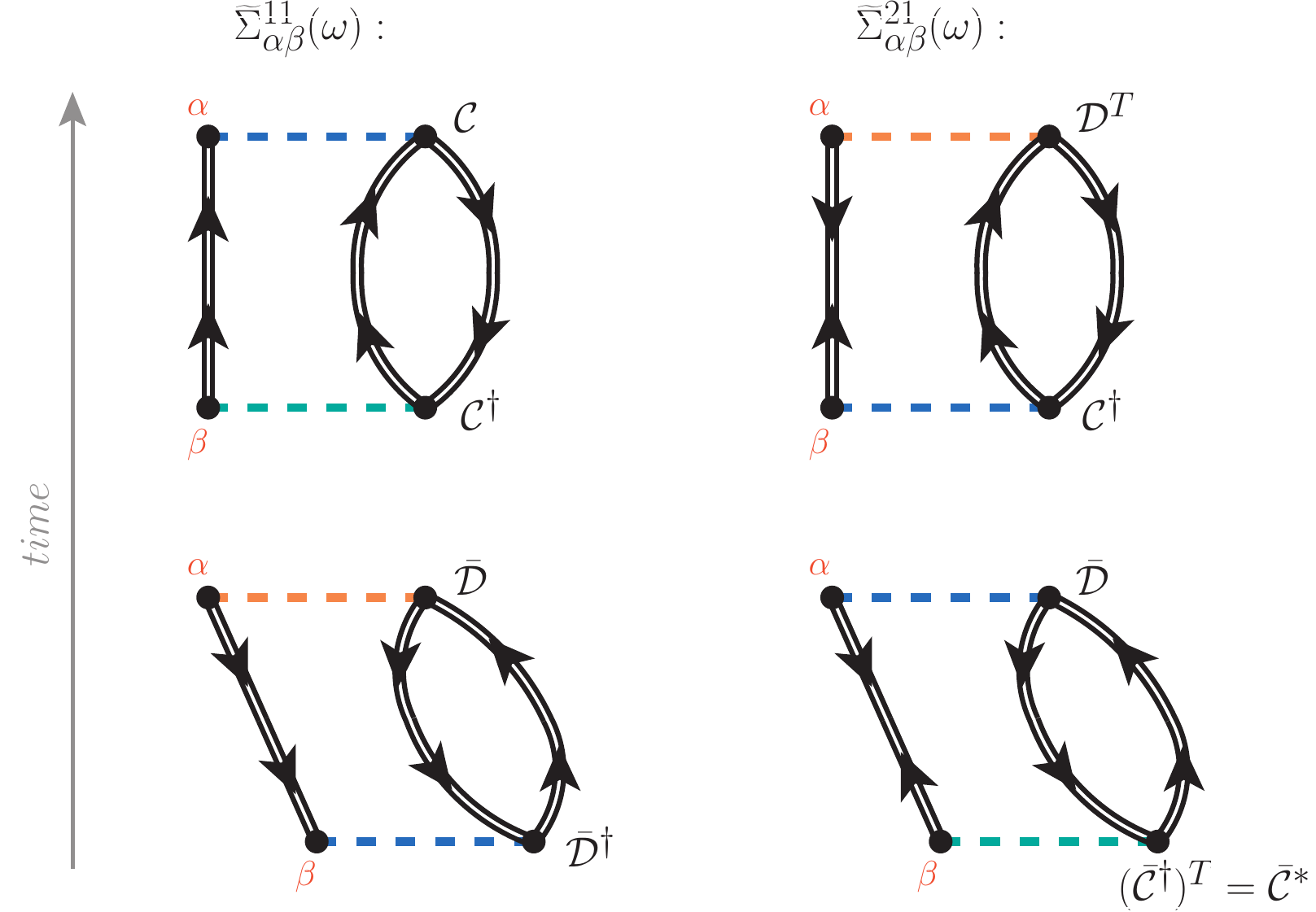}
  \caption{\emph{Left column:} The two time orderings through which the diagram of Fig.~\ref{Fig_SigADC2_11a} contributes to $\widetilde{\Sigma}^{11}(\omega)$. The top (bottom) diagram corresponds to the forward-going (backward-going) propagation. The matrices $\mathcal{C}$ and $\mathcal{D}$ to which a given vertex contributes are indicated next to it. \emph{Right column:} Analogous time orderings for the corresponding  contributions to $\widetilde{\Sigma}^{21}(\omega)$~(Fig.~\ref{Fig_SigADC2_11b}). 
  The $\mathcal{C}$~($\mathcal{D}$) topologies that contribute to the anomalous index of $\widetilde{\Sigma}^{21}(\omega)$ are highlighted with green~(orange) vertices. A comparison between the vertices on the left- and right-hand sides elucidates the occurrence of the same couplings $\mathcal{C}$ and $\mathcal{D}$ across all Eqs.~\eqref{eq:Sig_tild}.}
 \label{fig:SigII_CD}
\end{figure} 

Another consideration concerns how the same residues~$\mathcal{C}$ and~$\mathcal{D}$ arise in all Gorkov self-energies and follow the pattern shown in Eqs.~\eqref{eq:Sig_tild}. The first column of Fig.~\ref{fig:SigII_CD} depicts the possible time orientations of diagram~\ref{Fig_SigADC2_11a}, indicating the corresponding coupling matrices it contributes to. Contributions to matrix~$\mathcal{C}$ come from two upward-going lines  and a downward-going one ending into an interaction vertex, which results into the $v\mathcal{U}^2\mathcal{V}$ product in Eq.~\eqref{eq:ADC2_C}. At the entrance of the diagram the same structure 
is found but reverse and complex conjugate, leading to a contribution to $\mathcal{C}^\dagger$. Analogously, contributions from the backward going diagram have structure of type $v\mathcal{V}^2\mathcal{U}$ and lead to Eq.~\eqref{eq:ADC2_D}.
\begin{figure*}[t]
  \centering
  \subfloat[(a)]{\label{Fig_Aa}\includegraphics[scale=0.45]{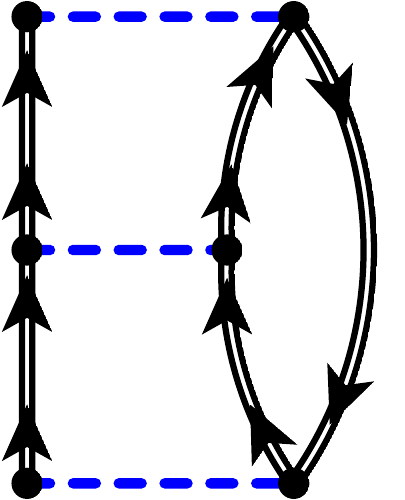}}
  \hspace{1.4cm}
  \subfloat[(b)]{\label{Fig_Ab}\includegraphics[scale=0.45]{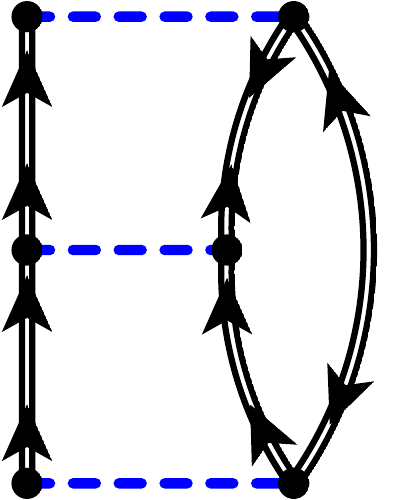}}
  \hspace{1.4cm}
  \subfloat[(c)]{\label{Fig_Ac}\includegraphics[scale=0.45]{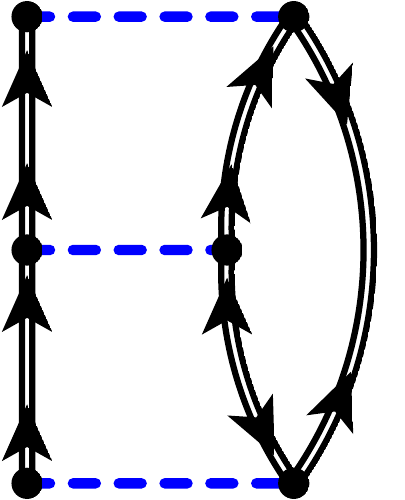}}
  \hspace{1.4cm}
  \subfloat[(d)]{\label{Fig_Ad}\includegraphics[scale=0.45]{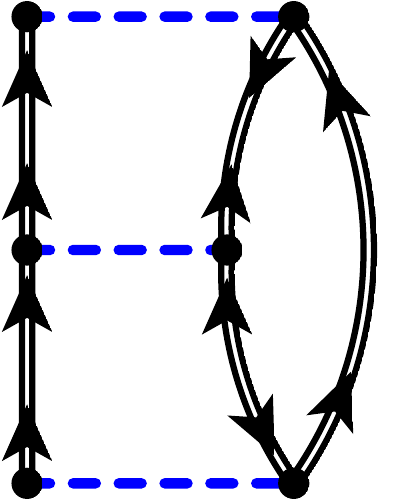}}
%  \newline  
%  \includegraphics[width=1.6\columnwidth]{Fig5_tmp2.pdf}
  \caption{Third-order skeleton diagrams corresponding to  $\widetilde{\Sigma}^{11}(\omega)$ with a particle-particle (pp) type intermediate interaction. The contributions to the other Nambu components of the self-energy with pp intermediate interactions originate from four analogous diagrams each, obtained by inverting one or both of the incoming and outgoing lines.}
\label{fig:ADC3_A}
\end{figure*} 
Whenever an anomalous self-energy is considered, one of the exit or entry lines has to be reversed, hence exchanging a $\mathcal{V}$ for a $\mathcal{U}$, which leads to inverting $\mathcal{C}$ with $\mathcal{D}$. This is shown in the second column of Fig.~\ref{fig:SigII_CD} for the corresponding contribution to $\widetilde{\Sigma}^{21}(\omega)$: 
%the top part of diagram \ref{fig:SigII_CD}c is exactly the same as the top part of~\ref{fig:SigII_CD}b, 
the top part of the upper-right diagram is exactly the same as the top part of the lower-left one, 
but it will enter as a transposed matrix in the Lehmann representation because it is an exit point of the self-energy in the first case and and entry point in the second. This property is general because the net  number of propagator (lines) flowing \emph{into} the interaction vertex is reversed exactly in the same way both for backward time propagation and for the inversion of a Nambu indices between normal to anomalous. It is easy to convince oneself that the same considerations apply to particle-number non-conserving interactions, as long as these are hermitian. Moreover, as for the case of quasiparticle antisymmetrization, the presence of anomalous propagators allows for any possible topological combination of lines and ensures that this correspondence is realised also for more complex diagrams, at any order in the Feynman expansion.
 Therefore, any portion of Feynman diagram contributing to a normal~(anomalous) \emph{forward} part of the self-energy will contribute identically to the \emph{backward} part of corresponding anomalous~(normal) case. It follows that exactly the same matrices~$\mathcal{C}$ and~$\mathcal{D}$ must appear in all four self-energies of Eqs.~\eqref{eq:Sig_tild}.
 
 The rigorous proof of this property is beyond the scope of the present work and is not elaborated on further. However, let us remind that relations~\eqref{eq:Sig_tild} naturally stem out from Nambu covariant theory of Ref.~\cite{Drissi2021ncpt}. In this case both the normal and anomalous contributions are embedded in a single propagator such that the~$\mathcal{C}$ and~$\mathcal{D}$ couplings are part of a unique coupling matrix.
For our purposes, we have verified by hand that Eqs.~\eqref{eq:Sig_tild} are satisfied by all diagrams discussed in the present work.

\subsection{\label{Sec:adc_3}Third-order skeleton diagrams }

Following the above discussion one concludes that it is sufficient to derive ADC($3$) expressions of the coupling and interaction matrices associated with one particular Gorkov self-energy. While the diagrams contributing to $\widetilde{\Sigma}^{11}(\omega)$ are presently employed, the other self-energies, Eqs.~(\ref{eq:Sig_tild12}-\ref{eq:Sig_tild22}) were checked to lead to the same results. 

\begin{figure*}[t!]
  \centering
  \subfloat[(a)]{\label{Fig_Ba}\includegraphics[scale=0.45]{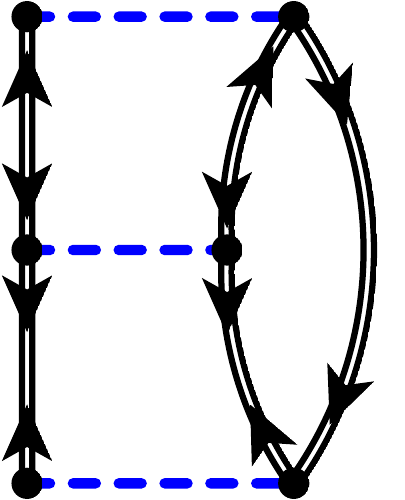}}
  \hspace{1.4cm}
  \subfloat[(b)]{\label{Fig_Bb}\includegraphics[scale=0.45]{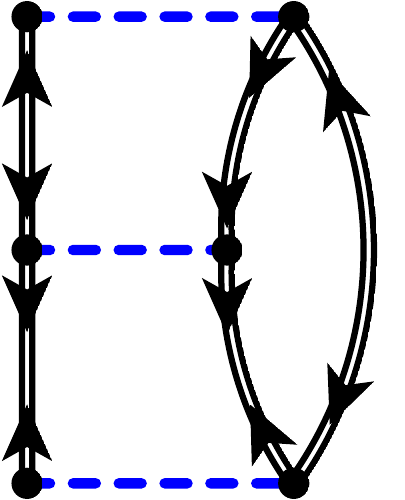}}
  \hspace{1.4cm}
  \subfloat[(c)]{\label{Fig_Bc}\includegraphics[scale=0.45]{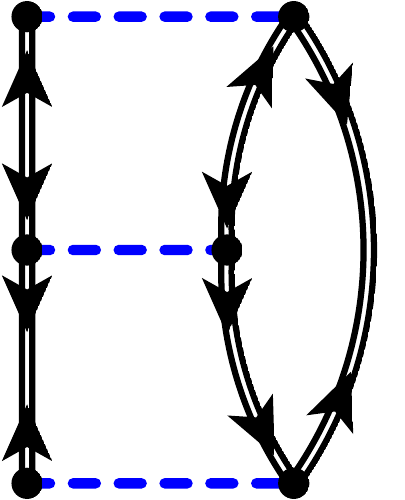}}
  \hspace{1.4cm}
  \subfloat[(d)]{\label{Fig_Bd}\includegraphics[scale=0.45]{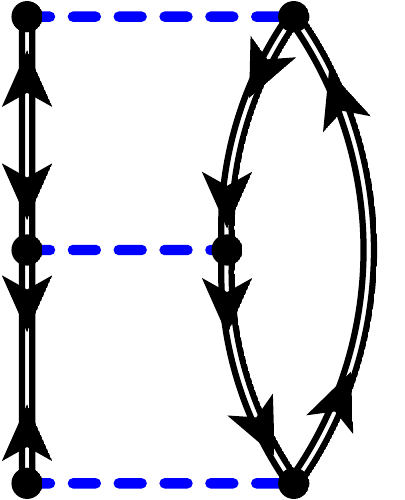}}
%  \newline  
%  \includegraphics[width=1.6\columnwidth]{Fig6_tmp2.pdf}
    \caption{Third-order skeleton diagrams contributing to $\widetilde{\Sigma}^{11}(\omega)$ with a hole-hole (hh) type intermediate interaction. Similarly to Fig.~\ref{fig:ADC3_A}, the contributions to the other Nambu components of the self-energy with hh intermediate interactions originate from four analogous diagrams each, obtained by inverting one or both of the incoming and outgoing lines.}
\label{fig:ADC3_B}
\end{figure*}

\begin{figure*}[t]
  \centering
  \subfloat[(a)]{\label{Fig_Ca}\includegraphics[scale=0.45]{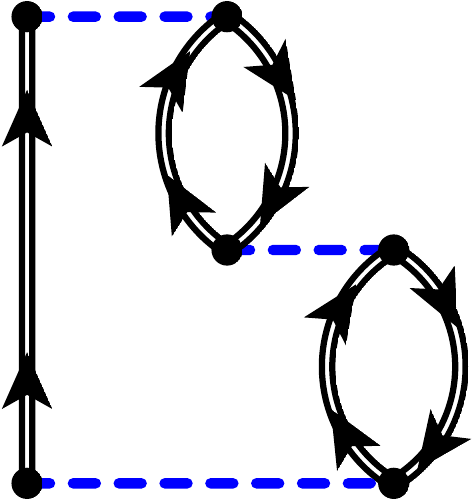}}
  \hspace{1.5cm}
  \subfloat[(b)]{\label{Fig_Cb}\includegraphics[scale=0.45]{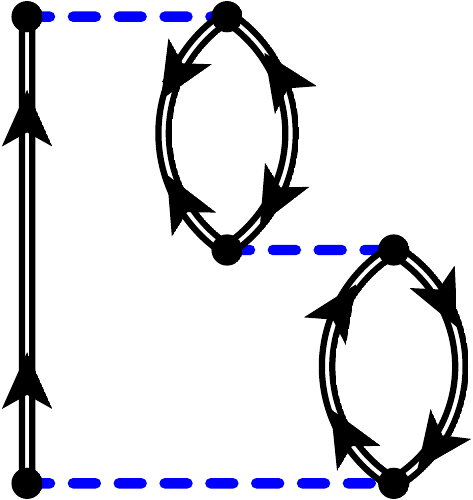}}
  \hspace{1.5cm}
  \subfloat[(c)]{\label{Fig_Cc}\includegraphics[scale=0.45]{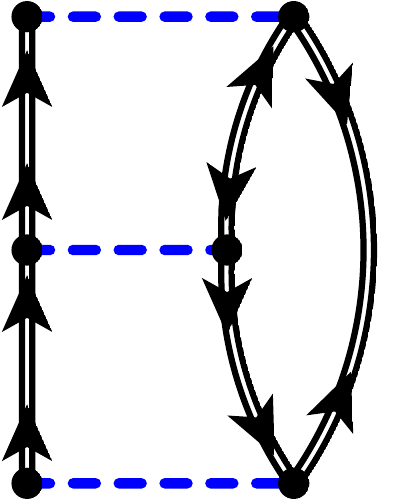}}
  \hspace{3.5cm}
  \newline   \vskip .7cm
  \hspace{1.cm}
  \subfloat[(d)]{\label{Fig_Cd}\includegraphics[scale=0.45]{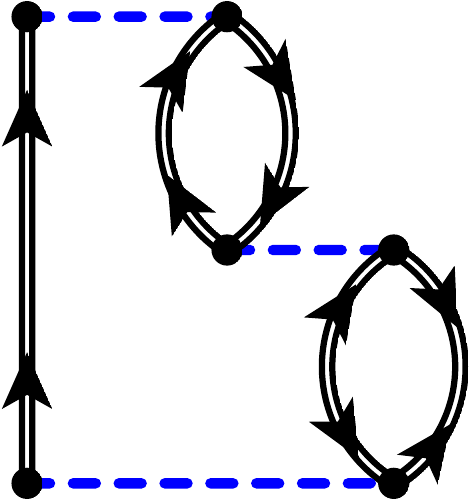}}
  \hspace{1.5cm}
  \subfloat[(e)]{\label{Fig_Ce}\includegraphics[scale=0.45]{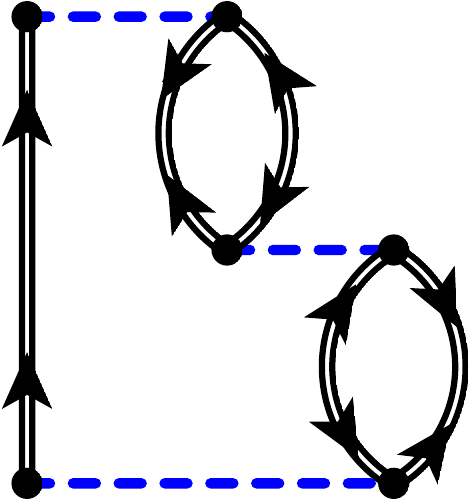}}
  \hspace{1.5cm}
  \subfloat[(f)]{\label{Fig_Cf}\includegraphics[scale=0.45]{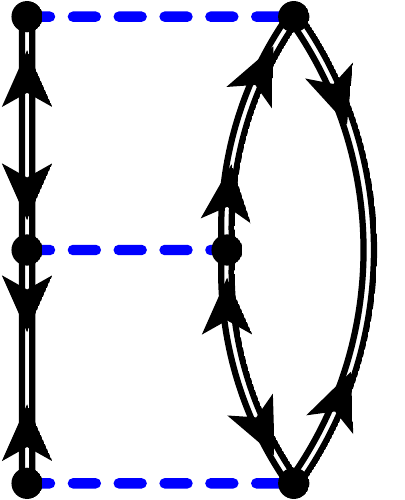}}
  \hspace{2.5cm}
  \newline   \vskip .7cm
  \hspace{1.5cm}
  \subfloat[(g)]{\label{Fig_Cg}\includegraphics[scale=0.45]{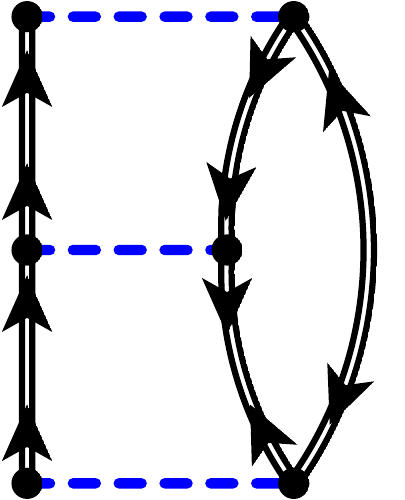}}
  \hspace{1.8cm}
  \subfloat[(h)]{\label{Fig_Ch}\includegraphics[scale=0.45]{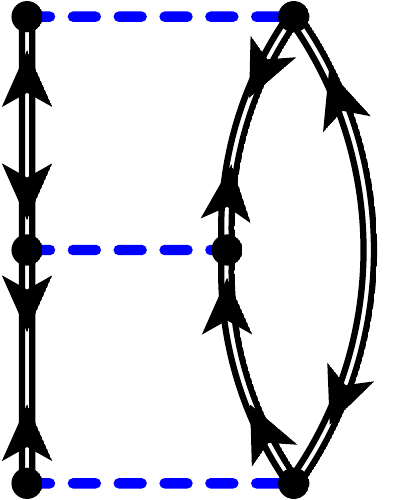}}
  \hspace{1.8cm}
  \subfloat[(i)]{\label{Fig_Ci}\includegraphics[scale=0.45]{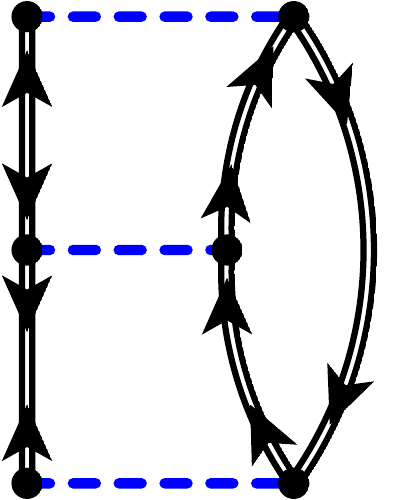}}
%  \newline
%  \includegraphics[width=1.6\columnwidth]{Fig7_tmp2.pdf}
  \caption{Third-order skeleton diagrams contributing to  $\widetilde{\Sigma}^{11}(\omega)$ with a particle-hole (ph) type intermediate interaction. Similarly to Figs.~\ref{fig:ADC3_A} and~Fig.~\ref{fig:ADC3_B},  the contributions to the other Nambu components of the self-energy with ph intermediate interactions originate from nine analogous diagrams each, obtained by inverting one or both of the incoming and outgoing lines.}
\label{fig:ADC3_C}
\end{figure*}

There exist 17 possible third-order skeleton diagrams that must be grouped in three classes on the basis of their connection through Pauli exchanges of propagator lines. 
These are depicted respectively in Figs.~\ref{fig:ADC3_A}, \ref{fig:ADC3_B} and~\ref{fig:ADC3_C}.
Each middle vertex in these diagrams acts as a seed for the all-orders Tamm-Dancoff  resummations generated by ADC(3). 

Diagram~\ref{Fig_Aa} is the diagram that makes two-particle and two-hole interact in the ISCs in the usual Dyson-ADC(3) formalism, respectively for forward and backward time propagation. Adding diagrams~\ref{Fig_Ab}, \ref{Fig_Ac} and~\ref{Fig_Ad} guarantees the antisymmetrization with respect to the third, non interacting quasiparticle. The frequency integrals needed to work out the algebraic expressions of these diagrams are discussed in App.~\ref{App:Freq_ints} and lead to the same contributions as in Eqs.~\eqref{eq:ADC2}, plus  second-order corrections to the coupling amplitudes and first-order correction to the energy matrix.

Let us first define the tensor
\begin{align}
  t^{k_3 k_4}_{k_1 k_2} \equiv{}& \sum_{\alpha \beta \gamma \delta} \frac{ \bar{\mathcal V}^{k_1}_{\alpha} \bar{\mathcal V}^{k_2}_{\beta}  \, v_{\alpha \beta, \gamma \delta} \, \mathcal{U}^{k_3}_\gamma \mathcal{U}^{k_4}_\delta }{- \left( \omega_{k_1} + \omega_{k_2}  + \omega_{k_3}  + \omega_{k_4} \right)}
\label{eq:BCC_t2}
\end{align}
that is closely related to the lowest-order double amplitude in Bogoliubov coupled cluster (BCC) theory~\cite{Signoracci2015BCC}.
Note that  BCC expressions are typically derived performing first the normal ordering of the Hamiltonian with respect to the Bogoliubov vacuum and expressing it in terms of Bogoliubov quasi-particle operators whereas the original matrix elements of $V$ appear in Eq.~\eqref{eq:BCC_t2}. In the special case of a HFB mean field, $\mathcal{U}$ and $\mathcal{V}$ amplitudes account for the normal ordering and $t^{k_3 k_4}_{k_1 k_2} $ does indeed reduce to the lowest order BCC double amplitude. Consequently, Eq.~\eqref{eq:BCC_t2} extends the concept of BCC amplitudes to account for the strength fragmentation of a dressed propagator.
With this tensor at hand, the contributions to the coupling amplitudes resulting from the diagrams displayed in Fig.~\ref{fig:ADC3_A} read as
\begin{subequations}\label{eq:3rd_CDpp}
\begin{align}
  \mathcal{C}^{(\rm{IIa})}_{\alpha, r} ={}& \!\frac 1{\sqrt{6}} \mathcal{P}_{123}  \! \sum_{\substack{\mu \,  \nu \,  \lambda \\ k_4 \, k_5}}  \frac{v_{\alpha \lambda, \mu \nu}}2 \left(\bar{\mathcal V}^{k_4}_{\mu} \bar{\mathcal V}^{k_5}_{\nu} \right)^{\!*} \! t^{k_1 k_2}_{k_4 k_5} \,  \bar{\mathcal V}^{k_3}_{\lambda}  , \label{eq:3rd_CIIa} \\
  \mathcal{C}^{(\rm{IIb})}_{\alpha, r} ={}& \!\frac 1{\sqrt{6}} \mathcal{P}_{123}  \! \sum_{\substack{\mu \,  \nu \,  \lambda \\ k_4 \, k_5}}   v_{\alpha \lambda, \mu \nu} \left(\bar{\mathcal V}^{k_4}_\nu \mathcal{U}^{k_5}_\lambda \right)^{\!*}  t^{k_1 k_2}_{k_4 k_5} \,  \mathcal{U}^{k_3}_\mu  , \label{eq:3rd_CIIb} \\
   \bar{\mathcal{D}}^{(\rm{IIa})}_{r, \alpha} ={}& \!\frac 1{\sqrt{6}} \mathcal{P}_{123} \! \sum_{\substack{\mu \,  \nu \,  \lambda \\ k_4 \, k_5}}   t^{k_4 k_5}_{k_1 k_2}  \,  \mathcal{U}^{k_3}_\lambda \left(  \mathcal{U}^{k_4}_\mu \mathcal{U}^{k_5}_\nu \right)^{\!*} \frac{ v_{ \mu \nu , \alpha \lambda} }2     , \label{eq:3rd_DIIa}   \\   
   \bar{\mathcal{D}}^{(\rm{IIb})}_{r, \alpha} ={}& \!\frac 1{\sqrt{6}} \mathcal{P}_{123} \! \sum_{\substack{\mu \,  \nu \,  \lambda \\ k_4 \, k_5}}   t^{k_4 k_5}_{k_1 k_2}  \,  \bar{\mathcal V}^{k_3}_\mu \left(  \mathcal{U}^{k_4}_\nu \bar{\mathcal V}^{k_5}_\lambda \right)^{\!*} \, v_{ \mu \nu , \alpha \lambda}     . \label{eq:3rd_DIIb}     
%CC%  \\ \nonumber ~  \\ \nonumber
%CC%   [\mathcal{D}^{(\rm{IIa})}]&{}^T_{\alpha, r} = \!\frac 1{\sqrt{6}} \mathcal{P}_{123} \! \sum_{\substack{\mu \,  \nu \,  \lambda \\ k_4 \, k_5}}  \frac{ v_{\alpha \lambda, \mu \nu} }2  \, \left(  \bar{\mathcal U}^{k_4}_\mu \bar{\mathcal U}^{k_5}_\nu \right)^{\!*}  \! t^{k_4 k_5}_{k_1 k_2}  \, \bar{\mathcal U}^{k_3}_\lambda
%CC%    \\ \nonumber   
%CC%     [\mathcal{D}^{(\rm{IIb})}]&{}^T_{\alpha, r} = \!\frac 1{\sqrt{6}} \mathcal{P}_{123} \! \sum_{\substack{\mu \,  \nu \,  \lambda \\ k_4 \, k_5}}  v_{\alpha \lambda, \mu \nu}  \left(  \bar{\mathcal U}^{k_4}_\nu  \mathcal{V}^{k_5}_\lambda  \right)^{\!*} \!  t^{k_4 k_5}_{k_1 k_2} \,   \mathcal{V}^{k_3}_\mu 
\end{align}
\end{subequations}
The first-order corrections to the energy matrix differ according to whether they refer to forward or backward poles of the self-energy, i.e. to the first or second term on the right-hand side of Eqs.~\eqref{eq:Sig_tild}, respectively,
\begin{align}
 \label{eq:3rd_EIa} 
  \mathcal{E}^{(\rm{Ia})}_{r,r'} ={}&   \left\{  \begin{array}{l}
        \frac 1{6} \mathcal{P}_{123}\mathcal{P}_{456} \left( \mathcal{E}^{(pp)}_{k_1 k_2, k_4 k_5} \, \delta_{k_3,k_6} \right)  \\ ~ \\   \qquad \qquad   \qquad  \qquad     \hbox{for forward poles,} \\
        ~\\~\\
        \frac 1{6} \mathcal{P}_{123}\mathcal{P}_{456} \left( \mathcal{E}^{(hh)}_{k_1 k_2, k_4 k_5} \, \delta_{k_3,k_6} \right)  \\ ~ \\   \qquad   \qquad  \qquad   \qquad    \hbox{for backward poles,}
    \end{array} \right.
\end{align}
where
\begin{align}
  \mathcal{E}^{(pp)}_{k_1 k_2, k_4 k_5} ={}& \sum_{\alpha \beta \gamma \delta}  \,( \mathcal{U}^{k_1}_\alpha \mathcal{U}^{k_2}_\beta )^*  \,    v_{\alpha \beta, \gamma \delta}  \,  \mathcal{U}^{k_4}_\gamma \mathcal{U}^{k_5}_\delta ,  \label{eq:3rd_Epp}  \\
  \mathcal{E}^{(hh)}_{k_1 k_2, k_4 k_5} ={}&  \sum_{\alpha \beta \gamma \delta}  \, \bar{\mathcal V}^{k_1}_\alpha \bar{\mathcal V}^{k_2}_\beta   \,    v_{\alpha \beta, \gamma \delta}  \,  ( \bar{\mathcal V}^{k_4}_\gamma \bar{\mathcal V}^{k_5}_\delta )^*  .  \label{eq:3rd_Ehh} 
\end{align}

The corresponding hh (pp) interaction contributions to the forward-going (backward-going) self-energies arise from the four diagrams in Fig.~\ref{fig:ADC3_B}. They are analogous to the diagrams of Fig.~\ref{fig:ADC3_A} except for inverting the orientation of all lines entering and leaving the intermediate interaction vertex. 
These diagrams lead to the following corrections to the  coupling amplitudes
\begin{subequations}\label{eq:3rd_CDhh}
\begin{align}
  \mathcal{C}^{(\rm{IIc})}_{\alpha, r} ={}& \!\frac 1{\sqrt{6}} \mathcal{P}_{123}  \! \sum_{\substack{\mu \,  \nu \,  \lambda \\ k_4 \, k_5}}  \frac{v_{\alpha \lambda, \mu \nu}}2 \left(\bar{\mathcal V}^{k_4}_{\mu} \bar{\mathcal V}^{k_5}_{\nu} \right)^{\!*} \! t^{k_4 k_5}_{k_1 k_2} \,  \bar{\mathcal V}^{k_3}_{\lambda}  , \label{eq:3rd_CIIc} \\
  \mathcal{C}^{(\rm{IId})}_{\alpha, r} ={}& \!\frac 1{\sqrt{6}} \mathcal{P}_{123}  \! \sum_{\substack{\mu \,  \nu \,  \lambda \\ k_4 \, k_5}}   v_{\alpha \lambda, \mu \nu} \left(\bar{\mathcal V}^{k_4}_\nu \mathcal{U}^{k_5}_\lambda \right)^{\!*}  t^{k_4 k_5}_{k_1 k_2} \,  \mathcal{U}^{k_3}_\mu  , \label{eq:3rd_CIId} \\
   \bar{\mathcal{D}}^{(\rm{IIc})}_{r, \alpha} ={}& \!\frac 1{\sqrt{6}} \mathcal{P}_{123} \! \sum_{\substack{\mu \,  \nu \,  \lambda \\ k_4 \, k_5}}   t^{k_1 k_2}_{k_4 k_5}  \,  \mathcal{U}^{k_3}_\lambda \left(  \mathcal{U}^{k_4}_\mu \mathcal{U}^{k_5}_\nu \right)^{\!*} \frac{ v_{ \mu \nu , \alpha \lambda} }2 , \label{eq:3rd_DIIc}   \\   
   \bar{\mathcal{D}}^{(\rm{IId})}_{r, \alpha} ={}&  \!\frac 1{\sqrt{6}} \mathcal{P}_{123} \! \sum_{\substack{\mu \,  \nu \,  \lambda \\ k_4 \, k_5}}   t^{k_1 k_2}_{k_4 k_5}  \,  \bar{\mathcal V}^{k_3}_\mu \left(  \mathcal{U}^{k_4}_\nu \bar{\mathcal V}^{k_5}_\lambda \right)^{\!*} \, v_{ \mu \nu , \alpha \lambda}  , \label{eq:3rd_DIId}   
%CC%  \\ \nonumber ~  \\ \nonumber
%CC%  [\mathcal{D}^{(\rm{IIc})}]&{}^T_{\alpha, r} =   \!\frac 1{\sqrt{6}} \mathcal{P}_{123} \! \sum_{\substack{\mu \,  \nu \,  \lambda \\ k_4 \, k_5}}  \frac{ v_{ \alpha \lambda,  \mu \nu} }2    \left(  \bar{\mathcal U}^{k_4}_\mu \bar{\mathcal U}^{k_5}_\nu \right)^{\!*} \! t^{k_1 k_2}_{k_4 k_5}  \,  \bar{\mathcal U}^{k_3}_\lambda 
%CC% \\  \nonumber    
%CC%  [\mathcal{D}^{(\rm{IId})}]&{}^T_{\alpha, r} =  \!\frac 1{\sqrt{6}} \mathcal{P}_{123} \! \sum_{\substack{\mu \,  \nu \,  \lambda \\ k_4 \, k_5}}    v_{ \alpha \lambda,  \mu \nu} \left(  \bar{\mathcal U}^{k_4}_\nu \mathcal{V}^{k_5}_\lambda \right)^{\!*}  t^{k_1 k_2}_{k_4 k_5}  \,  \mathcal{V}^{k_3}_\mu
\end{align}
\end{subequations}
whereas the corresponding  first-order corrections to the energy matrix are
\begin{align}
 \label{eq:3rd_EIb} 
  \mathcal{E}^{(\rm{Ib})}_{r,r'} ={}&  \left\{  \begin{array}{l}
        \frac 1{6} \mathcal{P}_{123}\mathcal{P}_{456} \left( \mathcal{E}^{(hh)}_{k_1 k_2, k_4 k_5} \, \delta_{k_3,k_6} \right)  \\ ~ \\   \qquad \qquad   \qquad  \qquad     \hbox{for forward poles,} \\
        ~\\~\\
        \frac 1{6} \mathcal{P}_{123}\mathcal{P}_{456} \left( \mathcal{E}^{(pp)}_{k_1 k_2, k_4 k_5} \, \delta_{k_3,k_6}\right)  \\ ~ \\   \qquad   \qquad  \qquad   \qquad    \hbox{for backward poles.}
    \end{array} \right.
\end{align}
The equivalence between the $\mathcal{E}$ and $\mathcal{E}^T$ denominators in Eqs.~\eqref{eq:Sig_tild} is restored only after adding Eqs.~\eqref{eq:3rd_EIa} and~\eqref{eq:3rd_EIb} together.   Hence, it is mandatory that diagrams in Figs.~\ref{fig:ADC3_A} and~\ref{fig:ADC3_B} are all computed together on the same footing. The topological relation between the two classes of diagrams, i.e. the inversion of lines in the intermediate interaction, is reflected into the fact that Eqs.~\eqref{eq:3rd_CDpp} and~\eqref{eq:3rd_CDhh} transform into each other under the exchange $t^{k_1 k_2}_{k_4 k_5} \leftrightarrow t^{k_4 k_5}_{k_1 k_2}$.  Inserting all contributions into Eqs.~\eqref{eq:Sig_tild} implies self-energy terms including mixed products of Eqs.~\eqref{eq:3rd_CDpp} and~\eqref{eq:3rd_CDhh}. These are rightful time orderings arising from fourth- and higher-order diagrams and therefore not depicted in Figs.~\ref{fig:ADC3_A},~\ref{fig:ADC3_B}  and~\ref{fig:ADC3_C}.

The remaining third-order skeleton diagrams involve a particle-hole type intermediate interaction and are displayed in  Fig.~\ref{fig:ADC3_C}. Performing the energy integral and making the antisymmetrization with respect to all ISC quasiparticle indices explicit through the use of the operator
\begin{align}
 &\mathcal{A}_{ij\ell} \,  f(k_i, k_j, k_\ell)  \nonumber \\
    & \qquad  \equiv  f(k_i, k_j, k_\ell)+  f(k_j, k_\ell, k_i) + f(k_\ell, k_i, k_j)  \nonumber \\
    & \qquad  \quad -  f(k_j, k_i, k_\ell) -  f(k_\ell, k_j, k_i) - f(k_i, k_\ell,  k_j)   \, ,
\label{eq:Aijk_def}
\end{align}
the nine diagrams of Fig.~\ref{fig:ADC3_C} introduce three additional terms to each coupling matrix
\begin{subequations}\label{eq:3rd_CDph}
\begin{align}
   \mathcal{C}^{(\rm{IIe})}_{\alpha, r} ={}& \!\frac 1{\sqrt{6}} \mathcal{A}_{123}  \! \sum_{\substack{\mu \,  \nu \,  \lambda \\ k_7 \, k_8}}   v_{\alpha \lambda, \mu \nu} \left(\bar{\mathcal V}^{k_7}_\nu \mathcal{U}^{k_8}_\lambda \right)^{\!*}  \mathcal{U}^{k_1}_\mu \, t^{k_8 k_2}_{k_7 k_3}    , \label{eq:3rd_CIIe} \\
   \mathcal{C}^{(\rm{IIf})}_{\alpha, r} ={}& \!\frac 1{\sqrt{6}} \mathcal{A}_{123}  \! \sum_{\substack{\mu \,  \nu \,  \lambda \\ k_7 \, k_8}}   v_{\alpha \lambda, \mu \nu} \left( \mathcal{U}^{k_7}_\lambda  \bar{\mathcal V}^{k_8}_\mu \right)^{\!*}  \mathcal{U}^{k_1}_\nu \, t^{k_8 k_2}_{k_7 k_3}   , \label{eq:3rd_CIIf} \\
   \mathcal{C}^{(\rm{IIg})}_{\alpha, r} ={}& \!\frac 1{\sqrt{6}} \mathcal{A}_{123}   \! \sum_{\substack{\mu \,  \nu \,  \lambda \\ k_7 \, k_8}}   v_{\alpha \lambda, \mu \nu} \left( \bar{\mathcal V}^{k_7}_\mu  \bar{\mathcal V}^{k_8}_\nu \right)^{\!*}  \bar{\mathcal V}^{k_1}_\lambda \, t^{k_8 k_2}_{k_7 k_3}   , \label{eq:3rd_CIIg} \\
    \bar{\mathcal{D}}^{(\rm{IIe})}_{r, \alpha} ={}& \!\frac 1{\sqrt{6}} \mathcal{A}_{123}  \! \sum_{\substack{\mu \,  \nu \,  \lambda \\ k_7 \, k_8}}     \bar{\mathcal V}^{k_1}_\nu   t^{k_2 k_8}_{k_3 k_7}   \left( \bar{\mathcal V}^{k_7}_\lambda  \mathcal{U}^{k_8}_\mu \right)^{\!*}     v_{ \mu \nu , \alpha \lambda}   , \label{eq:3rd_DIIe} \\
    \bar{\mathcal{D}}^{(\rm{IIf})}_{r, \alpha} ={}&  \!\frac 1{\sqrt{6}} \mathcal{A}_{123}  \! \sum_{\substack{\mu \,  \nu \,  \lambda \\ k_7 \, k_8}}    \bar{\mathcal V}^{k_1}_\mu   t^{k_2 k_8}_{k_3 k_7}   \left( \mathcal{U}^{k_7}_\nu \bar{\mathcal V}^{k_8}_\lambda \right)^{\!*}     v_{ \mu \nu , \alpha \lambda}   , \label{eq:3rd_DIIf} \\
    \bar{\mathcal{D}}^{(\rm{IIg})}_{r, \alpha} ={}&  \!\frac 1{\sqrt{6}} \mathcal{A}_{123}  \! \sum_{\substack{\mu \,  \nu \,  \lambda \\ k_7 \, k_8}}  \mathcal{U}^{k_1}_\lambda   t^{k_2 k_8}_{k_3 k_7}   \left( \mathcal{U}^{k_7}_\mu \mathcal{U}^{k_8}_\nu \right)^{\!*}   v_{ \mu \nu , \alpha \lambda}    , \label{eq:3rd_DIIg} 
%
%CC%  \\ \nonumber ~  \\ \nonumber
%CC%    [\mathcal{D}^{(\rm{IIe})}]&{}^T_{\alpha, r} = \!\frac 1{\sqrt{6}} \mathcal{A}_{123}  \! \sum_{\substack{\mu \,  \nu \,  \lambda \\ k_7 \, k_8}}   v_{\alpha \lambda, \mu \nu}  \left( \mathcal{V}^{k_7}_\lambda  \bar{\mathcal U}^{k_8}_\mu \right)^{\!*}  \,   \mathcal{V}^{k_1}_\nu  \,  t^{k_2 k_8}_{k_3 k_7}
%CC%      \\  \nonumber    
%CC%    [\mathcal{D}^{(\rm{IIf})}]&{}^T_{\alpha, r} =  \!\frac 1{\sqrt{6}} \mathcal{A}_{123}  \! \sum_{\substack{\mu \,  \nu \,  \lambda \\ k_7 \, k_8}}   v_{\alpha \lambda, \mu \nu}  \left( \bar{\mathcal U}^{k_7}_\nu    \mathcal{V}^{k_8}_\lambda \right)^{\!*} \,   \mathcal{V}^{k_1}_\mu \,  t^{k_2 k_8}_{k_3 k_7}
%CC%  \\  \nonumber    
%CC%    [\mathcal{D}^{(\rm{IIg})}]&{}^T_{\alpha, r} = \!\frac 1{\sqrt{6}} \mathcal{A}_{123}  \! \sum_{\substack{\mu \,  \nu \,  \lambda \\ k_7 \, k_8}}   v_{\alpha \lambda, \mu \nu}    \left( \bar{\mathcal U}^{k_7}_\mu \bar{\mathcal U}^{k_8}_\nu \right)^{\!*}     \bar{\mathcal U}^{k_1}_\lambda\,  t^{k_2 k_8}_{k_3 k_7}
\end{align}
\end{subequations}
whereas the particle-hole contribution to the ISC energy interaction matrix is given by
\begin{align}
  \mathcal{E}^{(\rm{Ic})}_{r,r'} ={}& \frac 1{6} \mathcal{A}_{123} \mathcal{A}_{456}  \left( \delta_{k_1,k_4} \, \mathcal{E}^{(ph)}_{k_2 k_3 , k_5 k_6} \right) 
\label{eq:3rd_EIc} 
\end{align}
with
\begin{align}
  \mathcal{E}^{(ph)}_{k_2 k_3 , k_5 k_6} ={}& \sum_{\alpha \beta \gamma \delta}  \,( \mathcal{U}^{k_2}_\alpha \bar{\mathcal V}^{k_3}_\beta )^*  \,    v_{\alpha \delta, \beta \gamma}  \,  \mathcal{U}^{k_5}_\gamma \bar{\mathcal V}^{k_6}_\delta \, .   \label{eq:3rd_Eph} 
\end{align}

\subsection{\label{Sec:Comb_diags}Non-skeleton contributions}

Sections~\ref{Sec:adc_1_2} and~\ref{Sec:adc_3} exhaust all the diagrams that enter \emph{fully self-consistent} computations up to ADC(3). In this case, the self-energy is purely a functional of the fully dressed propagator, $\mathbf{G}(\omega)$, and all above equations are expressed in terms of its spectroscopic amplitudes and poles, Eqs.~\eqref{eq:def_UV},~\eqref{eq:def_UVbar} and~\eqref{eq:wk}. If, instead, the many-body expansion is based on the unperturbed reference propagator $\mathbf{G}^{(0)}(\omega)$ additional \emph{composite}, i.e. non-skeleton, diagrams need to be included. Thus, the present section along with App.~\ref{App:SigInf_3rd_ord} introduce all remaining composite diagrams up to third order.

The unperturbed propagator~\eqref{eq:G0} has a spectral representation analogous to Eqs.~\eqref{eq:Gkv_props}
\begin{subequations}
\label{eq:G0_props_Lehman}
\begin{align}
  G^{(0) 11}_{\alpha \beta}(\omega) ={}& \sum_k \left\{   \frac{        \hfb{U}^k_\alpha    \; \;        \hfb{U}^k_\beta{}^*     }{\omega - \hfe{k} + i\eta} +    \frac{  \bar{\hfb{V}}^k_\alpha{}^*        \;\,  \bar{\hfb{V}}^k_\beta }{\omega + \hfe{k} - i\eta}     \right\}  \, ,  \\
  G^{(0) 12}_{\alpha \beta}(\omega) ={}& \sum_k \left\{   \frac{        \hfb{U}^k_\alpha    \; \;        \hfb{V}^k_\beta{}^*     }{\omega - \hfe{k} + i\eta} +    \frac{  \bar{\hfb{V}}^k_\alpha{}^*        \;\,  \bar{\hfb{U}}^k_\beta }{\omega + \hfe{k} - i\eta}     \right\}  \, ,  \\
  G^{(0) 21}_{\alpha \beta}(\omega) ={}& \sum_k \left\{   \frac{        \hfb{V}^k_\alpha    \; \;        \hfb{U}^k_\beta{}^*     }{\omega - \hfe{k} + i\eta} +    \frac{  \bar{\hfb{U}}^k_\alpha{}^*        \;    \bar{\hfb{V}}^k_\beta }{\omega + \hfe{k} - i\eta}     \right\}  \, , \\
  G^{(0) 22}_{\alpha \beta}(\omega) ={}& \sum_k \left\{   \frac{        \hfb{V}^k_\alpha    \; \;        \hfb{V}^k_\beta{}^*     }{\omega - \hfe{k} + i\eta} +     \frac{  \bar{\hfb{U}}^k_\alpha{}^*       \;\,  \bar{\hfb{U}}^k_\beta }{\omega + \hfe{k} - i\eta}     \right\}  \, , 
\end{align}
\end{subequations}
where we used the notation $\hfe{k}$, $\hfb{U}^k$ and $\hfb{V}^k$ to stress that these are not correlated spectroscopic quantities but unperturbed ones. For the present purpose, these are the solution of the HFB eigenvalue problem associated with $\Omega_U$
\begin{align}
  \sum_\beta & \!\!
  \left( \begin{array}{ccc}      t_{\alpha \beta} \!+ \!u_{\alpha \beta} \!- \!\mu\delta_{\alpha \beta}     &\,&         u^{an.}_{\alpha \bar\beta}  \\ ~\\
                  - (u^{an.}_{\bar\alpha \beta})^*   & &\!\! - t_{\bar\beta \bar\alpha} \! -\! u_{\bar\beta \bar\alpha}\! + \! \mu\delta_{\alpha \beta}    \end{array} \right) \!\!
  \left( \begin{array}{c} \hfb{U}^{k}_\beta  \\ ~ \\ \hfb{V}^{k}_\beta ~ \end{array} \right) \nonumber \\
  & \qquad \quad= \hfe{k}   \left( \begin{array}{c} \hfb{U}^{k}_\alpha  \\ ~ \\ \hfb{V}^{k}_\alpha ~ \end{array} \right) \, .
\label{eq:OmU_HFB_eqs}
\end{align}
Since the composite diagrams discussed in this section assume a HFB reference state, their contributions to ADC interactions and amplitudes are expressed in terms of the unperturbed state generated by Eq.~\eqref{eq:OmU_HFB_eqs}.

\subsubsection{\label{Sec:Comb_cHFB}Static self-energy}

The composite diagrams contributing to $\mathbf{\Sigma}^{(\infty)}_{\alpha \beta}(\omega)$ can be obtained by expanding  Gorkov Eq.~\eqref{eq:Gkv_eqs} up to second order and by inserting the results into the diagrams of Fig.~\ref{fig:SigI}. The resulting equations for the static self-energies are rather cumbersome and are detailed in App.~\ref{App:SigInf_3rd_ord}.  However, these are not needed in the vast majority of applications since their self-consistent counter part, Eqs.~\eqref{eq:SigI_expr}, is easier to compute and contains all of them implicitly. 

\subsubsection{\label{Sec:Comb_3rd}Third-order terms}

\begin{figure*}[t]
  \centering
  \subfloat[(a)]{\label{Fig_Ua}\includegraphics[scale=0.23]{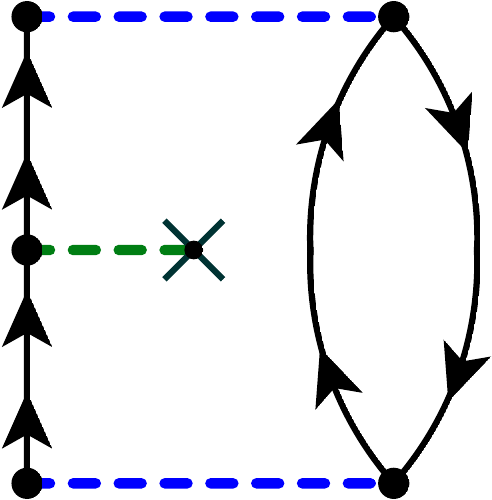}}
  \hspace{1.15cm}
  \subfloat[(b)]{\label{Fig_Ub}\includegraphics[scale=0.23]{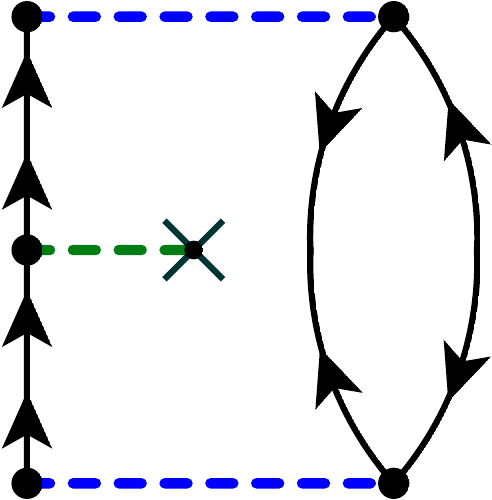}}
  \hspace{1.15cm}
  \subfloat[(c)]{\label{Fig_Uc}\includegraphics[scale=0.23]{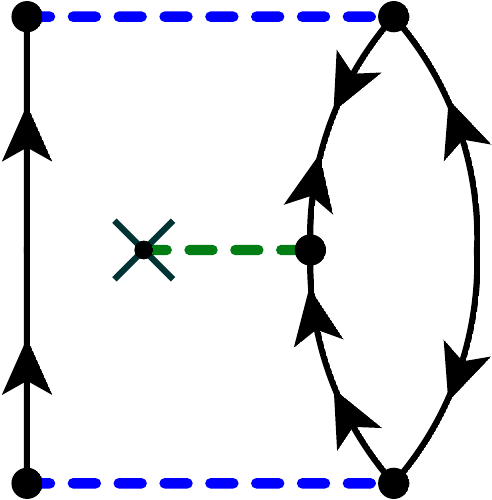}}
  \hspace{1.15cm}
  \subfloat[(d)]{\label{Fig_Ud}\includegraphics[scale=0.23]{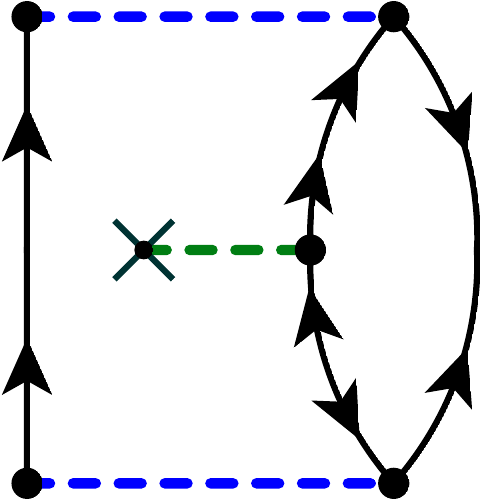}}
  \hspace{1.15cm}
  \subfloat[(e)]{\label{Fig_Ue}\includegraphics[scale=0.23]{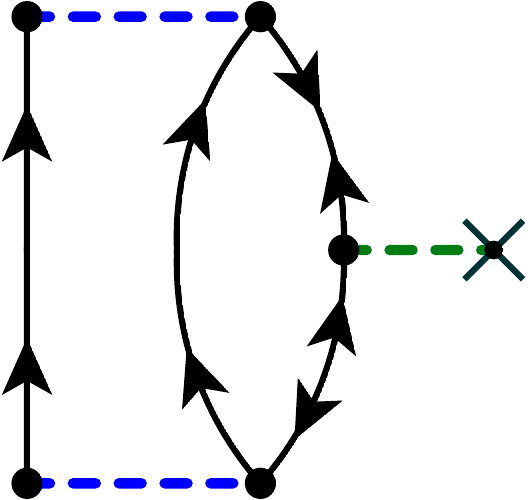}}
%\hspace{3.5cm}
  \newline   \vskip .5cm
  \hspace{.3cm}
  \subfloat[(f)]{\label{Fig_Uf}\includegraphics[scale=0.23]{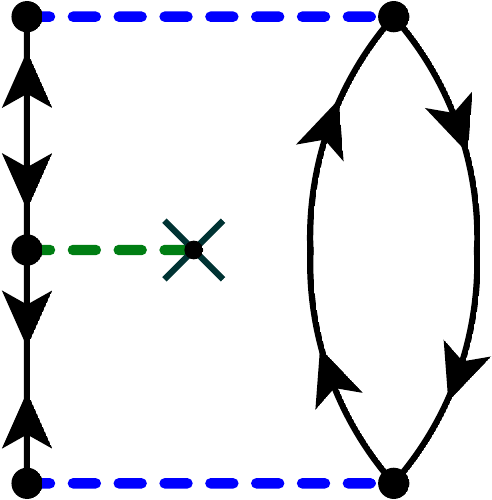}}
  \hspace{1.15cm}
  \subfloat[(g)]{\label{Fig_Ug}\includegraphics[scale=0.23]{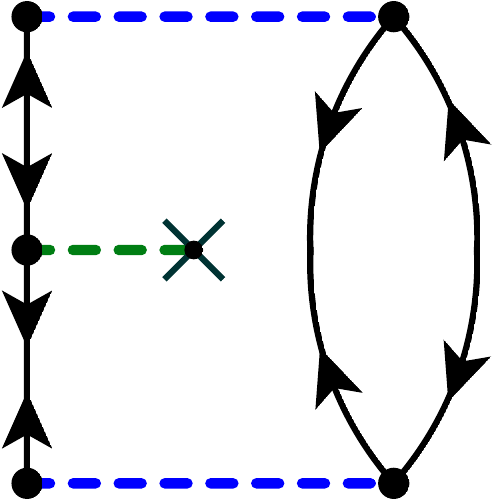}}
  \hspace{1.15cm}
  \subfloat[(h)]{\label{Fig_Uh}\includegraphics[scale=0.23]{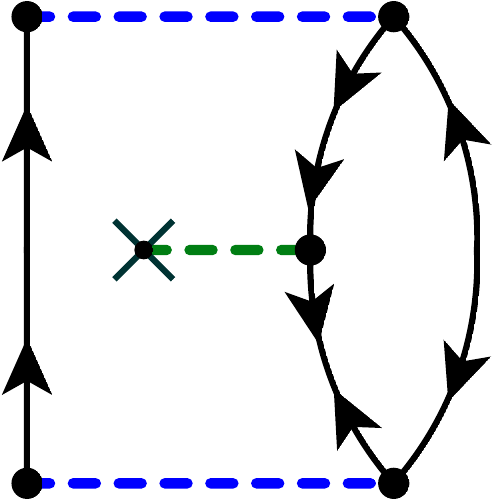}}
  \hspace{1.15cm}
  \subfloat[(i)]{\label{Fig_Ui}\includegraphics[scale=0.23]{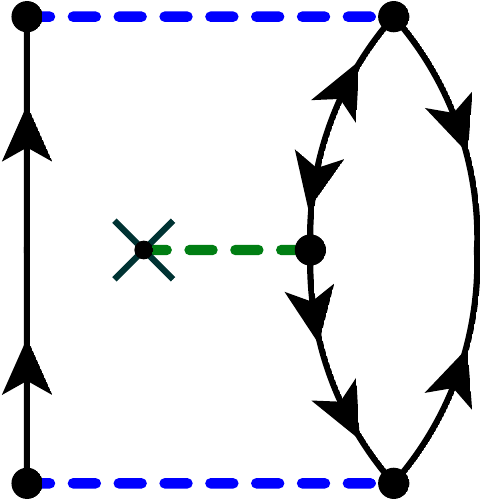}}
  \hspace{1.15cm}
  \subfloat[(j)]{\label{Fig_Uj}\includegraphics[scale=0.23]{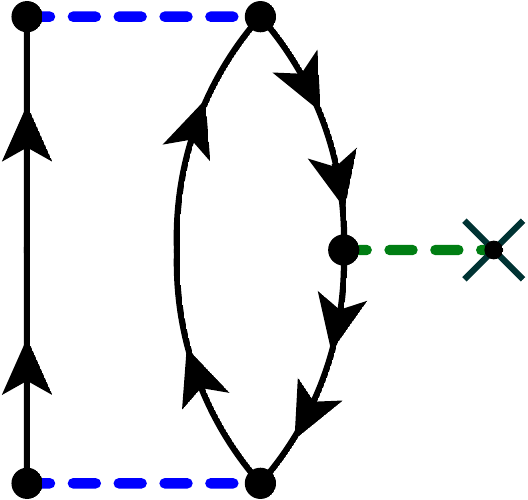}}
%\hspace{3.5cm}
  \newline   \vskip .5cm
  \hspace{.6cm}
  \subfloat[(k)]{\label{Fig_Uk}\includegraphics[scale=0.23]{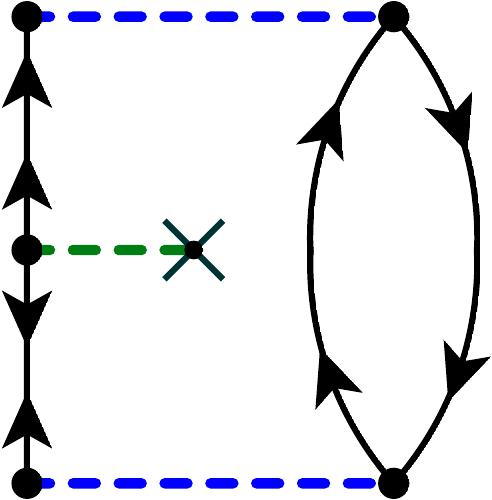}}
  \hspace{1.15cm}
  \subfloat[(l)]{\label{Fig_Ul}\includegraphics[scale=0.23]{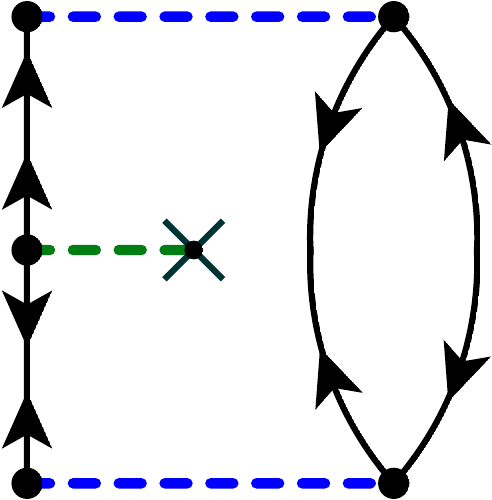}}
  \hspace{1.15cm}
  \subfloat[(m)]{\label{Fig_Um}\includegraphics[scale=0.23]{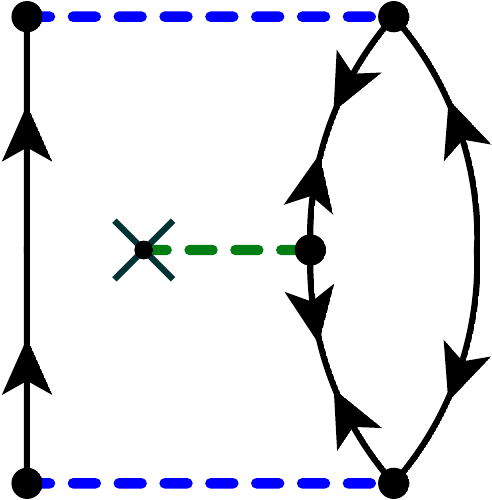}}
  \hspace{1.15cm}
  \subfloat[(n)]{\label{Fig_Un}\includegraphics[scale=0.23]{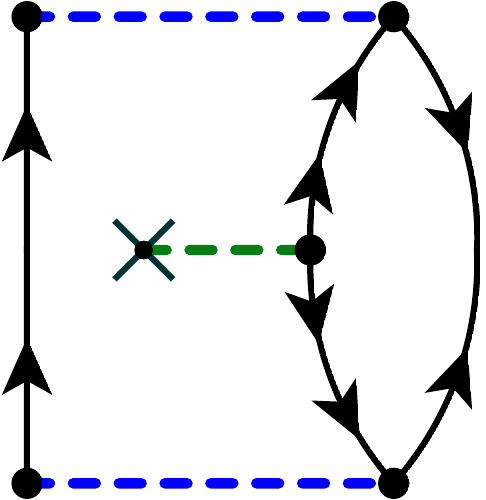}}
  \hspace{1.15cm}
  \subfloat[(o)]{\label{Fig_Uo}\includegraphics[scale=0.23]{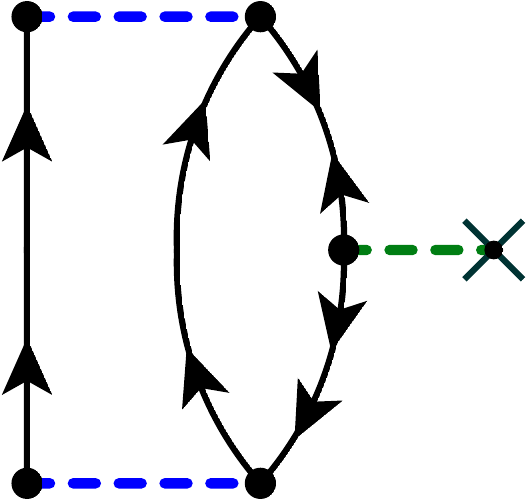}}
%\hspace{3.5cm}
  \newline   \vskip .5cm
  \hspace{.3cm}
  \subfloat[(p)]{\label{Fig_Up}\includegraphics[scale=0.23]{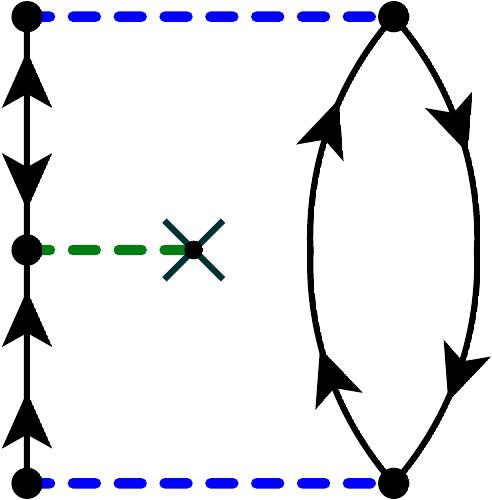}}
  \hspace{1.15cm}
  \subfloat[(q)]{\label{Fig_Uq}\includegraphics[scale=0.23]{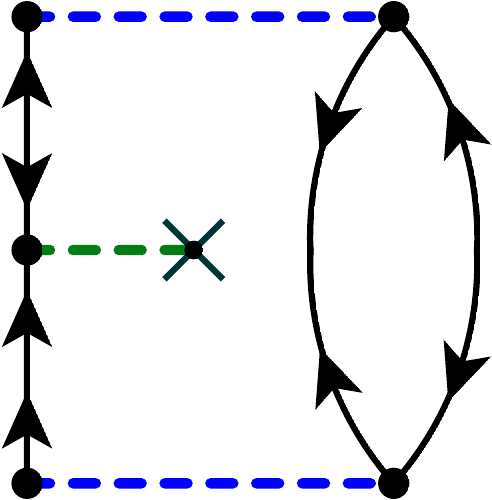}}
  \hspace{1.15cm}
  \subfloat[(r)]{\label{Fig_Ur}\includegraphics[scale=0.23]{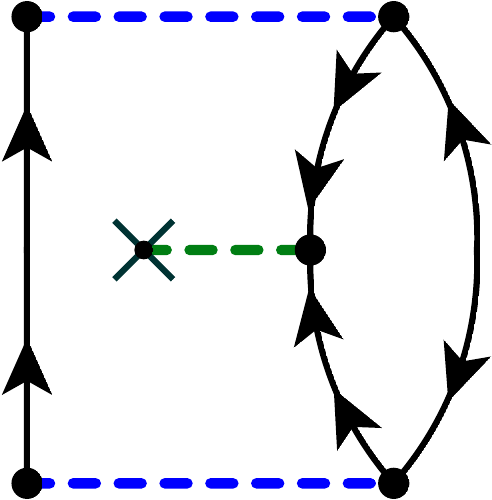}}
  \hspace{1.15cm}
  \subfloat[(s)]{\label{Fig_Us}\includegraphics[scale=0.23]{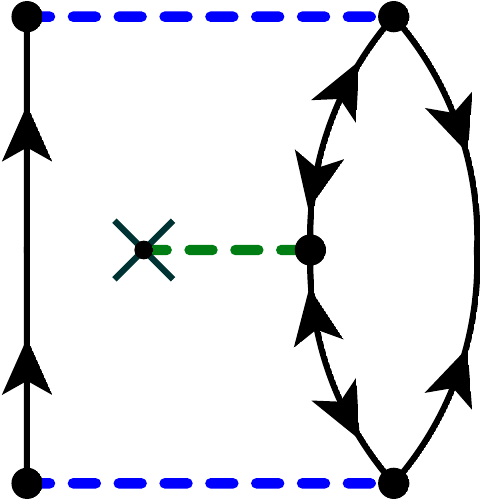}}
  \hspace{1.15cm}
  \subfloat[(t)]{\label{Fig_Ut}\includegraphics[scale=0.23]{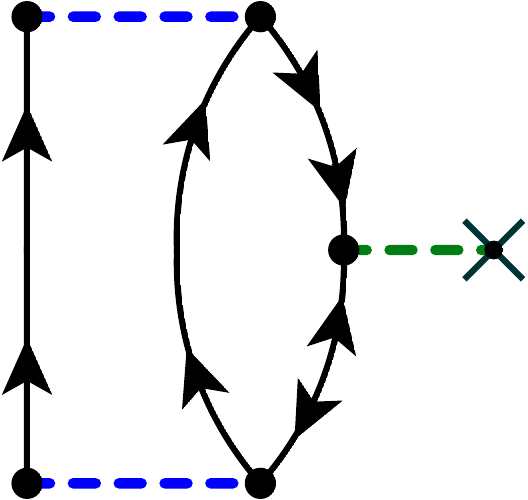}}
%  \newline
%  \includegraphics[width=1.8\columnwidth,height=1.2\columnwidth]{Fig8_tmp2.pdf}
  \caption{Third-order composite (non-skeleton) diagrams contributing to $\widetilde{\Sigma}^{11}(\omega)$ involving an external one-body potential $U$ of Eq.~\eqref{eq:def_U} as self-energy insertion.
  Dashed lines represent matrix elements of the two-particle interaction $V$, single lines are reference mean-field propagators from Eqs.~\eqref{eq:G0_props_Lehman} and dashed lines with a cross denote the one body potential $U$.   Similarly to Figs.~\ref{fig:ADC3_A}-\ref{fig:ADC3_C}, all other Nambu components of the self-energy receive contributions from analogous diagrams that are  obtained inverting one or both of the incoming and outgoing lines.}
\label{fig:ADC3_U}
\end{figure*}

The energy-dependent $\widetilde{\mathbf{\Sigma}}(\omega)$ at second order receives no contributions from self-energy insertions. Thus, the only composite diagrams appear at order three and involve the insertion of a static one-body potential to the known diagrams of Fig.~\ref{fig:SigIIab}.
This leads to the ten diagrams displayed in Fig.~\ref{fig:ADC3_U} for a generic external potential~$U$. In the following, we provide the contributions from these diagrams in terms of the matrix elements of $U$ and the amplitudes of Eq.~\eqref{eq:G0_props_Lehman}, with the understanding that these need to be substituted with those of $V^{HFB}-U$ introduced by the perturbation $\Omega_I$ from Eq.~\eqref{eq:Om_def}.

The top two rows in Fig.~\ref{fig:ADC3_U} cover all diagrams containing self-energies insertions originating from the normal component of $U$, i.e. the term associated with matrix elements $u_{\alpha\beta}$ in Eq.~\eqref{eq:def_U}. They contribute to the coupling matrices $ \mathcal{C}$ and $ \mathcal{D}$ through the normal singlet BCC amplitude
\begin{align}
\label{eq:BCC_t1n}
 t^{k_1}_{k_2} \equiv{}& \sum_{\alpha \beta} \; \frac{  \bar{\hfb{V}}^{k_1}_{\alpha}  \, u_{\alpha \beta}  \, \hfb{U}^{k_2}_\beta }{- \left( \omega_{k_1} + \omega_{k_2}  \right)} \, ,
\end{align}
and to the energy matrix through particle and hole interactions
\begin{subequations}\label{eq:3rd_Un}
\begin{align}
 \mathcal{E}^{(p)}_{k_1  k_2}  \equiv{}& \sum_{\alpha \beta} \; \left(         \hfb{U}^{k_1}_{\alpha} \right)^{\!*} \, u_{\alpha \beta} \,                 \hfb{U}^{k_2}_\beta                     \, , \label{eq:3rd_Up}  \\
 \mathcal{E}^{(h)}_{k_1  k_2}  \equiv{}& \sum_{\alpha \beta} \;       \bar{\hfb{V}}^{k_1}_{\alpha}                   \, u_{\alpha \beta} \, \left(  \bar{\hfb{V}}^{k_2}_\beta  \right)^{\!*}  \, . \label{eq:3rd_Uh} 
\end{align}
\end{subequations}
All together, this leads to the following ADC(3) contributions to the coupling matrices
\begin{subequations}\label{eq:3rd_Uext_norm}
\begin{align}
   \mathcal{C}^{(\rm{IIh})}_{\alpha, r} ={}& \!\frac 1{\sqrt{6}} \mathcal{A}_{123}  \! \sum_{\substack{\mu \,  \nu \,  \lambda \\ k_7}}   v_{\alpha \lambda, \mu \nu} \left(\bar{\hfb{V}}^{k_7}_\mu  \right)^{\!*} [t^{k_7}_{k_1}-t^{k_1}_{k_7}]  \, \hfb{U}^{k_2}_\nu \bar{\hfb{V}}^{k_3}_\lambda , \label{eq:3rd_CIIh}   \\
   \mathcal{C}^{(\rm{IIi})}_{\alpha, r} ={}& \!\frac 1{\sqrt{6}} \mathcal{P}_{123}  \! \sum_{\substack{\mu \,  \nu \,  \lambda \\ k_7}}   v_{\alpha \lambda, \mu \nu} \left( \hfb{U}^{k_7}_\lambda  \right)^{\!*} [t^{k_7}_{k_1}-t^{k_1}_{k_7}]  \, \hfb{U}^{k_2}_\mu \hfb{U}^{k_3}_\nu , \label{eq:3rd_CIIi}   \\
    \bar{\mathcal{D}}^{(\rm{IIh})}_{r, \alpha} ={}& \!\frac 1{\sqrt{6}} \mathcal{A}_{123}  \! \sum_{\substack{\mu \,  \nu \\  \lambda \, k_7}}  \left(   \hfb{U}^{k_7}_\mu \right)^{\!*}    [t^{k_7}_{k_1}-t^{k_1}_{k_7}]   \, \bar{\hfb{V}}^{k_2}_\nu   \hfb{U}^{k_3}_\lambda    v_{ \mu \nu , \alpha \lambda}  \, , \label{eq:3rd_DIIh} \\
    \bar{\mathcal{D}}^{(\rm{IIi})}_{r, \alpha} ={}& \!\frac 1{\sqrt{6}} \mathcal{P}_{123}  \! \sum_{\substack{\mu \,  \nu \\  \lambda \, k_7}}  \left(   \bar{\hfb{V}}^{k_7}_\lambda \right)^{\!*}    [t^{k_7}_{k_1}-t^{k_1}_{k_7}]   \, \bar{\hfb{V}}^{k_2}_\mu  \bar{\hfb{V}}^{k_3}_\nu    v_{ \mu \nu , \alpha \lambda}  \,   , \label{eq:3rd_DIIi} 
%
%CC%  \\ \nonumber ~  \\ \nonumber
%CC%    [\mathcal{D}^{(\rm{IIh})}]{}^T_{\alpha, r} &=  \!\frac 1{\sqrt{6}} \mathcal{A}_{123}  \! \sum_{\substack{\mu \,  \nu \,  \lambda \\ k_7}}   v_{\alpha \lambda, \mu \nu}  \left( \bar{\hfb{U}}^{k_7}_\mu   \right)^{\!*} [t^{k_7}_{k_1}-t^{k_1}_{k_7}]  \, \hfb{V}^{k_2}_\nu   \bar{\hfb{U}}^{k_3}_\lambda  ,
%CC% \\ \nonumber
%CC%    [\mathcal{D}^{(\rm{IIi})}]{}^T_{\alpha, r} &= \!\frac 1{\sqrt{6}} \mathcal{P}_{123}  \! \sum_{\substack{\mu \,  \nu \,  \lambda \\ k_7}}   v_{\alpha \lambda, \mu \nu}    \left( \hfb{V}^{k_7}_\lambda \right)^{\!*}  [t^{k_7}_{k_1}-t^{k_1}_{k_7}] \, \hfb{V}^{k_2}_\mu \hfb{V}^{k_3}_\nu   ,
\end{align}
\end{subequations}
and to the corresponding one-body energy interaction matrices
%\begin{align}
% \label{eq:3rd_EId} 
%  \mathcal{E}^{(\rm{Id})}_{r,r'} ={}&  \left\{  \begin{array}{l}
%          \mathcal{P}_{(14)(25)(36)}  \left[ \left(  \mathcal{E}^{(p)}_{k_1  k_4}  - \mathcal{E}^{(h)}_{k_4  k_1}  \right) \delta_{k_2 k_5}  \delta_{k_3 k_6} \right] \\ ~ \\   \qquad \qquad   \qquad  \qquad   \qquad \quad \hbox{for forward poles,} \\
%        ~\\~\\
%\mathcal{P}_{(14)(25)(36)}  \left[ \left( - \mathcal{E}^{(h)}_{k_1  k_4} + \mathcal{E}^{(p)}_{k_4  k_1}  \right) \delta_{k_2 k_5}  \delta_{k_3 k_6} \right]  \\ ~ \\   \qquad   \qquad  \qquad   \qquad  \qquad \quad \hbox{for backward poles,}
%    \end{array} \right.
%\end{align}
\begin{align}
 \label{eq:3rd_EId} 
  \mathcal{E}^{(\rm{Id})}_{r,r'} ={}& 
          \mathcal{P}_{(14)(25)(36)}  \left[ \left(  \mathcal{E}^{(p)}_{k_1  k_4}  - \mathcal{E}^{(h)}_{k_4  k_1}  \right) \delta_{k_2 k_5}  \delta_{k_3 k_6} \right] ,
\end{align}
where the cyclic permutation operator is intended to act on pairs of quasiparticle indices.

The remaining two rows of Fig.~\ref{fig:ADC3_U}  arise from the anomalous term in Eq.~\eqref{eq:def_U}. Introducing the anomalous single BCC amplitude in the two-particle channel
\begin{subequations}\label{eq:BCC_t1a}
\begin{align}
 t^{k_1  k_2}  \equiv{}& \sum_{\alpha \beta} \; \frac{ \left( u^{an.}_{\alpha \beta} \right)^{\!*}   \, \hfb{U}^{k_1}_{\alpha} \hfb{U}^{k_2}_\beta }{- 2 \left( \omega_{k_1} + \omega_{k_2}  \right)}  \, ,  \label{eq:BCC_t1a_pp} 
\end{align}
and in the two-hole channel
\begin{align}
 t_{k_1 k_2}  \equiv{}& \sum_{\alpha \beta} \;\; \frac{  \bar{\hfb{V}}^{k_1}_{\alpha}  \bar{\hfb{V}}^{k_2}_{\beta}  \, u^{an.}_{\alpha \beta} }{- 2 \left( \omega_{k_1} + \omega_{k_2}  \right)}  \, , \label{eq:BCC_t1a_hh} 
\end{align}
\end{subequations}
the anomalous one-body energy interaction matrix reads as
\begin{align}
 \mathcal{E}^{(an)}_{k_1 k_2}  \equiv{}& \sum_{\alpha \beta} \;  \frac 1 2 \left(   \hfb{U}^{k_1}_{\alpha} \right)^{\!*} \, u^{an.}_{\alpha \beta} \,                 \bar{\hfb{V}}^{k_2}_\beta      \,  \label{eq:3rd_Uan}  
 \end{align}
and acts by mixing the addition and removal components of a single quasiparticle.
With these definitions at hand, the remaining non-skeleton ADC(3) contributions are
 \begin{subequations}\label{eq:3rd_Uext_anm}
\begin{align}
   \mathcal{C}^{(\rm{IIj})}_{\alpha, r} ={}& \!\frac 1{\sqrt{6}} \mathcal{A}_{123}  \! \sum_{\substack{\mu \,  \nu \,  \lambda \\ k_7}}   v_{\alpha \lambda, \mu \nu} \left(\bar{\hfb{V}}^{k_7}_\mu  \right)^{\!*} [t^{k_7 k_1}\!-\!t_{k_7 k_1}]  \, \hfb{U}^{k_2}_\nu \bar{\hfb{V}}^{k_3}_\lambda , \label{eq:3rd_CIIj}   \\
   \mathcal{C}^{(\rm{IIk})}_{\alpha, r} ={}& \!\frac 1{\sqrt{6}} \mathcal{P}_{123}  \! \sum_{\substack{\mu \,  \nu \,  \lambda \\ k_7}}   v_{\alpha \lambda, \mu \nu} \left( \hfb{U}^{k_7}_\lambda  \right)^{\!*} [t^{k_7  k_1}\!-\!t_{k_7 k_1}]  \, \hfb{U}^{k_2}_\mu \hfb{U}^{k_3}_\nu , \label{eq:3rd_CIIk}   \\
    \bar{\mathcal{D}}^{(\rm{IIj})}_{r, \alpha} ={}& \!\frac 1{\sqrt{6}} \mathcal{A}_{123}  \! \sum_{\substack{\mu \,  \nu \\   \lambda \, k_7}}  \left(   \hfb{U}^{k_7}_\mu \right)^{\!*}    [t^{k_7 k_1}\!-\!t_{k_7 k_1}]   \, \bar{\hfb{V}}^{k_2}_\nu   \hfb{U}^{k_3}_\lambda    v_{ \mu \nu , \alpha \lambda}  \, , \label{eq:3rd_DIIj} \\
    \bar{\mathcal{D}}^{(\rm{IIk})}_{r, \alpha} ={}& \!\frac 1{\sqrt{6}} \mathcal{P}_{123}  \! \sum_{\substack{\mu \,  \nu \\   \lambda \, k_7}}  \left(   \bar{\hfb{V}}^{k_7}_\lambda \right)^{\!*}    [t^{k_7 k_1}\!-\!t_{k_7 k_1}]   \, \bar{\hfb{V}}^{k_2}_\mu  \bar{\hfb{V}}^{k_3}_\nu    v_{ \mu \nu , \alpha \lambda}  \, , \label{eq:3rd_DIIk} 
%
%CC%  \\ \nonumber ~  \\ \nonumber
%CC%     [\mathcal{D}^{(\rm{IIj})}_{\alpha, r}]{}^T &\!=  \!\frac 1{\sqrt{6}} \mathcal{A}_{123}  \! \sum_{\substack{\mu \,  \nu \,  \lambda \\ k_7}}   v_{\alpha \lambda, \mu \nu}  \left( \bar{\hfb{U}}^{k_7}_\mu   \right)^{\!*} [t^{k_7 k_1}\!-\!t_{k_7 k_1}]  \, \hfb{V}^{k_2}_\nu   \bar{\hfb{U}}^{k_3}_\lambda ,
%CC% \\ \nonumber
%CC%     [\mathcal{D}^{(\rm{IIk})}_{\alpha, r}]{}^T &\!= \!\frac 1{\sqrt{6}} \mathcal{P}_{123}  \! \sum_{\substack{\mu \,  \nu \,  \lambda \\ k_7}}   v_{\alpha \lambda, \mu \nu}    \left( \hfb{V}^{k_7}_\lambda \right)^{\!*}  [t^{k_7 k_1}\!-\!t_{k_7 k_1}] \, \hfb{V}^{k_2}_\mu \hfb{V}^{k_3}_\nu   ,
\end{align}
\end{subequations}
and
\begin{align}
 \label{eq:3rd_EIe} 
  \mathcal{E}^{(\rm{Ie})}_{r,r'} ={}& 
   \mathcal{P}_{(14)(25)(36)}  \left[ \left(  \mathcal{E}^{(an)}_{k_1  k_4}  + \mathcal{E}^{(an)\, *}_{k_4  k_1}  \right) \delta_{k_2 k_5}  \delta_{k_3 k_6} \right] .
\end{align}
%\begin{align}
% \label{eq:3rd_EIe} 
%  \mathcal{E}^{(\rm{Ie})}_{r,r'} ={}&  \left\{  \begin{array}{l}
%   \mathcal{P}_{(14)(25)(36)}  \left[ \left(  \mathcal{E}^{(an)}_{k_1  k_4}  + \mathcal{E}^{(an)\, *}_{k_4  k_1}  \right) \delta_{k_2 k_5}  \delta_{k_3 k_6} \right] \\ ~ \\   \qquad \qquad   \qquad  \qquad   \qquad \quad ~ \hbox{for forward poles,} \\
%        ~\\~\\
%   \mathcal{P}_{(14)(25)(36)}  \left[ \left(  \mathcal{E}^{(an)}_{k_4  k_1} + \mathcal{E}^{(an)\, *}_{k_1  k_4}  \right) \delta_{k_2 k_5}  \delta_{k_3 k_6} \right]  \\ ~ \\   \qquad   \qquad  \qquad   \qquad  \qquad \quad ~ \hbox{for backward poles.}
%    \end{array} \right.
%\end{align}

\section{\label{Sec:scgf_approx}Optimised Reference States and Approximations to Self-consistency}

The SCGF approach is based on using the dressed propagator $\mathbf{G}(\omega)$, Eq.~\eqref{eq:Gkv_props}, as the reference state upon which the self-energy is expanded. It generalises the unperturbed propagator $\mathbf{G}^{(0)}(\omega)$ to include full many-body correlations in the one-body Green's function. In this framework, only skeleton diagrams must be accounted for and the contributions discussed in Secs.~\ref{Sec:adc_1_2} and~\ref{Sec:adc_3} define the complete Gorkov-ADC(3) approach. 
Since $\mathbf{G}(\omega)$ is itself obtained by solving Gorkov equations, in practice one needs to compute the self-energy and diagonalize Eq.~\eqref{eq:ADC_mtx} iteratively until convergence.

Our experience from applications to nuclear structure is that the most important self-consistency effects arise from the cHFB terms, i.e. Eqs.~\eqref{eq:SigI_expr}~\cite{Barbieri2014QMBT}. These are rather straightforward to compute and require very modest computational resources, even for fully dressed propagators. On the other hand, the self-consistent computation of $\widetilde{\mathbf{\Sigma}}(\omega)$ becomes quickly prohibitive. 
If \hbox{$N_{\rm Bs}={\rm dim}(\{\alpha\})$} denotes the dimension of the single-particle basis, an unperturbed reference state $\mathbf{G}^{(0)}(\omega)$ implies a dimension ${\approx}2N_{\rm Bs}^3$ for the Gorkov eigenvalue problem~\eqref{eq:ADC_mtx} that generates half as many poles for $\mathbf{G}(\omega)$. At each subsequent self-consistency iteration the dimension of the ISC space grows as ${\rm dim}(\{r\}){\approx}(N_{\rm Bs})^{3^n}$, with $n$ being the number of iterations.
Therefore, it is mandatory to devise proper approximations  of the dressed propagator, $\mathbf{G}^{\rm red}(\omega)$, that limit the growth in the number of poles. 
Typical approaches proposed in the literature aim at a low-dimensional representation of the propagator either by binning of the spectral function in energy and momentum or by projecting it onto Krylov subspaces~\cite{Muther1993BagelLRC,Dewulf1997BagelBin,VanNeck2001JCP2nd,Dewulf2002prc_NM,Soma2014GkII}. The second approach is highly preferable, if not mandatory, when working with discrete Lehmann representations such as Eqs.~\eqref{eq:Gkv_props} and~~\eqref{eq:Sig_tild}.
In the context of Dyson SCGF, we introduced two workable techniques that follow the latter strategy and further rely on the conservation of the lowest moments of the spectral function~\cite{Barbieri2009Ni56,Rocco2018escatt,Raimondi2019GR}.
In the following, these ideas are generalized to the case of Gorkov propagators.

In general, both the one-body spectral function and the dressed propagator~\eqref{eq:Gkv_props} are uniquely defined by the set of quasiparticle poles, $\omega_k$, and spectroscopic amplitudes  $(\mathcal{U}^k,  \mathcal{V}^k)$. Given the number $D$ of independent poles ($k=1, \ldots D$), we aim at replacing these objects with a smaller set
\begin{align}
\omega_k \, , \left( \begin{array}{c} \mathcal{U}^k \\ \mathcal{V}^k \end{array} \right) 
\quad\longrightarrow\quad
\omega'_{i} \, , \left( \begin{array}{c} \hfb{U}'^{i}  \\ \hfb{V}'^{i} ~ \end{array} \right) \; 
\label{eq:RedUV}
\end{align}
where $i=1,\ldots d$ and $d\ll D$.
The $p$-th moment of the spectral distribution is defined in Nambu space as 
\begin{align}
 \mathbf{S}^{(p)}_{\alpha \beta} \equiv \sum_k  (\omega_k)^p \,  \left( \begin{array}{c}  \bar{\mathcal U}^k_\alpha{}^*  \\~\\ \bar{\mathcal V}^k_\alpha{}^* \end{array} \right)   \left(  \bar{\mathcal U}^k_\beta \quad \bar{\mathcal V}^k_\beta  \right)
\label{eq:Sp_mom_def}
\end{align}
and similarly for $\omega'_i$ and $(\hfb{U}'^i,  \hfb{V}'^i)$. The new set of poles and amplitudes is determined by imposing that it preserves all matrix elements of the first $2n$ moments, $p=0, 1, \ldots 2n-1$, for a given positive integer $n$. It is easy to verify that this condition is met by choosing
 \begin{subequations}\label{eq:RedUVw_vs_Z}
\begin{align}
 \omega'_i ={}& e^i   \label{eq:Rede_vs_Z}  \, , \\
 \left( \begin{array}{c} \bar{\hfb{U}}'^{i}_\alpha{}^*   \\~\\ \bar{\hfb{V}}'^{i}_\alpha{}^*  ~ \end{array} \right) ={}& \sum_{p=1}^{n}  [\mathbf{S}^{(p-1)}  \mathbf{Z}^{(p)}_i]_\alpha \, ,
\label{eq:RedUV_vs_Z}
\end{align}
 \end{subequations}
where $e^i$ and $(\mathbf{Z}^{(1)}_i, \mathbf{Z}^{(2)}_i, \ldots \mathbf{Z}^{(n)}_i)$ are the solutions of the eigenvalue problem
\begin{align}
&  \!\!\!\left( \begin{array}{ccccc} 
      \mathbf{S}^{(1)}      &  \mathbf{S}^{(2)}      &  \cdots &  \qquad & \mathbf{S}^{(n)} \\ 
      \mathbf{S}^{(2)}      &  \mathbf{S}^{(3)}      &  \cdots &  ~ & \mathbf{S}^{(n+1)} \\
        \vdots  &  \vdots   &    \ddots &&  \vdots \\   ~ \\
      \mathbf{S}^{(n)}     &  \mathbf{S}^{(n+1)}  &  \cdots &  ~ & \mathbf{S}^{(2n-1)} \\
  \end{array} \right)
  \left( \begin{array}{c}  \mathbf{Z}^{(1)}_i \\ \mathbf{Z}^{(2)}_i \\ \vdots \\ ~\\  \mathbf{Z}^{(n)}_i   \end{array} \right) \nonumber \\
   &  = e^i  
   \left( \begin{array}{ccccc} 
      \mathbf{S}^{(0)}      &  \mathbf{S}^{(1)}      &  \cdots &  \qquad & \mathbf{S}^{(n-1)} \\ 
      \mathbf{S}^{(1)}      &  \mathbf{S}^{(2)}      &  \cdots &  ~ & \mathbf{S}^{(n)} \\
        \vdots  &  \vdots   &    \ddots &&  \vdots \\   ~ \\
      \mathbf{S}^{(n-1)}  &  \mathbf{S}^{(n)}  &  \cdots &  ~ & \mathbf{S}^{(2n-2)} \\
  \end{array} \right)
     \left( \begin{array}{c}  \mathbf{Z}^{(1)}_i \\ \mathbf{Z}^{(2)}_i \\ \vdots \\ ~\\  \mathbf{Z}^{(n)}_i   \end{array} \right) ,
\label{eq:Red_SZ_diag}
\end{align}
with the sums over single-particle basis indices being implicit for simplicity.  Note  that each component $\mathbf{Z}^{(p)}_i$  is a vector both in the single-particle basis and in Nambu space, so that the number of new amplitudes is \hbox{$d=2 n N_{\rm Bs}$.} For very large values of $n$ the spectral distribution will be approximated with a growing number of poles and the original poles and amplitudes,  $\omega_k$ and  $(\bar{\mathcal U}^k,  \bar{\mathcal V}^k)$, will eventually be recovered exactly. However, the most important physical information is already preserved even for the simplest case, i.e. $n$=1. For example, the lowest spectral moment is
\begin{align}
   \mathbf{S}^{(0)}_{\alpha \beta} ={}& \left( \begin{array}{ccc} m_{\alpha \beta}  &~&  \tilde{\rho}_{\beta \alpha}^* \\ ~ \\   \tilde{\rho}_{\alpha \beta} && \rho_{\alpha \beta} \end{array} \right) 
\label{eq:S0_comps}
\end{align}
where the matrix
\begin{align}
 m_{\alpha \beta} \equiv{}&S^{(0),11}_{\alpha \beta} =  \sum_k  \,     \bar{\mathcal U}^k_\alpha{}^* \, \bar{\mathcal U}^k_\beta =   \langle \Psi_0 |  \bar{c}_\beta \bar{c}^\dagger_\alpha |\Psi_0\rangle  \, ,
\label{eq:S0_11_mtx}
\end{align}
is related to the density matrix through $m_{\alpha \beta}+\rho_{\bar\beta \bar\alpha }=\delta_{\alpha \beta}$ and informs on the distribution of unoccupied single-particle states~\cite{Rios2017SumRls}. Hence, preserving Eq.~\eqref{eq:S0_comps} will automatically preserve the pairing gaps, the density matrix and all one-body observables, including the average particle number~\eqref{eq:DefPsi0N} and the point-particle density distributions.  Likewise, the Koltun energy sum rule~\eqref{eq:Koltun} can be expressed in terms of the  $p=0,1$ moments
\begin{align}
  \Omega_0 ={}& \frac{1}{2 }  \sum_{\alpha \beta} \left[  ( t_{\alpha \beta} - \mu \delta_{\alpha \beta} )  S^{(0),22}_{ \beta \alpha } - \delta_{\alpha \beta} S^{(1),22}_{ \beta \alpha } \right]  \, ,
\label{eq:S0S1_koltun}
\end{align}
and the effective single-particle energies, obtained diagonalizing  $h^{\rm cent.}_{\alpha \beta} \equiv \mu \, \delta_{\alpha \beta} + S^{(1),11}_{\bar\beta \bar\alpha}-S^{(1),22}_{\alpha \beta}$, remain unchanged~\cite{Duguet2012espe}.

In spite of its efficiency, the effective dressed single-particle propagator for $n$=1 has already a number of poles, $d=2 N_{\rm Bs}$, that is doubled with respect to the standard mean-field reference. This implies  an eightfold increase in the dimension of the corresponding Gorkov eigenvalue problem~\eqref{eq:ADC_mtx}.  The  self-consistent approach based on conserving moments~\eqref{eq:Sp_mom_def} is therefore viable only for sufficiently small model spaces. 

A better approximation would consist in maintaining the same number of poles as  $\mathbf{G}^{(0)}(\omega)$ while not giving up the ability of conserving key quantities from the previous scheme.
For this purpose let us compute moments of the spectral functions with respect to the poles of $\mathbf{G}(\omega)$ and sum over both forward- and backward-going excitations from Eq.~\eqref{eq:Gkv_props}
\begin{align}
\mathbf{Q}^{(p)}_{\alpha \beta} ={}& \sum_k  \left\{ \frac{-1} {(\omega_k)^p} \,  \left( \begin{array}{c} \mathcal{U}^k_\alpha \\ \mathcal{V}^k_\alpha \end{array} \right)   \left(\mathcal{U}^k_\beta{}^* , \mathcal{V}^k_\beta{}^*  \right)  \right. \nonumber \\
 & \qquad +    \left. \frac {1} {(\omega_k)^p} \,  \left( \begin{array}{c} \bar{\mathcal V}^k_\alpha{}^* \\ \bar{\mathcal U}^k_\alpha{}^* \end{array} \right)   \left( \bar{\mathcal V}^k_\beta , \bar{\mathcal U}^k_\beta \right) \right\} \, .
\label{eq:Qp_mom_def}
\end{align}
Note that $\mathbf{Q}^{(p=1)}_{\alpha \beta}=\mathbf{G}_{\alpha \beta}(\omega=0)$. The set of amplitudes~\eqref{eq:RedUV} preserving the moments~\eqref{eq:Qp_mom_def} is obtained following Eqs.~\eqref{eq:Red_SZ_diag} and~\eqref{eq:RedUV_vs_Z} with $\mathbf{S}^{(p)}$ replaced by $\mathbf{Q}^{(p)}$ but choosing $\omega'_i = 1/{e^i}$. Equation~\eqref{eq:Qp_mom_def} leads to the general structure
\begin{align}
   \mathbf{Q}^{(p)}_{\alpha \beta} ={}& \left( \begin{array}{ccc}    {Q}^{(p),11}_{\alpha \beta}  &~&  {Q}^{(p),12}_{\alpha \beta}  \\ ~ \\     \left({Q}^{(p),12}_{ \beta \alpha}\right)^*   &&   (-)^p \left({Q}^{(p),11}_{\bar\beta \bar\alpha}\right)^* \end{array} \right)  \, ,
\label{eq:Qp_comps}
\end{align}
which is analogous the HFB eigenmatrix~\eqref{eq:SigI_expr}. Inserting it into Eq.~\eqref{eq:Red_SZ_diag}, it can be shown that solutions come in pairs with eigenvalues of opposite sign $(1/{e^i}, -1/{e^i})$ and consequently the approximated spectra distribution only has $d=n N_{\rm Bs}$ independent poles.
As for the case of the direct spectral function, Eq.~\eqref{eq:Sp_mom_def}, imposing the preservation of the first $2n$ moments $\mathbf{Q}^{(p)}$ lead to an approximation to the spectral distribution that approaches the complete set of $D$ quasiparticles as $n$ increases. However, the greatest advantage is for $n=1$ since it defines an approximation to $\mathbf{G}(\omega)$ (the self-consistent propagator) that has the same number of poles of the standard reference mean-field $\mathbf{G}^{(0)}(\omega)$.

The most relevant difference between moments~\eqref{eq:Sp_mom_def} and~\eqref{eq:Qp_mom_def} is that the orthonormality of eigenvectors~\eqref{eq:ADC_norm} implies that $ \mathbf{Q}^{(0)}_{\alpha\beta}=\mathbf{I} \delta_{\alpha\beta}$ is the identity matrix. Hence, the lowest moments $\mathbf{Q}^{(0)}$ non longer preserves the exact (normal and anomalous) reduced density matrices. On the other hand, the eigenvalue equation~\eqref{eq:Red_SZ_diag} for $n=1$ becomes
\begin{align}
\mathbf{Q}^{(1)}  \mathbf{Z}_i = e_i  \mathbf{Z}_i  \, ,
\label{eq:Q1_eigv_prob}
\end{align}
where $\mathbf{Z}^{(1)}_i{}^T = (\hfb{U}'^i,  \hfb{V}'^i)$ according of Eq.~\eqref{eq:RedUV_vs_Z}. 
In addition, relations~\eqref{eq:Qp_comps} for $\mathbf{Q}^{(1)}$ imply that this is a HFB-like problem.
Therefore, the poles $\omega'_i$ and $(\hfb{U}'^i,  \hfb{V}'^i)$ are \emph{both} a workable approximation of the self-consistent propagator $\mathbf{G}(\omega)$ and the solution of 
an unperturbed mean-field Hamiltonian $\Omega_U^{(OpRS)}$, which we dub \emph{optimized reference state} (OpRS) Hamiltonian.

To correctly define the external potential  $U^{(OpRS)}$ that generates the new reference state one has to remember that the eigenvalues $e_i$ are associated with moments of inverse poles, $1/\omega_k$, through Eq.~\eqref{eq:Qp_mom_def}. Thus, $\Omega_U^{(OpRS)}$ is identified with the inverse matrix of $\mathbf{Q}^{(1)}$
\begin{align}
 & \!\!\!\! \left( \begin{array}{ccc}    \Omega_{U\, \alpha \beta}^{(OpRS) 11}  &~&   \Omega_{U \, \alpha \beta}^{(OpRS) 12}  \\ ~ \\     \Omega_{U\, \alpha \beta}^{(OpRS) 21}   &&   \Omega_{U\, \alpha \beta}^{(OpRS) 22} \end{array} \right)  
  \equiv  \left [\mathbf{Q}^{(1)}_{\alpha \beta}\right]^{-1}  \nonumber \\
 &= \sum_i 
  \left( \begin{array}{ccc}    \hfb{U}'^i_\alpha    &~&   \bar{\hfb{V}}'^i_\alpha{}^*   \\ ~ \\      \hfb{V}'^i_\alpha    &~&   \bar{\hfb{U}}'^i_\alpha{}^*  \end{array} \right)  
  \!\! \left( \begin{array}{ccc}    \frac 1{e_i}  &~&     \\ ~ \\       && -\frac{1}{e_i}  \end{array} \right)  
  \!\! \left( \begin{array}{ccc}    \hfb{U}'^i_\beta{}^*   &~&   \hfb{V}'^i_\beta{}^*   \\ ~ \\      \bar{\hfb{V}}'^i_\beta    &~&   \bar{\hfb{U}}'^i_\beta  \end{array} \right)    \, ,
\label{eq:OpRS_vs_Q1}
\end{align}
where the sum runs over all positive eigenvalues ($e_i>0$) of Eq.~\eqref{eq:Q1_eigv_prob}. Comparing the matrix elements of Eq.~\eqref{eq:OpRS_vs_Q1} with the left-hand side of Eq.~\eqref{eq:OmU_HFB_eqs} defines the potential~$U^{(OpRS)}$.

Equation~\eqref{eq:OpRS_vs_Q1} has been systematically used to define an \emph{optimized reference state} for SCGF computations in nuclear structure~\cite{Soma2020Front}. 
Although the physical quantities in Eqs.~\eqref{eq:S0_comps},~\eqref{eq:S0_11_mtx} and~\eqref{eq:S0S1_koltun} are no longer preserved exactly, it is our experience that the Koltun sum rule and density distributions are still closely reproduced even with such OpRS and even for strongly correlated systems such as atomic nuclei. Thus, this has become our method of choice to implement self-consistency. In doing so, it is crucial to define $\mathbf{Q}^{(p)}$  in terms of the inverse quasiparticle energies because the conservation of lowest moments constrains more efficiently the extremes of the eigen spectrum rather than its central part. By employing powers of $1/\omega_k$ in Eq.~\eqref{eq:Qp_mom_def} one ensures that greater weight is given to preserve the structure of the self-consistent propagator $\mathbf{G}(\omega)$ near the Fermi surface.

In spite of being a somewhat poorer approximation than the one generated by Eqs.~\eqref{eq:Sp_mom_def}, the OpRS has the great advantage to be associated with an external mean-field potential $\Omega_U$ and propagator. Hence, the approximation made in replacing the true self-consistent propagator $\mathbf{G}(\omega)$ with $\mathbf{G}^{(OpRS)}(\omega)$ can always be corrected systematically by computing the non-skeleton diagrams of~Sec.~\ref{Sec:Comb_diags} for $U \equiv \Sigma^{HFB}-U^{(OpRS)}$.
 
\section{\label{Sec:Implem} Implementation of the ADC($n$) truncation hierarchy}

Having derived all diagramamatic contributions up to third order, it useful to briefly summarize the specific terms that enter the various truncation levels of the ADC($n$) method.
At each order $n$, the Gorkov propagator~\eqref{eq:Gkv_props} is obtained by diagonalizing Eq.~\eqref{eq:ADC_mtx}, with the eigenvectors normalised according to Eq.~\eqref{eq:ADC_norm} and  the chemical potential tuned to reproduce the correct number of particles on average (see Eq.~\eqref{eq:DefPsi0N} and Ref.~\cite{Soma2014GkII}). The matrix elements of the Gorkov eigenvalue problem are given as follows:
\begin{enumerate}
\item \emph{ADC(1)}. Only the cHFB sector of the Gorkov matrix contributes at first order, while all couplings to ISCs vanish: \hbox{\;$\mathcal{C}^{[ADC(1)]}=\bar{\mathcal D}^{[ADC(1)]}=\mathcal{E}^{[ADC(1)]}=0$.} One computes the matrix elements of $\mathbf{\Sigma}^{(\infty)}$ from Eqs.~\eqref{eq:SigI_expr} and adds the one-body interaction $T$ shifted by the chemical potential $\mu$.  In this case, a self-consistent computation reduces to the standard HFB problem.
\item \emph{ADC(2)}. The cHFB sector remains the same as for ADC(1). The coupling matrices $\mathcal{C}^{[ADC(2)]}$ and $\bar{\mathcal D}^{[ADC(2)]}$ and energy denominator $\mathcal{E}^{[ADC(2)]}$ are given by Eqs.~\eqref{eq:ADC2}.   Since all Feynman diagrams up to second order are of skeleton type, the ADC(2) is uniquely defined by these equations.
 \item \emph{Self-consistent ADC(3)}.  For a self-consistent computation the reference Gorkov propagator is replaced by a dressed one, which includes the fragmentation of single-particle strength. In this case only skeleton diagrams must be included. The ADC(3) equations remain unchanged in the cHFB sector but the couplings to ISCs receive additional terms from Eqs.~\eqref{eq:3rd_CDpp} through~\eqref{eq:3rd_EIc}. Specifically:
\begin{subequations}\label{eq:ADC3sc}
 \begin{align}
\mathcal{C}^{[ADC(3)\!-\!SC]}_{\alpha, r}={}&\mathcal{C}^{[ADC(2)]}_{\alpha, r}+\mathcal{C}^{(\rm{IIa})}_{\alpha, r}+\mathcal{C}^{(\rm{IIb})}_{\alpha, r} + \mathcal{C}^{(\rm{IIc})}_{\alpha, r} 
\nonumber \\& +\mathcal{C}^{(\rm{IId})}_{\alpha, r} + \mathcal{C}^{(\rm{IIe})}_{\alpha, r} + \mathcal{C}^{(\rm{IIf})}_{\alpha, r} + \mathcal{C}^{(\rm{IIg})}_{\alpha, r} \! ,\\
\hspace{-1.cm}
\bar{\mathcal D}^{[ADC(3)\!-\!SC]}_{r, \alpha}={}&\bar{\mathcal D}^{[ADC(2)]}_{r, \alpha}\!+\bar{\mathcal{D}}^{(\rm{IIa})}_{r, \alpha}\!+\bar{\mathcal{D}}^{(\rm{IIb})}_{r, \alpha}\!+\bar{\mathcal{D}}^{(\rm{IIc})}_{r, \alpha} 
\nonumber \\ &+\! \bar{\mathcal{D}}^{(\rm{IId})}_{r, \alpha}  \!+\! \bar{\mathcal{D}}^{(\rm{IIe})}_{r, \alpha} \!+\! \bar{\mathcal{D}}^{(\rm{IIf})}_{r, \alpha} + \bar{\mathcal{D}}^{(\rm{IIg})}_{r, \alpha}  \! , \\
%\hspace{-.1cm}
\mathcal{E}^{[ADC(3)\!-\!SC]}_{r, r'} ={}& \mathcal{E}^{[ADC(2)]}_{r, r'} + \mathcal{E}^{(\rm{Ia})}_{r,r'} + \mathcal{E}^{(\rm{Ib})}_{r,r'} + \mathcal{E}^{(\rm{Ic})}_{r,r'} .
\end{align}
\end{subequations}
\item \emph{Full ADC(3)}. Whenever a mean-field propagator is used to define the reference state for the Feynman-Gorkov expansion, composite diagrams must also be included. The first non-skeleton terms appear at third order and therefore contribute from ADC(3) onward. These are detailed in Eqs.~\eqref{eq:3rd_Uext_norm},~\eqref{eq:3rd_EId},~\eqref{eq:3rd_Uext_anm} and~\eqref{eq:3rd_EIe} and lead to the final Gorkov-ADC(3) equations:
\begin{subequations}\label{eq:ADC3full}
 \begin{align}
\mathcal{C}^{[ADC(3)\!-\!Full]}_{\alpha, r}={}&\mathcal{C}^{[ADC(3)\!-\!SC]}_{\alpha, r}+\mathcal{C}^{(\rm{IIh})}_{\alpha, r}+\mathcal{C}^{(\rm{IIi})}_{\alpha, r} 
\nonumber \\& \qquad  + \mathcal{C}^{(\rm{IIj})}_{\alpha, r}  +\mathcal{C}^{(\rm{IIk})}_{\alpha, r}  ,\\
%
%\hspace{-1.cm}
\bar{\mathcal D}^{[ADC(3)\!-\!Full]}_{r, \alpha}={}&\bar{\mathcal D}^{[ADC(3)\!-\!SC]}_{r, \alpha}\!+\bar{\mathcal{D}}^{(\rm{IIh})}_{r, \alpha}\!+\bar{\mathcal{D}}^{(\rm{IIi})}_{r, \alpha}
\nonumber \\ & \qquad+\! \bar{\mathcal{D}}^{(\rm{IIj})}_{r, \alpha}  \!+\! \bar{\mathcal{D}}^{(\rm{IIk})}_{r, \alpha}  \! , \\
%\hspace{-.1cm}
\mathcal{E}^{[ADC(3)\!-\!Full]}_{r, r'} ={}& \mathcal{E}^{[ADC(3)\!-\!SC]}_{r, r'} + \mathcal{E}^{(\rm{Id})}_{r,r'} + \mathcal{E}^{(\rm{Ie})}_{r,r'} .
\end{align}
\end{subequations}
Note that the latter corrections have to be computed for the \emph{residual} one-body interaction \hbox{$U \equiv \Sigma^{HFB}-U^{(MF)}$}, where $\Sigma^{HFB}$ is the HFB potential and $U^{(MF)}$ is the mean field that defines the reference state. Therefore, corrections~\eqref{eq:ADC3full}  vanish for the special case of an HFB reference state.
\end{enumerate}
It should be noted that for each of the above truncations the cHFB sector in Eq.~\eqref{eq:ADC_mtx} needs to be evaluated at least at the corresponding order in perturbation theory. In practical applications it is sufficient to exploit Eqs.~\eqref{eq:SigI_expr}, which are \emph{complete} to all orders if expressed in terms of the fragmented amplitudes $\mathcal{U}$ and  $\mathcal{V}$. This leads to a first level of self-consistency, referred to as `sc0', where only the cHFB part of the eigenmatrix~\eqref{eq:ADC_mtx} is updated iteratively~\cite{Soma2014GkII}. Complete self-consistency requires iterating also the coupling amplitudes and interactions, $\mathcal{C}$, $\mathcal{D}$, and $\mathcal{E}$: these are computed from the spectroscopic amplitudes and poles of the dressed propagator, whenever possible, or otherwise resorting to the corresponding OpRS discussed in Sec.~\ref{Sec:scgf_approx}.

The same working equations for the case of a spherical system in angular momentum coupling scheme are derived in App.~\ref{App:Jcoupl}, with the final eigenvalue problem given by Eqs.~\eqref{eq:Jcoup_ADC_mtx} and \eqref{eq:Jcoup_CD}. The J-coupled contributions corresponding to the above truncation schemes are collected in Secs.~\ref{App:Jcoup_cHFB}, \ref{App:Jcoup_adc2}, \ref{App:Jcoup_adc3} and~\ref{App:Jcoup_comp}, respectively.
Extensive details about the numerical implementation of Gorkov-ADC($n$) can be found in Ref.~\cite{Soma2014GkII}.

\section{\label{Sec:Concl}Conclusions}

The major outcome of the present work is the full development of the Gorkov method at the ADC($3$) truncation level for which all algebraic details necessary to perform a numerical implementation are provided.

The full set of working equations for a two-body Hamiltonian has been presented in Sec.~\ref{Sec:ADCn} for a general single-particle basis set and are summarised for the different ADC($n$) truncations in Sec.~\ref{Sec:Implem}. This allows for the largest possible range of applications to a variety of systems. Our results could be easily adapted to complex geometries as in molecules or specified to cylindrical and deformed bases (which will be highly relevant to deformed nuclei). The particular case of spherical symmetry can be exploited for simple atoms and semi-magic nuclear isotopes and is presently derived in full in App.~\ref{App:Jcoupl}. This development is expected to open the way to new advances in ab initio nuclear structure studies.

In spite of the large number of Gorkov diagrams that needs to be considered to build the ADC(3) approximation, i.e. 17 skeleton plus 20 composite ones at third order,  most contributions can be gathered together in a handful of final terms. Combining Goldstone (time-ordered) contributions across different diagrams leads to important simplifications~\cite{Arthuis:2018yoo}, i.e. it is not only to simplify the working equations but also to satisfy Pauli antisymmetrization. Further simplifications of our results from Sec.~\ref{Sec:ADCn} could be achieved by identifying specific terms with unperturbed Bogoliubov coupled cluster amplitudes~\cite{Signoracci2015BCC}.  The close relation between SCGF and coupled cluster methods was already observed at the level of the standard (particle number conserving) formulations~\cite{Nooijen1992CCGF,Nooijen1993CCGF}.
The ADC(3) many-body truncation is normally expected to be equivalent to triple corrections in CC~\cite{Trofimov2005IE_nD,Trofimov2002adc_ci_cc_comp}. Furthermore, the identification of BCC amplitudes pointed out in Sec.~\ref{Sec:ADCn} can be the basis for further improvements of the ADC(3) scheme, as discussed in Ref.~\cite{Barb2017LNP}.

In the effort to clarify all possible aspects of a future implementation of the method, the systematic procedures to handle self-consistency were discussed. In particular, the necessary approximations in applying SCGF to the energy-dependent self-energy can be rationalized as a process of learning an optimal reference state that encodes the most important features of many-body correlations. The proposed OpRS approach provides a way to maintain a faithful implementation, i.e. to correct systematically for the approximation itself, up to order ADC(3) by simply adding the composite diagrams corrections.

The numerical implementation of ADC(3) is notoriously  more difficult than ADC(2) and implies much higher needs for computational resources. Although it is not clear a priori how large model spaces can be reached, the present-day availability of massively parallel computing resources provides a significant advantage with respect to our early applications at the ADC(2) level~\cite{Soma2013Rapid,Soma2014Ca}.

With the right formalism in place and the hierarchy of approximations ADC(1) (i.e., HFB), ADC(2), ADC(3), it will be possible to better assess the limits and merits of the Gorkov approach. In particular, the implications of symmetry breaking can be addressed in a more systematic way.

The formalism set out in this work does not consider the implication of three-body forces, which are a crucial component of the nuclear Hamiltonian.  This is not a major hindrance for ab initio nuclear physics because the vast majority of state-of-the-art applications find it sufficient to include three-nucleon forces as effective two-body interactions.  The details of the full ADC($n$) Gorkov formalism based on two- and three-body forces, including anomalous and interaction-irreducible contributions, will be the subject of a future work.

\begin{acknowledgments}
CB acknowledges useful discussions with D. Van Neck on the effective propagators of Sec.~\ref{Sec:scgf_approx}.
This work was supported by the United Kingdom Science and Technology Facilities Council (STFC), through Grants No. ST/L005743/1 and ST/P005314/1.
\end{acknowledgments}

\appendix

\section{\label{App:Jcoupl}Coupling of angular momenta}

The Gorkov-ADC($3$) equations derived in Sec.~\ref{Sec:ADCn} are generally applicable to any 
many-fermion system, given an orthonormal one-body basis $\{|\alpha\rangle\}$ and its dual defined in Eq.~\eqref{eq:dualA}.
However, this form rarely represents the optimal choice for practical implementations. In most cases, computational requirements
can be drastically reduced by adopting an appropriate basis and exploiting the symmetries of the problem under consideration. 
We now discuss the particular case of a rotationally invariant Hamiltonian and (spherical) ground state with total
angular momentum $J=0$. This class of systems includes a number of atoms and ions in Quantum Chemistry and 
the vast majority of semi-magic isotopes in Nuclear Physics.

Most \emph{ab initio} implementations in nuclear physics exploit spherical single-particle basis states with an isospin spinor
${\cal X}_q(\tau)$ plus a spherical spherical harmonic $Y_\ell(\hat{r})$ and spin ${\cal X}_{\frac 1 2}(\sigma)$ coupled to 
angular momentum $j$ and its $z$-axis projection $m$:
\begin{align}
\label{eq:sph_sp}
\phi_\alpha(\vec{r},\sigma,\tau) ={}& f_{n_\alpha  \ell_\alpha  j_\alpha  q_\alpha} (r) \left[   Y_{\ell_\alpha}(\hat{r}) \otimes {\cal X}_{\frac 1 2}(\sigma)   \right]^{ j_\alpha}_{m_\alpha} {\cal X}_{q_\alpha}(\tau) \, ,
\end{align}
where $\vec{r}$, $\sigma$ and $\tau$ are the spatial, spin and isospin coordinates  and 
the most general radial function $f_a(r)$ may depend on all quantum numbers except $m_\alpha$ due to rotational symmetry.
Here, $q_\alpha$ stands for any quantum number (or set of numbers) that is needed to label different  distinguishable particles. Typically, it is not needed for a single-fermion system such as the electron gas while it is customary to use the nucleon charge to distinguish among protons and neutrons in atomic nuclei. However, the latter representation is inefficient for our purposes since the Gorkov formulation conserves only particle-number parity but not the total number of particles. Hence, the total charge is also not conserved. A~more general and practical choice  is a vector of quantum numbers encoding the particle-number parity of all types of particles in the systems. For example, in the case of Eq.~\eqref{eq:sph_sp}, $q_\alpha$ will be a `one-hot' vector with all zeros except for the element that identifies the given particle, which is set to 1.
In the following this representation of $q_\alpha$ will be used, always implicitly intended as a vector in $\{0,1\}^{N_f}$, with $N_f$ the number of different fermion species. 
For spin $\frac 1 2$, the combination of spatial parity $\pi_\alpha$ and $j_\alpha$ uniquely defines the orbital angular momentum~$\ell_\alpha$. Thus, it is convenient  to label our basis in terms of $\pi_\alpha$ and the particle-number parities $q_\alpha$ as these have corresponding good quantum numbers
for the many-body states of Eq.~\eqref{eq:Psi_k}.
To summarize, the collective index $\alpha$ denotes the set of quantum numbers
\begin{subequations}\label{eq:sp_bas}
\begin{align}
\label{eq:sp_bas_alpha}
\alpha \equiv{}& (n_\alpha ,  \pi_\alpha  , j_\alpha ,  q_\alpha , m_\alpha) = (a,  m_\alpha) \, ,
\end{align}
where we introduce a latin letter index,
\begin{align}
\label{eq:sp_bas_a}
 a \equiv{}& (n_\alpha ,  \pi_\alpha  , j_\alpha ,  q_\alpha) \, ,
\end{align}
\end{subequations}
to group the quantum numbers that are not contracted in the coupling of angular momenta. 

The dual basis  $\{|\bar{\alpha}\rangle\}$ can be made explicit by identifying the antiunitary transformation $\mathcal{T}$
with the time-reversal operator. Applying it to state~\eqref{eq:sph_sp} gives
\begin{align}
\label{eq:Time_rev}
 \mathcal{T} \, \phi_{a, m_\alpha}(\vec{r},\sigma,\tau)   ={}&   (-1)^{\ell_\alpha + j_\alpha - m_\alpha} \, \phi_{a, -m_\alpha}(\vec{r},\sigma,\tau) \, ,
\end{align}
from where one identifies the conjugate quantum number of $\alpha$ to be $\widetilde\alpha=(a,-m_\alpha)$. Since the parity $\pi_\alpha=(-1)^{\ell_\alpha}$ only 
introduces a global real phase, we define the antiunitary transformation simply as $\eta_\alpha\equiv(-1)^{ j_\alpha + m_\alpha}$.  It can be shown that  
\begin{subequations}
\label{eq:c_tensor_j} 
\begin{align}
 c^\dagger_\alpha  ={}& c^\dagger_{a,  m_\alpha} \, ,  \\   
 \bar{c}_\alpha  ={}& \eta_\alpha \, c_{\widetilde\alpha}  = (-1)^{ j_\alpha + m_\alpha} \, c_{a, -m_\alpha} \, ,   
\end{align}
\end{subequations}
are the $m_\alpha$-th components of irreducible tensor operators of rank $j_\alpha$.

We use a similar notation to Eq.~\eqref{eq:sp_bas} for  Gorkov quasiparticle indices $k$ and define the subset $\gbs{k}$
of rotationally invariant quantum numbers such that%
\footnote{The $\,\gbs{~}\,$ notation used here is unrelated to the definition of quantum number of the dual basis, Eq.~\eqref{eq:dualA}.  This should not cause confusion since the Gorkov quasiparticle and ISC indices, $k$ and $r$, do not posses a dual basis. Moreover, the distinction between direct and dual single-particle bases disappear from the angular momentum coupled equations discussed in the rest of this Appendix.}
\begin{align}
\label{eq:Gkv_k_bas}
 k ={}& (n_k, \pi_k, j_k, q_k, m_k) \equiv (\gbs{k}, m_k) \, .
\end{align}
Quantities $n_k$,  $\pi_k$,  $j_k$ and  $m_k$ denote the principal quantum number, parity, total angular momentum and its projection of the
many-body states in Eq.~\eqref{eq:Psi_k}. Similarly, 
\begin{align}
q_k \equiv \left|  \langle\Psi_k | N |\Psi_k \rangle - \text{N} \right| \mod  2 
\label{eq:def_qk}
\end{align}
is the difference between the particle-number parities of $|\Psi_k\rangle$ and $|\Psi_0\rangle$. Eq.~\eqref{eq:def_qk} is intended element-wise for all types of distinguishable fermions.
The ISC indices~\eqref{eq:r_indx} will require coupling to total angular momentum $J_r$ and projection $M_r$. In this case we choose the convention of coupling the first two indices to the intermediate angular momentum $J_{12}$, as detailed in Sec.~\ref{App:Jcoup_adc}, and define the general index $r=(\gbs{r},M_r)$ with~$\gbs{r}\equiv[(\gbs{k}_1,\gbs{k}_2, J_{12}, \gbs{k}_3), J_r]$.

Let us now take the assumption that the target ground state $|\Psi_0\rangle$ is spherical with good angular momentum and parity $J^\pi=0^+$. Considering the definition of the 
spectroscopic amplitudes~\eqref{eq:def_UV} and applying the Wigner-Eckart theorem with the tensor operators~\eqref{eq:c_tensor_j}, one finds
\begin{subequations}\label{eq:UV_red}
\begin{align}
 \mathcal{U}^{k}_\alpha ={}&  \langle \Psi_0 |   c_\alpha | \Psi_k \rangle \nonumber \\ ={}&  \eta_{\widetilde\alpha} \langle \Psi_0 |   \bar{c}_{\widetilde\alpha} | \Psi_k \rangle  \nonumber \\
   ={}&  (-1)^{j_\alpha-m_\alpha}  \CG{j_k}{j_\alpha}{m_k}{-m_\alpha}{0}{0} \langle \Psi_0 ||   \bar{c}_a || \Psi_k \rangle \nonumber \\
   ={}&   \delta_{j_\alpha,  j_k}  \delta_{m_\alpha,  m_k}   \frac{  \langle \Psi_0 ||   \bar{c}_a || \Psi_k \rangle }{ \hat{j}_\alpha }  \nonumber \\
   \equiv{}&   \dpjq{a,  \gbs{k}}   \delta_{m_\alpha,  m_k}   \,   \mathcal{U}^{\gbs{k}}_a \, ,
\label{eq:Ubar_red}  \\
 \mathcal{V}^{k}_\alpha ={}&  \langle \Psi_0 |   \bar{c}^\dagger_\alpha | \Psi_k \rangle  \nonumber \\ ={} &   \eta_\alpha  \langle \Psi_0 |   c^\dagger_{\widetilde\alpha} | \Psi_k \rangle  \nonumber \\
 %={}&  (-1)^{j_\alpha-m_\alpha}  \CG{j_k}{j_\alpha}{m_k}{-m_\alpha}{0}{0} \langle \Psi_0 ||   c^\dagger_a || \Psi_k \rangle \nonumber \\
  ={}&   \delta_{j_\alpha,  j_k}  \delta_{m_\alpha,  m_k}   \frac{ \langle \Psi_0 ||   c^\dagger_a || \Psi_k \rangle}{\hat{j}_\alpha} (-1)^{2 j_\alpha} \nonumber \\
   \equiv{}&    \dpjq{a , \gbs{k}}  \delta_{m_\alpha,  m_k}   \,  \mathcal{V}^{\gbs{k}}_a \, ,
\label{eq:Vbar_red}  
\end{align}
where the notation $\hat{j}\equiv\sqrt{2j+1}$ is used and $\CG{j_1}{j_2}{m_1}{m_2}{j_3}{m_3}$ denotes the usual Clebsh-Gordan coefficient. Applying transformations~\eqref{eq:def_UVbar} yields
\begin{align}
 \bar{\mathcal U}^{k}_\alpha ={}&  \quad \eta_\alpha  \mathcal{U}^{k}_{\widetilde{\alpha}}  =   \dpjq{a,  \gbs{k}}  \delta_{m_\alpha, -m_k}   \,   (-1)^{j_\alpha+m_\alpha} \,   \mathcal{U}^{\gbs{k}}_a \, ,
\label{eq:U_red}  \\
 \bar{\mathcal V}^{k}_\alpha ={}&   - \eta_\alpha  \mathcal{V}^{k}_{\widetilde{\alpha}} =  \dpjq{a,  \gbs{k}}   \delta_{m_\alpha, -m_k}   \, (-1)^{j_\alpha-m_\alpha}  \, \mathcal{V}^{\gbs{k}}_a \, .
\label{eq:V_red}  
\end{align}
\end{subequations}
In Eqs.~\eqref{eq:UV_red}, we have applied the conservation of particle-number parity and of parity in coordinate space and have introduced a compact notation for multiple Kronecker $\delta$s
on the conserved symmetries:
\begin{align}
   \dpjq{a,   \gbs{k}} \equiv{}&   \delta_{\pi_\alpha,  \pi_k} \,  \delta_{j_\alpha,  j_k}  \,  \delta_{q_\alpha,  q_k}  \,  ,
\label{eq:Kro_dlt}
\end{align}
where, again, $\delta_{q_\alpha  q_k}$ is element-wise on the vectors $q_\alpha$ and  $q_k$. 
Note that Eq.~\eqref{eq:Kro_dlt} does not imply an equality on the principal quantum numbers, thus \hbox{$\delta_{\gbs{k}_1,\gbs{k}_2}=\delta_{n_{k_1},n_{k_2}}\dpjq{\gbs{k}_1,   \gbs{k}_2}$}.
Equations~\eqref{eq:UV_red} show that all spectroscopic amplitudes, barred and non barred, amount down to the same two reduced quantities $\mathcal{U}^{\gbs{k}}_a$ and $\mathcal{V}^{\gbs{k}}_a$ plus some phase factor.

The Hamiltonian~\eqref{eq:H} is also independent of the third component of total angular momentum due to rotational symmetry. 
The most general one-body operator that we encounter is given by Eq.~\eqref{eq:def_U} and breaks particle number symmetry. Its matrix elements can then factored as
\begin{subequations}\label{eq:UJab}
\begin{align}
                  u_{\alpha \beta} ={}&  \dpjq{a,  b}   \delta_{m_\alpha,  m_\beta} \, u_{a  b} \, , \label{eq:UJab_norm} \\
 u^{an.}_{\alpha \beta} ={}&  \dpjq{a,  b}   \delta_{m_\alpha,  -m_\beta} \,  (-1)^{j_\beta-m_\beta} \; \widetilde{u}_{a  b} \, ,  \label{eq:UJab_anom}
\end{align}
\end{subequations}
where the $\delta^{(\pi j q)}$s reflect the fact that conservation of spatial parity and of odd or even particle number is still assumed.
For the two-body interaction $V$ we adopt the usual angular momentum coupling convention for its properly normalised matrix elements,
\begin{align}
  v^J_{a b, g d} = {}&\frac 1{\sqrt{1+\delta_{a, b}}} \frac 1{\sqrt{1+\delta_{g, d}}} \nonumber \\
     &~~ \times \sum_{\substack{m_\alpha m_\beta \\ \; m_\gamma m_\delta}} \CG{j_\alpha}{j_\beta}{m_\alpha}{m_\beta}{J}{M} \; v_{\alpha \beta, \gamma \delta} \; \CG{j_\gamma}{j_\delta}{m_\gamma}{m_\delta}{J}{M} \, .
\label{eq:VJabgd}
\end{align}
Moreover, the ADC(3) contributions discussed in Sec.~\ref{App:Jcoup_adc3} are conveniently expressed in terms of the particle-hole coupling, which is related to Eq.~\eqref{eq:VJabgd} through the Pandya transformation:
\begin{align}
  v^{(ph)\,J}_{a b^{-1}, g d^{-1}} = {}&\sum_{J_2} (-1)^{j_\beta + j_\gamma + J_2} (2J_2+1) \CSJ{j_\alpha}{j_\beta}{J}{j_\gamma}{j_\delta}{J_2}  \nonumber \\
      & \quad \times \sqrt{1+\delta_{a, d}} \, v^{J_2}_{a d, b g} \, \sqrt{1+\delta_{b, g}} \, .
% \\  v^{(ph)\,J}_{a b^{-1}, g d^{-1}} = {}& - \sum_{J_2} (-1)^{2J_2} (2J_2+1) \CSJ{j_\alpha}{j_\beta}{J}{j_\gamma}{j_\delta}{J_2}  \sqrt{1+\delta_{a d}} v^{J_2}_{a d, g b} \sqrt{1+\delta_{b g}} \, . 
\label{eq:Vph_abgd}
\end{align}

\subsection{\label{App:Jcoup_cHFB}First-order self-energy}

From the definitions~\eqref{eq:UV_red} one can easily derive the angular momentum form of the reduced density matrices:
\begin{align}
\rho_{\alpha \beta} ={}&  \dpjq{a,  b} \delta_{m_\alpha,  m_\beta} \, \sum_\gbs{k}    \dpjq{a,  \gbs{k}} \; (  \mathcal{V}^\gbs{k}_a)^* \,  \mathcal{V}^\gbs{k}_b  \nonumber \\
  ={}&   \dpjq{a,  b}  \delta_{m_\alpha,  m_\beta} \, \rho_{a  b}  \, ,
\label{eq:rhoJ_ab} 
\end{align}
and
\begin{align}
 \tilde{\rho}_{\alpha \beta} ={}& \dpjq{a,  b} \delta_{m_\alpha,  m_\beta} \,  (-1)^{2j_\alpha} \sum_\gbs{k}    \dpjq{a,  \gbs{k}} \; (  \mathcal{V}^\gbs{k}_a)^* \,  \mathcal{U}^\gbs{k}_b  \nonumber \\
 ={}&   \dpjq{a,  b}  \delta_{m_\alpha,  m_\beta} \, \tilde{\rho}_{a  b}  \, ,
\label{eq:kappaJ_ab}
\end{align}
Inserting the latter into Eqs.~\eqref{eq:SigI_expr} leads to the following expression for the energy-independent terms of the self-energy
\begin{widetext}
\begin{subequations}\label{eq:Sig1J_ab}
\begin{align}
\Sigma^{(\infty)\,11}_{\alpha \beta} ={}&  \dpjq{a,b}  \delta_{m_\alpha, m_\beta}  \sum_{g \, d \, J}  \dpjq{g , d} \frac{2J+1}{2 j_\alpha +1} \sqrt{1+\delta_{a, g} }  v^J_{a g , b d} \sqrt{1+\delta_{b, d} } \, \rho_{d g}  % \nonumber \\
              \equiv   \dpjq{a,b}  \delta_{m_\alpha, m_\beta}  \Lambda_{a b} \label{eq:Sig1J_ab11} \; , \\
\Sigma^{(\infty)\,12}_{\alpha \beta} ={}&  \dpjq{a,b}  \delta_{m_\alpha, m_\beta} \frac{(-1)^{2j_\alpha}}2 \sum_{g \, d }  \dpjq{g , d} (-1)^{2j_\gamma} \frac{\hat{j}_\gamma}{\hat{j}_\alpha} \sqrt{1+\delta_{a, b} }  v^{J=0}_{a b, g d} \sqrt{1+\delta_{g, d} } \, \tilde{\rho}_{g d}  % \nonumber \\
              \equiv   \dpjq{a,b}  \delta_{m_\alpha, m_\beta}  \tilde{h}_{a b}  \label{eq:Sig1J_ab12}  \; , \\
\Sigma^{(\infty)\,21}_{\alpha \beta} ={}&   \dpjq{a,b}  \delta_{m_\alpha, m_\beta}  ( \tilde{h}_{b a} )^* \label{eq:Sig1J_ab21}  \; , \\
\Sigma^{(\infty)\,22}_{\alpha \beta} ={}&  -  \dpjq{a,b}  \delta_{m_\alpha, m_\beta} \Lambda_{b a}     = -  \dpjq{a,b}  \delta_{m_\alpha, m_\beta} (\Lambda_{a b})^* \label{eq:Sig1J_ab22}  \; .
\end{align}
\end{subequations}
\end{widetext}

\subsection{\label{App:Jcoup_adc}Dynamic self-energy}

The dynamic part of the self-energy involves the amplitudes $\mathcal{C}_{\alpha,r}$ and  $\mathcal{D}_{r,\alpha}$  that couple single-particle states to the ISCs defined by Eqs.~\eqref{eq:r_indx}. 
Their contributions have been presented in Eqs.~\eqref{eq:ADC2} and Secs.~\ref{Sec:adc_3} and~\ref{Sec:Comb_3rd}  at different level of many-body truncation and are
fully antisymmetric with respect to the Gorkov indices in $r$. However, it is convenient to decompose each term (generally indicated by the symbol ``$\cdot$'' in the following) through cyclic permutations of partially antysimmetrized amplitudes
\begin{subequations}
\label{eq:defM123}
\begin{align}
  \mathcal{C}^{(\cdot)}_{\alpha,r} ={}& \frac 1{\sqrt 3} \mathcal{P}_{123} \,   \mathcal{M}^{(\cdot)}_{\alpha, k_1k_2 k_3} \, , % \mathcal{C}_{\alpha, r} \, , %= \frac 1{\sqrt{3}} \{  \mathcal{M}_{\alpha, k_1k_2 k_3} +   \mathcal{M}_{\alpha, k_3 k_1 k_2} +   \mathcal{M}_{\alpha, k_2 k_3 k_1} \}
 \\
  \mathcal{D}^{(\cdot)}_{r,\alpha} ={}& \frac 1{\sqrt 3} \mathcal{P}_{123} \,  \mathcal{N}^{(\cdot)}_{k_1k_2 k_3, \alpha} \, ,  % \mathcal{D}_{\alpha, r} \, , %= \frac 1{\sqrt{3}} \{  \mathcal{N}_{k_1k_2 k_3, \alpha} +   \mathcal{N}_{k_3 k_1 k_2, \alpha} +   \mathcal{N}_{k_2k_3 k_1, \alpha} \}
\end{align}
\end{subequations}
where $\mathcal{M}_{\alpha, k_1k_2k_3}=-\mathcal{M}_{\alpha, k_2k_1k_3}$ and $\mathcal{N}_{k_1k_2k_3, \alpha}=-\mathcal{N}_{k_2k_1k_3, \alpha}$  are antisymmetric with respect to the exchange of the first two indices in $r=( k_1,k_2,k_3)$.
With this choice, it is efficient to first couple Gorkov quasiparticles $k_1$ and $k_2$ to an intermediate angular momentum $J_{12}$ and then adding  $k_3$ to obtain the total angular momentum of the ISC, with quantum numbers $J_r$ and $M_r$.
We adopt this convention and define the angular-momentum coupled amplitudes through the relations
\begin{widetext}
\begin{subequations}\label{eq:MN_Jcoup_conv}
\begin{align}
  \sum_{\substack{m_{k_1}  m_{k_2} \\ m_{k_3}} }    \mathcal{M}_{\alpha, r}  \,    \CG{j_{k_1}}{j_{k_2}}{m_{k_1}}{m_{k_2}}{J_{12}}{M_{12}}  \, \CG{J_{12}}{j_{k_3}}{M_{12}}{m_{k_3}}{J_r}{M_r} \,   \equiv{}&
            \dpjq{a ,  \gbs{r}}   \delta_{m_\alpha, M_r}  \sqrt{\frac{1+ \delta_{\gbs{k}_1, \gbs{k}_2 }}2} \, \mathcal{M}_{a,\gbs{r}} \, ,
\label{eq:M_Jcoup}  \\
  \sum_{\substack{m_{k_1}  m_{k_2} \\ m_{k_3}} }    \mathcal{N}_{r, \alpha}  \,    \CG{j_{k_1}}{j_{k_2}}{m_{k_1}}{m_{k_2}}{J_{12}}{M_{12}}  \, \CG{J_{12}}{j_{k_3}}{M_{12}}{m_{k_3}}{J_r}{M_r} \,   \equiv{}&
              \dpjq{a, \gbs{r}}    \delta_{m_\alpha, M_r}  \sqrt{\frac{1+ \delta_{\gbs{k}_1, \gbs{k}_2 }}2} \,  \mathcal{N}_{\gbs{r}, a}  \, ,
\label{eq:N_Jcoup}
\end{align}
while the barred quantities follow through the equivalent of Eqs.~\eqref{eq:def_CDbar} for $\bar{\mathcal M}$ and $\bar{\mathcal N}$:
\begin{align}
  \sum_{\substack{m_{k_1}  m_{k_2} \\ m_{k_3}} }    \bar{\mathcal M}_{\alpha, r}  \,    \CG{j_{k_1}}{j_{k_2}}{m_{k_1}}{m_{k_2}}{J_{12}}{M_{12}}  \, \CG{J_{12}}{j_{k_3}}{M_{12}}{m_{k_3}}{J_r}{M_r} \,   ={}&
               \dpjq{a,  \gbs{r}}    \delta_{m_\alpha,  -M_r}  (-1)^{j_\alpha + m_\alpha} \sqrt{\frac{1+ \delta_{\gbs{k}_1, \gbs{k}_2 }}2} \, \mathcal{M}_{a,\gbs{r}} \, ,
\label{eq:Mbar_Jcoup}  \\
  \sum_{\substack{m_{k_1}  m_{k_2} \\ m_{k_3}} }    \bar{\mathcal N}_{r, \alpha}  \,    \CG{j_{k_1}}{j_{k_2}}{m_{k_1}}{m_{k_2}}{J_{12}}{M_{12}}  \, \CG{J_{12}}{j_{k_3}}{M_{12}}{m_{k_3}}{J_r}{M_r} \,   ={}&
               \dpjq{a,  \gbs{r}}  \delta_{m_\alpha, -M_r} (-1)^{j_\alpha - m_\alpha} \sqrt{\frac{1+ \delta_{\gbs{k}_1, \gbs{k}_2}}2} \,  \mathcal{N}_{\gbs{r}, a}   \, .
\label{eq:Nbar_Jcoup}
\end{align}
\end{subequations}
%\end{widetext}
Eqs.~\eqref{eq:MN_Jcoup_conv} apply to each separate contribution of the ADC($n$) expansion, as well as the corresponding fully antisymmetrised amplitudes $\mathcal{C}_{\alpha,r}$ and  $\mathcal{D}_{r,\alpha}$.
Note that the collective Kronecker $\delta$s entering the above equations involve the total angular momentum $J_r$,  parity $\pi_r\equiv\pi_{k_1}\pi_{k_2}\pi_{k_3}$ and particle-number parities
\hbox{$q_r\equiv |q_{k_1}+q_{k_2}+q_{k_3}| \mod 2$} (element-wise for all types of particles) of the ISC $r$ but \emph{do not} impose charge conservation.
 As we see below, some  $\mathcal{M}_{\alpha,r}$ and  $\mathcal{N}_{r,\alpha}$ amplitudes may have more stringent selection rules on $q_r$ that arise from charge conservation in the two-body interaction  but these are reshuffled by the permutations in Eqs.~\eqref{eq:defM123},
so that only the conservation of even and odd particle-number applies to the final $\mathcal{C}_{\alpha,r}$ and  $\mathcal{D}_{r,\alpha}$.

%\begin{widetext}
The angular-momentum coupled form of the cyclic permutation operator is given by
\begin{align}
   \sum_{\substack{m_{k_1}  m_{k_2} \\ m_{k_3}} }    \sum_{\substack{m_{k_4}  m_{k_5} \\ m_{k_6}} }  &  \,    \CG{j_{k_1}}{j_{k_2}}{m_{k_1}}{m_{k_2}}{J_{12}}{M_{12}}  \, \CG{J_{12}}{j_{k_3}}{M_{12}}{m_{k_3}}{J_r}{M_r} \,    \mathcal{P}_{r, r'} \,
      \CG{j_{k_4}}{j_{k_5}}{m_{k_4}}{m_{k_5}}{J_{45}}{M_{45}}  \, \CG{J_{45}}{j_{k_6}}{M_{45}}{m_{k_6}}{J_{r'}}{M_{r'}}   \nonumber \\
         ={}&  \left[    \delta_{\gbs{k}_1,  \gbs{k}_4 } \,  \delta_{\gbs{k}_2,  \gbs{k}_5 } \, \delta_{\gbs{k}_3,  \gbs{k}_6}  \,  \delta_{J_{12}, J_{45}}    +     \delta_{\gbs{k}_3,  \gbs{k}_4 } \,  \delta_{\gbs{k}_1,  \gbs{k}_5 } \, \delta_{\gbs{k}_2,  \gbs{k}_6}     
              (-1)^{j_{k_1} + j_{k_2} + 2j_{k_3} + J_{12}}  \hat{J}_{12}  \hat{J}_{45}   \CSJ{ j_{k_2} }{ j_{k_1} }{ J_{12} }{ j_{k_3} }{ J_r  }{ J_{45} }     \right.   \nonumber \\
          &+~  \left.    \delta_{\gbs{k}_2,  \gbs{k}_4 } \,  \delta_{\gbs{k}_3,  \gbs{k}_5 } \, \delta_{\gbs{k}_1,  \gbs{k}_6}   
              (-1)^{2j_{k_1} + j_{k_2} + j_{k_3} + J_{45}}  \hat{J}_{12}  \hat{J}_{45}   \CSJ{ j_{k_1} }{ j_{k_2} }{ J_{12} }{ j_{k_3} }{ J_r  }{ J_{45} }   \right]  \,  \delta_{J_r J_{r'}}  \delta_{M_r  M_{r'}}  \nonumber \\
         \equiv{}&      \dpjq{ \gbs{r},  \gbs{r}'}  \delta_{M_r , M_{r'}} \, \sqrt{\frac{1+  \delta_{\gbs{k}_1, \gbs{k}_2 }}2} \; \mathcal{P}_{\gbs{r} ,\gbs{r}'} \; \sqrt{\frac{1+ \delta_{\gbs{k}_4, \gbs{k}_5 }}2} 
\label{eq:P123_Jcoup}
\end{align}
in terms of Wigner 6-j coefficients.

For the first-order corrections to the energy denominators, a similar combination of cyclic permutations is employed
\begin{align}
  \mathcal{E}^{(I\cdot)}_{r, r'} \equiv{}& \frac 1 3  \mathcal{P}_{123} \,   \mathcal{F}^{(I\cdot)}_{k_1k_2 k_3, k_4k_5 k_6}   \mathcal{P}_{456} \, ,
\label{eq:def_Frr}
\end{align}
whereas the angular-momentum coupling for the partially antisymmetrized energy is defined as follows
\begin{align}
  \sum_{\substack{m_{k_1}  m_{k_2} \\ m_{k_3}} }    \sum_{\substack{m_{k_4}  m_{k_5} \\ m_{k_6}} }    \,    \CG{j_{k_1}}{j_{k_2}}{m_{k_1}}{m_{k_2}}{J_{12}}{M_{12}}  \,  \CG{J_{12}}{j_{k_3}}{M_{12}}{m_{k_3}}{J_r}{M_r} \,    \mathcal{F}_{r, r'} \,
    &  \CG{j_{k_4}}{j_{k_5}}{m_{k_4}}{m_{k_5}}{J_{45}}{M_{45}}  \, \CG{J_{45}}{j_{k_6}}{M_{45}}{m_{k_6}}{J_{r'}}{M_{r'}}   \nonumber \\
       \equiv{}&      \dpjq{ \gbs{r},  \gbs{r}'}   \delta_{M_r,  M_{r'}} \, \sqrt{\frac{1+  \delta_{\gbs{k}_1, \gbs{k}_2 }}2}  \; \mathcal{F}_{\gbs{r} ,\gbs{r}'} \;   \sqrt{\frac{1+ \delta_{\gbs{k}_4, \gbs{k}_5 }}2} \; .
\label{eq:E_Jcoup} 
\end{align}
%\end{widetext}
It is convenient to factor out the terms $\sqrt{(1+  \delta_{\gbs{k}_1, \gbs{k}_2})/2}$ from definitions~\eqref{eq:MN_Jcoup_conv} through~\eqref{eq:E_Jcoup} because these cancel out when restricting the sums over $\gbs{r}$ to ISCs ordered in the first two indices, $\gbs{k}_1 \lesssim\gbs{k}_2$ (see Sec.~\eqref{App:Gmtx_jcoup} below).

\subsubsection{\label{App:Jcoup_adc2} ADC(2) amplitudes}

The ADC(2) version of the $ \mathcal{M}$ and $ \mathcal{N}$ amplitudes is given by Eqs.~\eqref{eq:ADC2}.  
Following definitions~\eqref{eq:M_Jcoup} and~\eqref{eq:N_Jcoup}, one has
%\begin{widetext}
\begin{align}
    \mathcal{M}^{(I)}_{a,\gbs{r}} ={}&  \Delta(j_{k_1}, j_{k_2}, J_{12})  \Delta(J_{12}, j_{k_3}, J_r )  \, (-1)^{ j_\alpha  + j_{k_3} - J_{12}} \,  \frac{ \hat{J}_{12} }{ \hat{j}_\alpha }   \nonumber \\
     & \times \sum_{\substack{m \lesssim v \\ l} }
         \, \sqrt{ 1 + \delta_{a, l} } \,  v^{J_{12}}_{al,mv}
       \frac { \mathcal{U}^{\gbs{k}_1}_m  \dpjq{m, \gbs{k}_1 }   \mathcal{U}^{\gbs{k}_2}_v  \dpjq{v, \gbs{k}_2}  - (-1)^{j_{k_1} + j_{k_2} - J_{12}}     \mathcal{U}^{\gbs{k}_1}_v  \dpjq{v, \gbs{k}_1 }   \mathcal{U}^{\gbs{k}_2}_m  \dpjq{m, \gbs{k}_2}  } 
                {  \sqrt{ 1 + \delta_{m, v } } \sqrt{ 1 + \delta_{\gbs{k}_1, \gbs{k}_2 } }  } 
       \mathcal{V}^{\gbs{k}_3}_l    \dpjq{l, \gbs{k}_3 } 
\label{eq:MI_Jcoup}
\end{align}
and
\begin{align}
    \mathcal{N}^{(I)}_{\gbs{r}, a} ={}&  \Delta(j_{k_1}, j_{k_2}, J_{12})  \Delta(J_{12}, j_{k_3}, J_r ) \, (-1)^{ j_\alpha  - j_{k_3} - J_{12}} \,  \frac{ \hat{J}_{12} }{ \hat{j}_\alpha }   \nonumber \\
     & \times \sum_{\substack{m \lesssim v \\ l} }
       \frac { \mathcal{V}^{\gbs{k}_1}_m  \dpjq{m, \gbs{k}_1 }   \mathcal{V}^{\gbs{k}_2}_v  \dpjq{v, \gbs{k}_2}  - (-1)^{j_{k_1} + j_{k_2} - J_{12}}     \mathcal{V}^{\gbs{k}_1}_v  \dpjq{v, \gbs{k}_1 }   \mathcal{V}^{\gbs{k}_2}_m  \dpjq{m, \gbs{k}_2}  } 
                { \sqrt{ 1 + \delta_{\gbs{k}_1, \gbs{k}_2 } }  \sqrt{ 1 + \delta_{m, v } }  }  \,
        v^{J_{12}}_{mv,al} \, \sqrt{ 1 + \delta_{a, l} } \,  \mathcal{U}^{\gbs{k}_3}_l   \dpjq{l, \gbs{k}_3 } \, ,
\label{eq:NI_Jcoup}
\end{align}
where
\begin{align}
     \Delta(j_{1}, j_{2}, j_{3})  ={}&  \left\{  \begin{array}{lcl}
        1 & \quad & \hbox{if } |j_1-j_2| \leqslant j_3 \leqslant j_1+j_2 \, ,\\
        0& \quad & \hbox{otherwise} \,
    \end{array} \right.
\label{eq:Triang_ineq}
\end{align}
is the triangular condition. The sums over single-particle states $m$ and $v$ have been restricted to ordered $m\lesssim v$ by exploiting the antisymmetry of $v^{J_{12}}_{mv,al}$.  Here and in the rest of the this Appendix, we interpret the inequality $m \lesssim v$ as
$m \leqslant v$ if $(-1)^{j_{\mu} + j_{\nu} - J_{12}}=-1$ is statisfied and as $m<v$ otherwise.

The energy denominator for three freely propagating quasiparticles, Eq.~\eqref{eq:ADC2_E}, is diagonal in the ISCs indices and does not need to be antisymmetrized. Hence, we follow a different convention than Eq.~\eqref{eq:E_Jcoup} and define
\begin{align}
  \sum_{\substack{m_{k_1}  m_{k_2} \\ m_{k_3}} }    \sum_{\substack{m_{k_4}  m_{k_5} \\ m_{k_6}} }    \,   &  \CG{j_{k_1}}{j_{k_2}}{m_{k_1}}{m_{k_2}}{J_{12}}{M_{12}}  \,  \CG{J_{12}}{j_{k_3}}{M_{12}}{m_{k_3}}{J_r}{M_r} \,    \mathcal{E}^{(0)}_{r, r'} \,
      \CG{j_{k_4}}{j_{k_5}}{m_{k_4}}{m_{k_5}}{J_{45}}{M_{45}}  \, \CG{J_{45}}{j_{k_6}}{M_{45}}{m_{k_6}}{J_{r'}}{M_{r'}}   \nonumber \\
      ={}&       \delta_{\gbs{k}_1,  \gbs{k}_4 } \,  \delta_{\gbs{k}_2,  \gbs{k}_5 } \, \delta_{\gbs{k}_3,  \gbs{k}_6}  \,  \delta_{J_{12}, J_{45}} \,  \delta_{J_r, J_{r'}}  \delta_{M_r,  M_{r'}}  \,  \left( \omega_{\gbs{k}_1} + \omega_{\gbs{k}_2} + \omega_{\gbs{k}_3} \right)  \nonumber \\
       \equiv{}&      \dpjq{ \gbs{r},  \gbs{r}'}   \delta_{M_r,  M_{r'}}    \, \mathcal{E}^{(0)}_{\gbs{r} ,\gbs{r}'} \, .
\label{eq:E0_Jcoup} 
\end{align}
\end{widetext}

\subsubsection{\label{App:Jcoup_adc3} Self-consistent ADC(3) amplitudes}

ADC(3) contributions require summation on the unperturbed doublet amplitudes~\eqref{eq:BCC_t2}. Here, an angular-momentum coupling convention similar to the one used for two-body interactions is adopted and the rotationally-invariant matrix elements is defined as
\begin{align}
   t^{\gbs{k}_1 \, \gbs{k}_2 \, J}_{\gbs{k}_3 \, \gbs{k}_4} 
    \equiv{}& \!\! \sum_{\substack{m_{k_1} \, m_{k_2} ~\\~ m_{k_3} \, m_{k_4}} } \!\!\!\! \CG{j_{k_1}}{j_{k_2}}{m_{k_1}}{m_{k_2}}{J}{M}    t^{k_1 k_2}_{k_3 k_4}   \,   \CG{j_{k_3}}{j_{k_4}}{m_{k_3}}{m_{k_4}}{J}{\,-\!M} (-1)^{J-M}  \nonumber \\
    ={}& \Delta(j_{k_1}, j_{k_2}, J)  \Delta(j_{k_3}, j_{k_4}, J )  \nonumber \\
     \times & \!\!\sum_{a \, b \, g \, d} \!\!\! \frac { \mathcal{V}^{\gbs{k}_3}_a  \dpjq{a, \gbs{k}_3 } \,  \mathcal{V}^{\gbs{k}_4}_b  \dpjq{b, \gbs{k}_4}   \, v^{J}_{ab,gd}  \,      \mathcal{U}^{\gbs{k}_1}_g  \dpjq{g, \gbs{k}_1 } \,  \mathcal{U}^{\gbs{k}_2}_d  \dpjq{d, \gbs{k}_2}} {- (\omega_{\gbs{k}_1} + \omega_{\gbs{k}_2} + \omega_{\gbs{k}_3} + \omega_{\gbs{k}_4})} \; ,
\label{eq:T2_Jcoup}
\end{align}
where the upper (lower) indices are coupled together as they correspond to direct interactions among pairs of Gorkov quasiparticles.  Whenever the quasiparticles are coupled through a two-body interaction in the particle-hole channel, it becomes more efficient to
first perform a Pandya transformation similar to Eq.~\eqref{eq:Vph_abgd}. In particular, we choose the convention
\begin{align}
   t^{\gbs{k}_1 \, \gbs{k}_3 \, (ph) J}_{\gbs{k}_2 \, \gbs{k}_4}    ={}& \sum_{J_2} (-1)^{ j_{k_2} - j_{k_4} + J_2} (2 J_2 + 1) \nonumber \\
   &\qquad \qquad \times \CSJ{j_{k_3}}{j_{k_4}}{J}{j_{k_2}}{j_{k_1}}{J_2} \,   t^{\gbs{k}_1 \, \gbs{k}_3 \, J_2}_{\gbs{k}_2 \, \gbs{k}_4}  \; .
\label{eq:T2ph_Jcoup}
\end{align}

Given the above definitions, we are  in the position to state the angular momentum coupled form for the skeleton ADC(3) diagrams. 
Contributions from Eqs.~\eqref{eq:3rd_CDpp} and~\eqref{eq:3rd_CDhh} can be conveniently gathered as the only difference among them is in the inversion of the upper and lower indices in $ t^{k_1 k_2}_{k_3 k_4}$. 
After introducing a common factor for normalization and angular-momentum conditions,
\begin{align}
    K(\gbs{r}) ={}& \frac { \Delta(j_{k_1}, j_{k_2}, J_{12})  \Delta(J_{12}, j_{k_3}, J_r ) }{ \sqrt{ 1 + \delta_{\gbs{k}_1, \gbs{k}_2 } } } \, ,
\end{align}
the $\mathcal{M}_{a,\gbs{r}}$ amplitudes read as
\begin{widetext}
\begin{subequations}\label{eq:MII_Jcoup}
\begin{align}
    \mathcal{M}^{(\rm{IIa+c})}_{a,\gbs{r}} ={}&  (-1)^{ j_\alpha  + j_{k_3} - J_{12}}  \,  K(\gbs{r}) \,  \frac{ \hat{J}_{12} }{ \hat{j}_\alpha }   
      \sum_{\substack{m \, v \, l} } \sum_{  \gbs{k}_7 \,  \gbs{k}_8 }   \,   \sqrt{ 1 + \delta_{a, l } }  \, v^{J_{12}}_{al,mv}  \,   \sqrt{ 1 + \delta_{m, v } }   \nonumber \\   
     & \qquad \times    \frac 1 2   \left( \mathcal{V}^{\gbs{k}_7}_m  \dpjq{m, \gbs{k}_7 } \,  \mathcal{V}^{\gbs{k}_8}_v  \dpjq{v, \gbs{k}_8 } \right)^* \,
           \left[ t^{\gbs{k}_1\, \gbs{k}_2\, J_{12}}_{\gbs{k}_7 \, \gbs{k}_8} + (-1)^{2 J_{12}} \, t^{\gbs{k}_7 \, \gbs{k}_8\, J_{12}}_{\gbs{k}_1 \, \gbs{k}_2}   \right]   \mathcal{V}^{\gbs{k}_3}_l  \dpjq{l, \gbs{k}_3 } \, ,
\label{eq:MIIac_Jcoup}  \\
%    %\mathcal{M}^{(\rm{IIa+c})}_{a,\gbs{r}} 
%    ={}&  (-1)^{ j_\alpha  + j_{k_3} - J_{12}}  \,  K(\gbs{r}) \,  \frac{ \hat{J}_{12} }{ \hat{j}_\alpha }   
%      \sum_{\substack{m \leq v \\ l} } \sum_{  \gbs{k}_7 \leq  \gbs{k}_8 }    \,    \sqrt{ 1 + \delta_{a, l } }  \,  v^{J_{12}}_{al,mv}    \nonumber \\
%      & \quad \times   \frac{ \left( \mathcal{V}^{\gbs{k}_7}_m  \dpjq{m, \gbs{k}_7 } \,  \mathcal{V}^{\gbs{k}_8}_v  \dpjq{v, \gbs{k}_8 }  - (-1)^{j_\mu + j_\nu - J_{12}}   \,  \mathcal{V}^{\gbs{k}_8}_m  \dpjq{m, \gbs{k}_8 } \, \mathcal{V}^{\gbs{k}_7}_v  \dpjq{v, \gbs{k}_7 } \right)^* }
%                   { \sqrt{ 1 + \delta_{m v} }     \; ( 1 + \delta_{ \gbs{k}_7 ,  \gbs{k}_8 } ) }  
%    \left[     t^{\gbs{k}_1\, \gbs{k}_2\, J_{12}}_{\gbs{k}_7 \, \gbs{k}_8}   + (-1)^{2 J_{12}} \, t^{\gbs{k}_7 \, \gbs{k}_8\, J_{12}}_{\gbs{k}_1 \, \gbs{k}_2}      \right]   \,  \mathcal{V}^{\gbs{k}_3}_l  \dpjq{l, \gbs{k}_3 }   ,
%\label{eq:MIIac_Jcoup_bis}  \\
%\end{align}
%\begin{align}
    \mathcal{M}^{(\rm{IIb+d})}_{a,\gbs{r}} ={}&  (-1)^{ j_\alpha  + j_{k_3} - J_{12}}  \,  K(\gbs{r}) \,  \frac{ \hat{J}_{12} }{ \hat{j}_\alpha }   
      %\sum_{\substack{m \, v \, l \\  \gbs{k}_7 ,  \gbs{k}_8 }}   
      \sum_{m \, v \, l} \sum_{  \gbs{k}_7 \,  \gbs{k}_8 }     (-1)^{2 j_{k_3} + 2 j_{k_8} }  \,  v^{(ph)\, J_{12}}_{am^{-1},vl^{-1}}   \nonumber \\
     & \qquad \times   \left( \mathcal{V}^{\gbs{k}_7}_v  \dpjq{v, \gbs{k}_7 } \,  \mathcal{U}^{\gbs{k}_8}_l  \dpjq{l, \gbs{k}_8 } \right)^* \,
           \left[ t^{\gbs{k}_1\, \gbs{k}_2\, J_{12}}_{\gbs{k}_7 \, \gbs{k}_8} + (-1)^{2 J_{12}} \, t^{\gbs{k}_7 \, \gbs{k}_8\, J_{12}}_{\gbs{k}_1 \, \gbs{k}_2}   \right]   \mathcal{U}^{\gbs{k}_3}_m  \dpjq{m, \gbs{k}_3 } \, .
\label{eq:MIIbd_Jcoup2}
\end{align}
The particle-hole interactions from diagrams of Fig.~\ref{fig:ADC3_C} and Eqs.~\eqref{eq:3rd_CIIe},\eqref{eq:3rd_CIIf} and \eqref{eq:3rd_CIIg} are
\begin{align}
    \mathcal{M}^{(\rm{IIe'})}_{a,\gbs{r}} ={}&  (-1)^{ j_\alpha  + j_{k_3} - J_{12}}  \,  K(\gbs{r}) \,  \frac{ \hat{J}_{12} }{ \hat{j}_\alpha }   
      %\sum_{\substack{m \, v \, l \\  \gbs{k}_7 ,  \gbs{k}_8 }}   
      \sum_{m \, v \, l} \sum_{  \gbs{k}_7 \,  \gbs{k}_8 }     (-1)^{2j_{k_3} + 2j_{k_8}}  \,  v^{(ph)\, J_{12}}_{am^{-1},vl^{-1}}   \nonumber \\
     & ~~ \times   \left( \mathcal{V}^{\gbs{k}_7}_v  \dpjq{v, \gbs{k}_7 } \,  \mathcal{U}^{\gbs{k}_8}_l  \dpjq{l, \gbs{k}_8 } \right)^* \,
           \left[ (-1)^{j_{k_1} + j_{k_2} - j_{k_7} - j_{k_8}} \, t^{\gbs{k}_8\, \gbs{k}_1\, (ph) J_{12}}_{\gbs{k}_7 \, \gbs{k}_2} \, + \,  t^{\gbs{k}_7\, \gbs{k}_2\, (ph) J_{12}}_{\gbs{k}_8 \, \gbs{k}_1}   \right]   \mathcal{U}^{\gbs{k}_3}_m  \dpjq{m, \gbs{k}_3 }  \, ,
\label{eq:MIIe1_Jcoup}  \\
%\end{align}
%\begin{align}
    \mathcal{M}^{(\rm{IIf'})}_{a,\gbs{r}} ={}&  (-1)^{ j_\alpha  + j_{k_3} - J_{12}}  \,  K(\gbs{r}) \,  \frac{ \hat{J}_{12} }{ \hat{j}_\alpha }   
      %\sum_{\substack{m \, v \, l \\  \gbs{k}_7 ,  \gbs{k}_8 }}   
      \sum_{m \, v \, l} \sum_{  \gbs{k}_7 \,  \gbs{k}_8 }     (-1)^{1 + 2j_{k_3} + 2j_{k_7} } \,  v^{(ph)\, J_{12}}_{av^{-1},ml^{-1}}   \nonumber \\
     &~~ \times    \left( \mathcal{U}^{\gbs{k}_7}_l  \dpjq{l, \gbs{k}_7 } \, \mathcal{V}^{\gbs{k}_8}_m  \dpjq{m, \gbs{k}_8 }  \right)^* \,
           \left[ (-1)^{j_{k_1} + j_{k_2}  - J_{12} } \, t^{\gbs{k}_8\, \gbs{k}_1\, (ph) J_{12}}_{\gbs{k}_7 \, \gbs{k}_2} \, + \, (-1)^{j_{k_7} + j_{k_8} - J_{12} } \, t^{\gbs{k}_7\, \gbs{k}_2\, (ph) J_{12}}_{\gbs{k}_8 \, \gbs{k}_1}   \right]   \mathcal{U}^{\gbs{k}_3}_v  \dpjq{v, \gbs{k}_3 } \, ,
\label{eq:MIIf1_Jcoup} \\
%\end{align}
%\begin{align}
    \mathcal{M}^{(\rm{IIg})}_{a,\gbs{r}} ={}&  (-1)^{ j_\alpha  + j_{k_3} - J_{12}}  \,  K(\gbs{r}) \,  \frac{ \hat{J}_{12} }{ \hat{j}_\alpha }   
      %\sum_{\substack{m \, v \, l \\  \gbs{k}_7 ,  \gbs{k}_8 }}   
      \sum_{m \, v \, l} \sum_{  \gbs{k}_7 \,  \gbs{k}_8 }   \,   \sqrt{ 1 + \delta_{a, l } } \,  v^{J_{12}}_{al,mv}  \,   \sqrt{ 1 + \delta_{m, v } }  \nonumber \\
     & ~~ \times    \left( \mathcal{V}^{\gbs{k}_7}_m  \dpjq{m, \gbs{k}_7 } \, \mathcal{V}^{\gbs{k}_8}_v  \dpjq{v, \gbs{k}_8 }  \right)^* \,
           \left[ (-1)^{j_{k_1} + j_{k_2} - j_{k_7} - j_{k_8}} \, t^{\gbs{k}_8\, \gbs{k}_1\, (ph) J_{12}}_{\gbs{k}_7 \, \gbs{k}_2} \, + \, t^{\gbs{k}_7\, \gbs{k}_2\, (ph) J_{12}}_{\gbs{k}_8 \, \gbs{k}_1}   \right]   \mathcal{V}^{\gbs{k}_3}_l  \dpjq{l, \gbs{k}_3 } \, .
\label{eq:MIIg_Jcoup}
\end{align}
\end{subequations}
where we used the primed superscripts ($\rm{e'}$ and $\rm{f'}$) to indicate that Eqs.~\eqref{eq:MIIe1_Jcoup} and~\eqref{eq:MIIf1_Jcoup} are actually linear combinations of~\eqref{eq:3rd_CIIe} and~\eqref{eq:3rd_CIIf}.

Similarly, Eqs.~\eqref{eq:3rd_DIIa}-\eqref{eq:3rd_DIIb} and~\eqref{eq:3rd_DIIc}-\eqref{eq:3rd_DIId} for the $\mathcal{N}_{\gbs{r}, a}$ amplitudes give
\begin{subequations}\label{eq:NII_Jcoup}
\begin{align}
    \mathcal{N}^{(\rm{IIa+c})}_{\gbs{r}, a} ={}&  (-1)^{ j_\alpha  - j_{k_3} - J_{12}}  \,  K(\gbs{r}) \,  \frac{ \hat{J}_{12} }{ \hat{j}_\alpha }   
      \sum_{\substack{m \, v \, l} } \sum_{  \gbs{k}_7 \,  \gbs{k}_8 }   \,   \frac 1 2     \left[  t^{\gbs{k}_7 \, \gbs{k}_8\, J_{12}}_{\gbs{k}_1 \, \gbs{k}_2}  + (-1)^{2 J_{12}} \, t^{\gbs{k}_1\, \gbs{k}_2\, J_{12}}_{\gbs{k}_7 \, \gbs{k}_8}  \right]   \nonumber \\
     & \qquad \times     \left( \mathcal{U}^{\gbs{k}_7}_m  \dpjq{m, \gbs{k}_7 } \,  \mathcal{U}^{\gbs{k}_8}_v  \dpjq{v, \gbs{k}_8 } \right)^* \,
           \mathcal{U}^{\gbs{k}_3}_l  \dpjq{l, \gbs{k}_3 } \, 
                \sqrt{ 1 + \delta_{m, v } }  \,  v^{J_{12}}_{mv,al} \,   \sqrt{ 1 + \delta_{a, l } } \, ,
\label{eq:NIIac_Jcoup}  \\
%          \mathcal{N}^{(\rm{IIa+c})}_{\gbs{r}, a}
%        ={}&  (-1)^{ j_\alpha  - j_{k_3} - J_{12}}  \,  K(\gbs{r}) \,  \frac{ \hat{J}_{12} }{ \hat{j}_\alpha }   
%      \sum_{\substack{m \leq v \\ l} } \sum_{  \gbs{k}_7 \leq  \gbs{k}_8 }    \,     \left[  t^{\gbs{k}_7 \, \gbs{k}_8\, J_{12}}_{\gbs{k}_1 \, \gbs{k}_2}  + (-1)^{2 J_{12}} \, t^{\gbs{k}_1\, \gbs{k}_2\, J_{12}}_{\gbs{k}_7 \, \gbs{k}_8}  \right]   \nonumber \\
%      & \qquad \times    \frac{    \left( \mathcal{U}^{\gbs{k}_7}_m  \dpjq{m, \gbs{k}_7 } \,  \mathcal{U}^{\gbs{k}_8}_v  \dpjq{v, \gbs{k}_8}   - (-1)^{j_\mu + j_\nu - J_{12}}  \, \mathcal{U}^{\gbs{k}_7}_v  \dpjq{v, \gbs{k}_7 } \,  \mathcal{U}^{\gbs{k}_8}_m  \dpjq{m, \gbs{k}_8}   \right)^*  }
%                   {    ( 1 + \delta_{ \gbs{k}_7 ,  \gbs{k}_8 } ) \;  \sqrt{ 1 + \delta_{m v} }     }   \,  v^{J_{12}}_{mv,al} \,   \sqrt{ 1 + \delta_{a, l } }  \, ,
%\label{eq:NIIac_Jcoup_bis}  \\
%\end{align}
%\begin{align}
    \mathcal{N}^{(\rm{IIb+d})}_{\gbs{r}, a} ={}&  (-1)^{ j_\alpha  - j_{k_3} - J_{12}}  \,  K(\gbs{r}) \,  \frac{ \hat{J}_{12} }{ \hat{j}_\alpha }   
      %\sum_{\substack{m \, v \, l \\  \gbs{k}_7 ,  \gbs{k}_8 }}   
      \sum_{m \, v \, l} \sum_{  \gbs{k}_7 \,  \gbs{k}_8 }     \left[  t^{\gbs{k}_7 \, \gbs{k}_8\, J_{12}}_{\gbs{k}_1 \, \gbs{k}_2}  + (-1)^{2 J_{12}} \, t^{\gbs{k}_1\, \gbs{k}_2\, J_{12}}_{\gbs{k}_7 \, \gbs{k}_8}  \right]   \nonumber \\
     & \qquad \times   \left( \mathcal{U}^{\gbs{k}_7}_v  \dpjq{v, \gbs{k}_7 } \,  \mathcal{V}^{\gbs{k}_8}_l  \dpjq{l, \gbs{k}_8 } \right)^* \,  \mathcal{V}^{\gbs{k}_3}_m  \dpjq{m, \gbs{k}_3 }  \,  v^{(ph)\, J_{12}}_{vl^{-1},am^{-1}}   \, ,
\label{eq:NIIbd_Jcoup2}
\end{align}
while the coupling of particle-hole interactions from the diagrams of Fig.~\ref{fig:ADC3_C} and Eqs.~\eqref{eq:3rd_DIIe},\eqref{eq:3rd_DIIf} and \eqref{eq:3rd_DIIg} are
\begin{align}
     \mathcal{N}^{(\rm{IIe'})}_{\gbs{r}, a} ={}&  (-1)^{ j_\alpha  - j_{k_3} - J_{12}}  \,  K(\gbs{r}) \,  \frac{ \hat{J}_{12} }{ \hat{j}_\alpha }   
      %\sum_{\substack{m \, v \, l \\  \gbs{k}_7 ,  \gbs{k}_8 }}   
     (-1) \sum_{m \, v \, l}   \sum_{  \gbs{k}_7 \,  \gbs{k}_8 }    \left[   (-1)^{j_{k_1} + j_{k_2}  - J_{12}} \, t^{\gbs{k}_2\, \gbs{k}_7\, (ph) J_{12}}_{\gbs{k}_1 \, \gbs{k}_8}  +   (-1)^{j_{k_7} + j_{k_8}  - J_{12}}  \, t^{\gbs{k}_1\, \gbs{k}_8\, (ph) J_{12}}_{\gbs{k}_2 \, \gbs{k}_7}   \right] \nonumber \\  
       & \qquad\qquad \qquad\qquad \qquad\qquad \qquad\qquad \qquad\qquad \times   
            \left( \mathcal{V}^{\gbs{k}_7}_l  \dpjq{l, \gbs{k}_7 } \, \mathcal{U}^{\gbs{k}_8}_m  \dpjq{m, \gbs{k}_8 }  \right)^* \,
             \mathcal{V}^{\gbs{k}_3}_v  \dpjq{v, \gbs{k}_3 } \,  v^{(ph)\, J_{12}}_{ml^{-1},av^{-1}}  \, ,
\label{eq:NIIe1_Jcoup} \\
%\end{align}
%\begin{align}
   \mathcal{N}^{(\rm{IIf'})}_{\gbs{r}, a} ={}&  (-1)^{ j_\alpha  - j_{k_3} - J_{12}}  \,  K(\gbs{r}) \,  \frac{ \hat{J}_{12} }{ \hat{j}_\alpha }   
      %\sum_{\substack{m \, v \, l \\  \gbs{k}_7 ,  \gbs{k}_8 }}   
      \sum_{m \, v \, l} \sum_{  \gbs{k}_7 \,  \gbs{k}_8 }      \left[ (-1)^{j_{k_1} + j_{k_2} - j_{k_7} - j_{k_8}} \, t^{\gbs{k}_2\, \gbs{k}_7\, (ph) J_{12}}_{\gbs{k}_1 \, \gbs{k}_8} \, + \, t^{\gbs{k}_1\, \gbs{k}_8\, (ph) J_{12}}_{\gbs{k}_2 \, \gbs{k}_7}   \right]  \nonumber \\  
       & \qquad\qquad \qquad\qquad \qquad\qquad  \qquad\qquad \qquad\qquad \times  \,  \left( \mathcal{U}^{\gbs{k}_7}_v  \dpjq{v, \gbs{k}_7 } \,  \mathcal{V}^{\gbs{k}_8}_l  \dpjq{l, \gbs{k}_8 } \right)^* \, v^{(ph)\, J_{12}}_{vl^{-1},am^{-1}}  \mathcal{V}^{\gbs{k}_3}_m  \dpjq{m, \gbs{k}_3 }  \, ,
\label{eq:NIIf1_Jcoup}  \\
%\end{align}
%\begin{align}
    \mathcal{N}^{(\rm{IIg})}_{\gbs{r}, a} ={}&  (-1)^{ j_\alpha  - j_{k_3} - J_{12}}  \,  K(\gbs{r}) \,  \frac{ \hat{J}_{12} }{ \hat{j}_\alpha }   
      %\sum_{\substack{m \, v \, l \\  \gbs{k}_7 ,  \gbs{k}_8 }}   
      \sum_{m \, v \, l} \sum_{  \gbs{k}_7 \,  \gbs{k}_8 }     \left[  (-1)^{j_{k_1} + j_{k_2}  - j_{k_7} - j_{k_8} } \, t^{\gbs{k}_2\, \gbs{k}_7\, (ph) J_{12}}_{\gbs{k}_1 \, \gbs{k}_8} \, + \, t^{\gbs{k}_1\, \gbs{k}_8\, (ph) J_{12}}_{\gbs{k}_2 \, \gbs{k}_7}  \right]   \nonumber \\  
       & \qquad\qquad \qquad\qquad \qquad\qquad \times    
           \left( \mathcal{U}^{\gbs{k}_7}_m  \dpjq{m, \gbs{k}_7 } \, \mathcal{U}^{\gbs{k}_8}_v  \dpjq{v, \gbs{k}_8 }  \right)^* \,
             \sqrt{ 1 + \delta_{m, v } }  \,   v^{J_{12}}_{mv,al}  \,  \sqrt{ 1 + \delta_{a, l } } \;  \mathcal{U}^{\gbs{k}_3}_l  \dpjq{l, \gbs{k}_3 } \, .
\label{eq:NIIg_Jcoup}  % \\
    %\mathcal{N}^{(\rm{IIg})}_{\gbs{r}, a} 
%    ={}&  (-1)^{ j_\alpha  - j_{k_3} - J_{12}}  \,  K(\gbs{r}) \,  \frac{ \hat{J}_{12} }{ \hat{j}_\alpha }   
%       \sum_{\substack{m \leq v \\ l}}      \left[  t^{\gbs{k}_1\, \gbs{k}_8\, (ph) J_{12}}_{\gbs{k}_2 \, \gbs{k}_7} + (-1)^{j_{k_1} + j_{k_2}  + j_{k_7} + j_{k_8}  - 2 J_{12}} \, t^{\gbs{k}_2\, \gbs{k}_7\, (ph) J_{12}}_{\gbs{k}_1 \, \gbs{k}_8}   \right]    \nonumber \\
%     & \quad \times  \sum_{  \gbs{k}_7 \leq  \gbs{k}_8 }    \frac {  \left( \mathcal{U}^{\gbs{k}_7}_m  \dpjq{m, \gbs{k}_7 } \, \mathcal{U}^{\gbs{k}_8}_v  \dpjq{v, \gbs{k}_8 }  + (-1)^{j_\mu + j_\nu - J_{12}} \, \mathcal{U}^{\gbs{k}_8}_m  \dpjq{m, \gbs{k}_8 } \, \mathcal{U}^{\gbs{k}_7}_v  \dpjq{v, \gbs{k}_7 }  \right)^* }{   (1+ \delta{\gbs{k}_7 , \gbs{k}_8} ) \;\sqrt{ 1 + \delta_{m, v } }  }  \,
%            v^{J_{12}}_{mv,al}  \,  \sqrt{ 1 + \delta_{a, l } } \;  \mathcal{U}^{\gbs{k}_3}_l  \dpjq{l, \gbs{k}_3 } \, .
%\label{eq:NIIg_Jcoup_bis}
\end{align}
\end{subequations}

Finally, energy denominators for the skeleton ADC(3) self-energy are given by
\begin{subequations}\label{eq:Fadc3_Sklt_Jcoup}
\begin{align}
     \mathcal{F}^{(Ipp)}_{\gbs{r}, \gbs{r}'} ={}& K(\gbs{r}) \, K(\gbs{r}\,')  \, \delta_{\gbs{k}_3, \gbs{k}_6} \, \delta_{J_{12},J_{45}} \,
            \sum_{a\, b\, g\, d} \left( \mathcal{U}^{\gbs{k}_1}_a \mathcal{U}^{\gbs{k}_2}_b \right)^* \,  \sqrt{ 1 + \delta_{a, b} } \, v^{J_{12}}_{ab,gd} \,  \sqrt{ 1 + \delta_{g, d} } \; \, \mathcal{U}^{\gbs{k}_4}_g \mathcal{U}^{\gbs{k}_5}_d \,  ,
\label{eq:FIpp_Jcoup}  \\
%\end{align}
%\begin{align}
     \mathcal{F}^{(Ihh)}_{\gbs{r}, \gbs{r}'} ={}& K(\gbs{r}) \, K(\gbs{r}\,')  \, \delta_{\gbs{k}_3, \gbs{k}_6} \, \delta_{J_{12},J_{45}} \,
            \sum_{a\, b\, g\, d} \mathcal{V}^{\gbs{k}_1}_a \mathcal{V}^{\gbs{k}_2}_b  \,  \sqrt{ 1 + \delta_{a, b} } \, v^{J_{12}}_{ab,gd} \,  \sqrt{ 1 + \delta_{g, d} } \;   \left( \mathcal{V}^{\gbs{k}_4}_g \mathcal{V}^{\gbs{k}_5}_d \right)^* \,  ,
\label{eq:FIhh_Jcoup}  \\
%\end{align}
%\begin{align}
     \mathcal{F}^{(Iph)}_{\gbs{r}, \gbs{r}'} ={}& K(\gbs{r}) \, K(\gbs{r}\,')  \, \delta_{\gbs{k}_3, \gbs{k}_6} \, \delta_{J_{12},J_{45}} \,
    \left[    \sum_{\alpha \beta \gamma \delta}  \,( \mathcal{U}^{\gbs{k}_1}_a \mathcal{V}^{\gbs{k}_2}_b )^*  \,   v^{(ph)\,J_{12}}_{a b^{-1}, g d^{-1}}   \,  \mathcal{U}^{\gbs{k}_4}_g \mathcal{V}^{\gbs{k}_5}_d 
    \right.   \nonumber \\  &\qquad \qquad \qquad  \qquad \left. 
    - ~(-1)^{j_{k_1}+j_{k_2}-J_{12}}       \sum_{\alpha \beta \gamma \delta}  \,( \mathcal{U}^{\gbs{k}_2}_a \mathcal{V}^{\gbs{k}_1}_b )^*  \,   v^{(ph)\,J_{12}}_{a b^{-1}, g d^{-1}}   \,  \mathcal{U}^{\gbs{k}_4}_g \mathcal{V}^{\gbs{k}_5}_d  
    \right.  \nonumber \\  &\qquad \qquad \qquad \qquad  \left.  
   - ~(-1)^{j_{k_4}+j_{k_5}-J_{12}}          \sum_{\alpha \beta \gamma \delta}  \,( \mathcal{U}^{\gbs{k}_1}_a \mathcal{V}^{\gbs{k}_2}_b )^*  \,   v^{(ph)\,J_{12}}_{a b^{-1}, g d^{-1}}   \,  \mathcal{U}^{\gbs{k}_5}_g \mathcal{V}^{\gbs{k}_4}_d \right.   \nonumber \\
   &\qquad \qquad \qquad \qquad  \left.    + ~ (-1)^{j_{k_1}+j_{k_2}-j_{k_4}-j_{k_5}}      \sum_{\alpha \beta \gamma \delta}  \,( \mathcal{U}^{\gbs{k}_2}_a \mathcal{V}^{\gbs{k}_1}_b )^*  \,   v^{(ph)\,J_{12}}_{a b^{-1}, g d^{-1}}   \,  \mathcal{U}^{\gbs{k}_5}_g \mathcal{V}^{\gbs{k}_4}_d   \right] \, .
\label{eq:FIph_Jcoup} 
\end{align}
\end{subequations}
\end{widetext}

\subsubsection{\label{App:Jcoup_comp}Non-skeleton diagrams}

Using Eqs.~\eqref{eq:UJab}, the normal and anomalous singlet BCC amplitudes, Eqs.~\eqref{eq:BCC_t1n} and~\eqref{eq:BCC_t1a}, can be re-expressed as follows
\begin{align}
   t^{k_1}_{k_2} ={}& 
     \dpjq{\gbs{k}_1 , \gbs{k}_2}  \delta_{m_{k_1}  \! , \! - m_{k_2} }  \!\! \sum_{a \, b} 
         \frac{  \dpjq{\gbs{k}_1,  a} \hfb{V}^{\gbs{k}_1}_a  (-1)^{j_{k_1}+m_{k_1}} u_{a b}  \hfb{U}^{\gbs{k}_2}_b }  {-(\hfe{\gbs{k}_1} + \hfe{\gbs{k}_2})} \dpjq{a, b} \nonumber \\
     \equiv{}& \dpjq{\gbs{k}_1 , \gbs{k}_2}  \delta_{m_{k_1}  \! , \! - m_{k_2} }  (-1)^{j_{k_1}+m_{k_1}} \, t^{\gbs{k}_1}_{\gbs{k}_2} 
\label{eq:T1Jab_n}
\end{align}
and 
\begin{align}
   t^{k_1 \, k_2} ={}& 
     \dpjq{\gbs{k}_1 , \gbs{k}_2}  \delta_{m_{k_1}  \! , \! - m_{k_2} }  \!\! \sum_{a \, b} 
         \frac{  (\widetilde{u}_{a b})^* \;  \hfb{U}^{\gbs{k}_1}_a  \hfb{U}^{\gbs{k}_2}_b  (-1)^{j_{k_2}-m_{k_2}}  \dpjq{b, \gbs{k}_2} }   {- 2 (\hfe{\gbs{k}_1} + \hfe{\gbs{k}_2})} \dpjq{a, b} \nonumber \\
     \equiv{}& \dpjq{\gbs{k}_1 , \gbs{k}_2}  \delta_{m_{k_1}  \! , \! - m_{k_2} }  (-1)^{j_{k_1}+m_{k_1}} \, t^{\gbs{k}_1 \, \gbs{k}_2}  \, ,
\label{eq:T1Jab_a1} \\
   t_{k_1 \, k_2} ={}& 
     \dpjq{\gbs{k}_1 , \gbs{k}_2}  \delta_{m_{k_1}  \! , \! - m_{k_2} }  \!\! \sum_{a \, b} 
         \frac{   \dpjq{\gbs{k}_1,  a}  \hfb{V}^{\gbs{k}_1}_a  \hfb{V}^{\gbs{k}_2}_b  (-1)^{j_{k_1}+m_{k_1}}   \; \widetilde{u}_{a b} }   {- 2 (\hfe{\gbs{k}_1} + \hfe{\gbs{k}_2})} \dpjq{a, b} \nonumber \\
     \equiv{}& \dpjq{\gbs{k}_1 , \gbs{k}_2}  \delta_{m_{k_1}  \! , \! - m_{k_2} }  (-1)^{j_{k_1}+m_{k_1}} \, t_{\gbs{k}_1 \, \gbs{k}_2}  \, .
\label{eq:T1Jab_a2} 
\end{align}

One can now write down the composite $\mathcal{M}_{\gbs{r},a}$ amplitudes  at second order
\begin{widetext}
\begin{subequations}\label{eq:MII_Uext_Jcoup}
\begin{align}
    \mathcal{M}^{(\rm{IIh+i})}_{a,\gbs{r}} ={}&  (-1)^{ j_\alpha  + j_{k_3} - J_{12}}  \,  K(\gbs{r}) \,  \frac{ \hat{J}_{12} }{ \hat{j}_\alpha }   
      \sum_{\substack{m \, v \, l} }   \,   \sqrt{ 1 + \delta_{a, l } }  \, v^{J_{12}}_{al,mv}  \,   \sqrt{ 1 + \delta_{m, v } }    \nonumber \\   
     & \qquad \times    \left\{    \dpjq{m, \gbs{k}_1 } \sum_{  \gbs{k}_7 }    ( \hfb{V}^{\gbs{k}_7}_m )^*  \dpjq{\gbs{k}_1, \gbs{k}_7 }  \,  \left[t^{\gbs{k}_7}_{\gbs{k}_1} - (-1)^{2 j_{k_1}} \,  t^{\gbs{k}_1}_{\gbs{k}_7} \right]  \, \hfb{U}^{\gbs{k}_2}_v  \dpjq{v, \gbs{k}_2 } \,  \hfb{V}^{\gbs{k}_3}_l  \dpjq{l, \gbs{k}_3 }  \right. \nonumber \\
     &\qquad \qquad   \left.  + ~ \hfb{U}^{\gbs{k}_1}_m  \dpjq{m, \gbs{k}_1 }  \, \dpjq{v, \gbs{k}_2 } \sum_{  \gbs{k}_7 }     ( \hfb{V}^{\gbs{k}_7}_v )^*  \dpjq{\gbs{k}_2, \gbs{k}_7 }   \,  \left[t^{\gbs{k}_7}_{\gbs{k}_2} - (-1)^{2 j_{k_2}}  \, t^{\gbs{k}_2}_{\gbs{k}_7} \right] \,  \hfb{V}^{\gbs{k}_3}_l  \dpjq{l, \gbs{k}_3 }   \right.  \nonumber \\ 
     &\qquad \qquad   \left.  + ~  \hfb{U}^{\gbs{k}_1}_m  \dpjq{m, \gbs{k}_1}  \,  \hfb{U}^{\gbs{k}_2}_v  \dpjq{v, \gbs{k}_2} \, \dpjq{l, \gbs{k}_3 } \sum_{  \gbs{k}_7 }      ( \hfb{U}^{\gbs{k}_7}_l )^*   \dpjq{\gbs{k}_3, \gbs{k}_7 }  \,  \left[(-1)^{2 j_{k_3}}  \, t^{\gbs{k}_7}_{\gbs{k}_3} -  t^{\gbs{k}_3}_{\gbs{k}_7} \right]  
             \right\}  \, ,
\label{eq:MIIhi_Jcoup}  \\
    \mathcal{M}^{(\rm{IIj+k})}_{a,\gbs{r}} ={}&  (-1)^{ j_\alpha  + j_{k_3} - J_{12}}  \,  K(\gbs{r}) \,  \frac{ \hat{J}_{12} }{ \hat{j}_\alpha }   
      \sum_{\substack{m \, v \, l} }   \,   \sqrt{ 1 + \delta_{a, l } }  \, v^{J_{12}}_{al,mv}  \,   \sqrt{ 1 + \delta_{m, v } }    \nonumber \\   
     & \qquad \times    \left\{    \dpjq{m, \gbs{k}_1 } \sum_{  \gbs{k}_7 }  ( \hfb{V}^{\gbs{k}_7}_m )^*  \dpjq{\gbs{k}_1, \gbs{k}_7 }  \,  \left[t^{\gbs{k}_7 \, \gbs{k}_1} -   t_{\gbs{k}_7 \, \gbs{k}_1} \right]  \; \hfb{U}^{\gbs{k}_2}_v  \dpjq{v, \gbs{k}_2 } \;  \hfb{V}^{\gbs{k}_3}_l  \dpjq{l, \gbs{k}_3 }  \right. \nonumber \\
     &\qquad \qquad   \left.  + ~ \hfb{U}^{\gbs{k}_1}_m  \dpjq{m, \gbs{k}_1 }  \; \dpjq{v, \gbs{k}_2 } \sum_{  \gbs{k}_7 }   ( \hfb{V}^{\gbs{k}_7}_v )^*  \dpjq{\gbs{k}_2, \gbs{k}_7 }   \,  \left[t^{\gbs{k}_7 \, \gbs{k}_2} -  t_{\gbs{k}_7 \, \gbs{k}_2} \right] \;  \hfb{V}^{\gbs{k}_3}_l  \dpjq{l, \gbs{k}_3 }   \right.  \nonumber \\ 
     &\qquad \qquad   \left.  + ~  \hfb{U}^{\gbs{k}_1}_m  \dpjq{m, \gbs{k}_1}  \;  \hfb{U}^{\gbs{k}_2}_v  \dpjq{v, \gbs{k}_2} \; \dpjq{l, \gbs{k}_3 } \sum_{  \gbs{k}_7 }      (-1)^{2 j_{k_3}}   ( \hfb{U}^{\gbs{k}_7}_l )^*   \dpjq{\gbs{k}_3, \gbs{k}_7 }  \,  \left[  t^{\gbs{k}_7 \, \gbs{k}_3} -  t_{\gbs{k}_7 \, \gbs{k}_3} \right]   
             \right\}  \, ,
\label{eq:MIIjk_Jcoup} 
\end{align}
\end{subequations}
as well as the corresponding $\mathcal{N}_{\gbs{r},a}$ amplitudes
\begin{subequations}\label{eq:NII_Uext_Jcoup}
\begin{align}
    \mathcal{N}^{(\rm{IIh+i})}_{\gbs{r},a} ={}&  (-1)^{ j_\alpha  - j_{k_3} - J_{12}}  \,  K(\gbs{r}) \,  \frac{ \hat{J}_{12} }{ \hat{j}_\alpha }     \nonumber \\   
     &  \times   \sum_{m \, v \, l}     \left\{    \dpjq{m, \gbs{k}_1 } \sum_{  \gbs{k}_7 }    ( \hfb{U}^{\gbs{k}_7}_m )^*  \dpjq{\gbs{k}_1, \gbs{k}_7 }  \,  \left[ (-1)^{2 j_{k_1}} \,  t^{\gbs{k}_7}_{\gbs{k}_1} -  t^{\gbs{k}_1}_{\gbs{k}_7} \right]  \, \hfb{V}^{\gbs{k}_2}_v  \dpjq{v, \gbs{k}_2 } \,  \hfb{U}^{\gbs{k}_3}_l  \dpjq{l, \gbs{k}_3 }  \right. \nonumber \\
     & \quad   \left.  + ~ \hfb{V}^{\gbs{k}_1}_m  \dpjq{m, \gbs{k}_1 }  \, \dpjq{v, \gbs{k}_2 } \sum_{  \gbs{k}_7 }     ( \hfb{U}^{\gbs{k}_7}_v )^*  \dpjq{\gbs{k}_2, \gbs{k}_7 }   \,  \left[ (-1)^{2 j_{k_2}}  \, t^{\gbs{k}_7}_{\gbs{k}_2} - t^{\gbs{k}_2}_{\gbs{k}_7} \right] \,  \hfb{U}^{\gbs{k}_3}_l  \dpjq{l, \gbs{k}_3 }   \right.  \nonumber \\ 
     & \quad   \left.  + ~  \hfb{V}^{\gbs{k}_1}_m  \dpjq{m, \gbs{k}_1}  \,  \hfb{V}^{\gbs{k}_2}_v  \dpjq{v, \gbs{k}_2} \, \dpjq{l, \gbs{k}_3 } \sum_{  \gbs{k}_7 }      ( \hfb{V}^{\gbs{k}_7}_l )^*   \dpjq{\gbs{k}_3, \gbs{k}_7 }  \,  \left[ t^{\gbs{k}_7}_{\gbs{k}_3} - (-1)^{2 j_{k_3}}  \, t^{\gbs{k}_3}_{\gbs{k}_7} \right]  
             \right\}     \,    \sqrt{ 1 + \delta_{m, v } }   \, v^{J_{12}}_{mv,al}  \,  \sqrt{ 1 + \delta_{a, l } }  \, ,
\label{eq:NIIhi_Jcoup}  \\
    \mathcal{N}^{(\rm{IIj+k})}_{\gbs{r},a} ={}&  (-1)^{ j_\alpha  - j_{k_3} - J_{12}}  \,  K(\gbs{r}) \,  \frac{ \hat{J}_{12} }{ \hat{j}_\alpha }     \nonumber \\   
     &  \times \sum_{m \, v \, l }   \,  \left\{    \dpjq{m, \gbs{k}_1 } \sum_{  \gbs{k}_7 } (-1)^{2 j_{k_1}} ( \hfb{U}^{\gbs{k}_7}_m )^*  \dpjq{\gbs{k}_1, \gbs{k}_7 }  \,  \left[t^{\gbs{k}_7 \, \gbs{k}_1} -   t_{\gbs{k}_7 \, \gbs{k}_1} \right]  \; \hfb{V}^{\gbs{k}_2}_v  \dpjq{v, \gbs{k}_2 } \;  \hfb{U}^{\gbs{k}_3}_l  \dpjq{l, \gbs{k}_3 }  \right. \nonumber \\
     & \quad   \left.  + ~ \hfb{V}^{\gbs{k}_1}_m  \dpjq{m, \gbs{k}_1 }  \; \dpjq{v, \gbs{k}_2 } \sum_{  \gbs{k}_7 }  (-1)^{2 j_{k_2}}   ( \hfb{U}^{\gbs{k}_7}_v )^*  \dpjq{\gbs{k}_2, \gbs{k}_7 }   \,  \left[t^{\gbs{k}_7 \, \gbs{k}_2} -  t_{\gbs{k}_7 \, \gbs{k}_2} \right] \;  \hfb{U}^{\gbs{k}_3}_l  \dpjq{l, \gbs{k}_3 }   \right.  \nonumber \\ 
     & \quad   \left.  + ~  \hfb{V}^{\gbs{k}_1}_m  \dpjq{m, \gbs{k}_1}  \;  \hfb{V}^{\gbs{k}_2}_v  \dpjq{v, \gbs{k}_2} \; \dpjq{l, \gbs{k}_3 } \sum_{  \gbs{k}_7 }   ( \hfb{V}^{\gbs{k}_7}_l )^*   \dpjq{\gbs{k}_3, \gbs{k}_7 }  \,  \left[  t^{\gbs{k}_7 \, \gbs{k}_3} -  t_{\gbs{k}_7 \, \gbs{k}_3} \right]  
             \right\}  \,   \sqrt{ 1 + \delta_{m, v } }   \, v^{J_{12}}_{mv,al}  \,  \sqrt{ 1 + \delta_{a, l } }   \, .
\label{eq:NIIjk_Jcoup} 
\end{align}
\end{subequations}

%The energy denominators for the composite self-energy at third order propagate the single-particle interactions of Eqs.~\eqref{eq:3rd_Un} and ~\eqref{eq:3rd_Uan}:
%\begin{subequations}\label{eq:Jcoup_3rd_E}
%\begin{align}
% \mathcal{E}^{(p)}_{k_1  k_2}  ={}& \dpjq{\gbs{k}_1, \gbs{k}_2 } \delta_{m_1, m_2} \sum_{a \, b} \dpjq{\gbs{k}_1, a } \dpjq{a, b } \; \left(         \hfb{U}^{\gbs{k}_1}_a \right)^{\!*} \, u_{a b} \,                 \hfb{U}^{\gbs{k}_2}_b    \, ,  \label{eq:Jcoup_3rd_Ep}  \\
% \mathcal{E}^{(h)}_{k_1  k_2}  ={}& \dpjq{\gbs{k}_1, \gbs{k}_2 } \delta_{m_1, m_2} \sum_{a \, b} \dpjq{\gbs{k}_1, a } \dpjq{a, b } \;       \hfb{V}^{\gbs{k}_1}_a                   \, u_{a b} \, \left(  \hfb{V}^{\gbs{k}_2}_b  \right)^{\!*}  \,  \label{eq:Jcoup_3rd_Eh}  \\
% \mathcal{E}^{(an)}_{k_1 k_2}  ={}& \dpjq{\gbs{k}_1, \gbs{k}_2 } \delta_{m_1, m_2}  \frac{1}{2} \sum_{a \, b} \dpjq{\gbs{k}_1, a } \dpjq{a, b } \;  \left(   \hfb{U}^{\gbs{k}_1}_a \right)^{\!*} \, \widetilde{u}_{\alpha \beta} \,                 \hfb{V}^{\gbs{k}_2}_b      \,  \label{eq:Jcoup_3rd_Ean}  
%\end{align}
%\end{subequations}
The energy denominators for the composite self-energy at third order read as
\begin{subequations}\label{eq:Fadc3_Comp_Jcoup}
\begin{align}
     \mathcal{F}^{(Id)}_{\gbs{r}, \gbs{r}'} ={}& K(\gbs{r}) \, K(\gbs{r}\,') 
   \left( \delta_{\gbs{k}_1, \gbs{k}_4}\, \delta_{\gbs{k}_2, \gbs{k}_5} \,-\,
   (-1)^{j_{k_1} + j_{k_2}  - J_{12}}
   \delta_{\gbs{k}_2, \gbs{k}_4}\, \delta_{\gbs{k}_1, \gbs{k}_5} \right) \, 
   \delta_{J_{12}, J_{45}}
   \nonumber \\ &\qquad \qquad \qquad \qquad \times 
   \dpjq{\gbs{k}_3, \gbs{k}_6 } 
          \sum_{a\, b}  \, \dpjq{a, \gbs{k}_3 } \, \dpjq{b, \gbs{k}_3 } \left[   \left( \hfb{U}^{\gbs{k}_3}_a \right)^* \! u_{a b}  \,\hfb{U}^{\gbs{k}_6}_b +     \hfb{V}^{\gbs{k}_6}_a   u_{a b}  \left( \hfb{V}^{\gbs{k}_3}_b  \right)^* \right] \, ,
\label{eq:FId_Jcoup} \\
%\end{align}
%\begin{align}
     \mathcal{F}^{(Ie)}_{\gbs{r}, \gbs{r}'} ={}& K(\gbs{r}) \, K(\gbs{r}\,') 
   \left( \delta_{\gbs{k}_1, \gbs{k}_4}\, \delta_{\gbs{k}_2, \gbs{k}_5} \,-\,
   (-1)^{j_{k_1} + j_{k_2}  - J_{12}}
   \delta_{\gbs{k}_2, \gbs{k}_4}\, \delta_{\gbs{k}_1, \gbs{k}_5} \right) \, 
   \delta_{J_{12}, J_{45}}
   \nonumber \\ &\qquad \qquad \qquad \qquad \times 
   \dpjq{\gbs{k}_3, \gbs{k}_6 } \, \frac 1 2 \,
            \sum_{a\, b}  \, \dpjq{a, \gbs{k}_3 } \, \dpjq{b, \gbs{k}_3 } \left[   \left( \hfb{U}^{\gbs{k}_3}_a \right)^* \widetilde{u}_{a b}  \hfb{V}^{\gbs{k}_6}_b +     \hfb{U}^{\gbs{k}_6}_a \left( \widetilde{u}_{a b}   \hfb{V}^{\gbs{k}_3}_b  \right)^* \right] \,  .
\label{eq:FIe_Jcoup} 
\end{align}
\end{subequations}
\end{widetext}

\subsection{\label{App:Gmtx_jcoup}J-coupled Gorkov eigenvalue problem}

The angular momentum coupling conventions just described allow to decouple the Gorkov equations~\eqref{eq:ADC_mtx} into several independent eigenvalue problems, one for each
set $(\pi,J,q)$ of the parity, total angular momentum, and particle number parity of the many-body state~$|\Psi^{e(o)}_k \rangle$.
Moreover, we use antisymmetry to limit the sums over the ISCs to ordered combinations 
of the first two indices. For example,
\begin{align}
\sum_\gbs{r}  \ldots  \quad \equiv \quad  \sum_{\gbs{k}_1    \lesssim   \gbs{k}_2 ,\, J_{12} ,\,  \gbs{k}_3} \ldots  \; ,
\label{eq:eg_ics_sum}
\end{align}
where the configurations with $\gbs{k}_1=\gbs{k}_2$ are included only when $(-1)^{j_{k_1}+j_{k_2}-J_{12}}=-1$ and the sum on the right hand side runs implicitly over quantum numbers that satisfy the constraint  $(\pi_r,J_r,q_r)=(\pi,J,q)$.
With this choice, the $\sqrt{(1+ \delta_{\gbs{k}_1 \gbs{k}_2})/2}$ terms from Eqs.~\eqref{eq:MN_Jcoup_conv} through~\eqref{eq:E_Jcoup} drop out. 
The Gorkov Eqs.~\eqref{eq:ADC_mtx} for channel $(\pi_k,J_k,q_k)$ become
\begin{widetext}
\begin{align}
   \omega_k  \left( \begin{array}{c} \mathcal{U}^\gbs{k}_a \\ ~  \\  \mathcal{V}^\gbs{k}_a \\ ~ \\ \mathcal{W}^\gbs{k}_\gbs{r} \\ ~  \\  \mathcal{Z}^\gbs{k}_\gbs{r}  \end{array} \right) 
  ={}&
   \left( \begin{array}{ccccccc}
              t_{ab} - \mu \delta_{ab} + \Lambda_{ab}      &&         \tilde{h}_{ab}       &&          \mathcal{C}_{a,\gbs{r}'}   &&    \mathcal{D}^*_{\gbs{r}',a}   \\ ~ \\
             (\tilde{h}_{ba})^*           & & - t_{ab} + \mu \delta_{ab} - (\Lambda_{ab})^*      &&          \mathcal{D}_{\gbs{r}',a}   & &   (-1)^{2J_r}\mathcal{C}^*_{a,\gbs{r}'}    \\ ~ \\
                       \mathcal{C}^*_{b,\gbs{r}}         &&    \mathcal{D}^*_{\gbs{r},b}      &&     \mathcal{E}_{\gbs{r},\gbs{r}'}     &   \\ ~ \\
                        \mathcal{D}_{\gbs{r},b}   &&    (-1)^{2J_r} \mathcal{C}_{b,\gbs{r}}  & &    &&   - \mathcal{E}_{\gbs{r}',\gbs{r}} \\
 \end{array} \right) \, 
%   \left( \begin{array}{ccccccc}
%              T - \mu \mathbb{I} + \Sigma^{(\infty)\,11}       &&         \Sigma^{(\infty)\,12}       &&          \mathcal{C}   &&    \mathcal{D}^\dagger   \\ ~ \\
%             \Sigma^{(\infty)\,21}            & & - T + \mu \mathbb{I} + \Sigma^{(\infty)\,22}      &&          \mathcal{D}^T   & &   (-1)^{2J_r}\mathcal{C}^*    \\ ~ \\
%                       \mathcal{C}^\dagger         &&    \mathcal{D}^*      &&     \mathcal{E}     &   \\ ~ \\
%                        \mathcal{D}   &&    (-1)^{2J_r} \mathcal{C}^T  & &    &&   - \mathcal{E}^T \\
% \end{array} \right) \, 
  \left( \begin{array}{c}  \mathcal{U}^\gbs{k}_b  \\ ~ \\ \mathcal{V}^\gbs{k}_b  \\ ~ \\ \mathcal{W}^\gbs{k}_{\gbs{r}'}  \\ ~ \\  \mathcal{Z}^\gbs{k}_{\gbs{r}'} \end{array}  \right) \, ,
\label{eq:Jcoup_ADC_mtx}
\end{align}
\end{widetext}
with implicit sums over repeated indices $n_\beta$ and~$\gbs{r}'$.  The matrix elements of the coupling amplitudes are
%where the matrix elements of the coupling amplitudes are
\begin{subequations}
\label{eq:Jcoup_CD}
\begin{align}
  \mathcal{C}_{a, \gbs{r}} ={}&   \frac 1 {\sqrt{3}} \sum_{\gbs{r}'}   \mathcal{M}_{a, \gbs{r}'} \,  \mathcal{P}_{\gbs{r}' ,\gbs{r}}  \, ,  \label{eq:Jcoup_Car}  \\
  \mathcal{D}_{\gbs{r}, a} ={}&   \frac 1 {\sqrt{3}} \sum_{\gbs{r}'}   \mathcal{P}_{\gbs{r} ,\gbs{r}'}  \, \mathcal{N}_{\gbs{r}', a}  \, ,   \label{eq:Jcoup_Dra}
\end{align}
\end{subequations}
and the energy denominators
\begin{align}
\label{eq:Jcoup_Err} 
  \mathcal{E}_{\gbs{r}, \gbs{r}'} ={}& \mathcal{E}^{(0)}_{\gbs{r}, \gbs{r}'} + \frac 1 3 \sum_{\gbs{r}_2 \,\gbs{r}_3}   \mathcal{P}_{\gbs{r} ,\gbs{r}_2}  \mathcal{F}^{(I)}_{\gbs{r}_2, \gbs{r}_3} \mathcal{P}_{\gbs{r}_3, \gbs{r}'} \, .
\end{align}
The ISC components of the eigenvectors~\eqref{eq:Jcoup_ADC_mtx}, $\mathcal{W}^\gbs{k}_\gbs{r}$ and $\mathcal{Z}^\gbs{k}_\gbs{r}$,  correspond to the angular-momentum coupling of Eqs.~\eqref{eq:def_WZ}, respectively following the conventions~\eqref{eq:M_Jcoup} and~\eqref{eq:Nbar_Jcoup}. The normalization condition becomes
\begin{align}
  \sum_a  | \mathcal{U}^\gbs{k}_a  |^2  +  \sum_a | \mathcal{V}^\gbs{k}_a  |^2  +    \sum_\gbs{r} | \mathcal{W}^\gbs{k}_\gbs{r}  |^2  +   \sum_\gbs{r}  | \mathcal{Z}^\gbs{k}_\gbs{r}  |^2  = 1 \, ,
\label{eq:Jcoup_ADC_norm}
\end{align}
with the sums over $\gbs{r}$ as specified by Eq.~\eqref{eq:eg_ics_sum}.

Finally, the GMK sum rule for total energy is given by
\begin{align}
  E_0 = {}& \Omega_0 + \mu \langle N \rangle =
   \sum_{(\pi_k,j_k,q_k)}  \frac{2 j_k + 1}{2}  \nonumber \\ &\times \sum_{\substack{n_k \\ a, b}}    \dpjq{b,  \gbs{k}}  \dpjq{a,   \gbs{k}} \left[  t_{ab} + (\mu - \omega_\gbs{k}) \delta_{a b}  \right] \,   \mathcal{V}^\gbs{k}_b{}^* \,  \mathcal{V}^\gbs{k}_a \, .
 \label{eq:Jcoup_Koltun}
\end{align}

\section{\label{App:SigInf_3rd_ord} Static self-energy up to third order}

The complete static self-energy is given by the diagrams of Fig.~\ref{fig:SigI} and Eqs.~\eqref{eq:SigI_expr} and its computation requires the knowledge of the exact (dressed) propagators~\eqref{eq:Gkv_props}. In some cases it can be useful to access the separate contributions at each order in perturbation theory. The first-order terms are simply given by Eqs.~\eqref{eq:SigI_expr} but with the spectroscopic amplitudes replaced by the HFB wave functions from Eq.~\eqref{eq:OmU_HFB_eqs}. These terms read as
\begin{subequations}\label{eq:SigInf_I}
\begin{align}
  \Sigma^{(I)\,11}_{\alpha \beta} ={}& - \Sigma^{(\infty)\,22 (I)}_{\bar{\beta}  \bar{\alpha}} = \sum_{k \, \gamma \, \delta}    v_{\alpha \gamma, \beta \delta}\,  \bar{\hfb{V}}^k_\delta{}^* \bar{\hfb{V}}^k_\gamma  \, , \label{eq:SigInf_I_11} \\
  \Sigma^{(I)\,12}_{\alpha \beta} ={}&  \left( \Sigma^{(I)\,21}_{\beta \alpha} \right)^* = \frac 1 2 \sum_{k \, \gamma \, \delta}    v_{\alpha \bar{\beta}, \gamma \delta} \,  \bar{\hfb{V}}^k_\gamma{}^* \hfb{U}^k_\delta  \, . \label{eq:SigInf_I_12} 
\end{align}
\end{subequations}
In order to extract the specific contributions at second and third orders, the dressed propagator must be expanded according to Eqs.~\eqref{eq:Gkv_eqs}.
The first-order contribution is simply
\begin{align}
   \mathbf{G}^{(1)}_{\alpha \beta}(\omega)  ={}&  \sum_{\gamma \, \delta} \mathbf{G}^{(0)}_{\alpha \gamma}(\omega) \, \mathbf{\Sigma}^{(\infty) I}_{\gamma \delta} \, \mathbf{G}^{(0)}_{\delta \beta}(\omega) \, ,
\label{eq:G1_prop}
\end{align}
where the sums over Nambu indices is implicit in the matrix algebra. 
The second-order self-energy is obtained by inserting Eq.~\eqref{eq:G1_prop} into the frequency integrals for~$\mathbf{\Sigma}^{(\infty)}$ as
\begin{subequations}\label{eq:SigInf_II}
\begin{align}
  &\Sigma^{(\infty, 2)\,11}_{\alpha \beta} =  \sum_{\gamma \,  \delta }   \int_{C\uparrow}  \frac{d\,\omega}{2\pi i}\,  v_{\alpha \gamma, \beta \delta} \,  G^{(1) 11}_{\delta \gamma}(\omega)  \nonumber \\
     & \qquad=   \sum_{\substack{\gamma \,  \delta \\ k_1 \, k_2}}  v_{\alpha \gamma, \beta \delta}   
       \frac{  \hfb{U}^{k_1}_\delta  Q^{\rm fb}_{k_1 \, k_2}  \bar{\hfb{V}}^{k_2}_\gamma  + \left(  \hfb{U}^{k_1}_\gamma  Q^{\rm fb}_{k_1 k_2}  \bar{\hfb{V}}^{k_2}_\delta \right)^* }{ - (\hfe{k_1} + \hfe{k_2}) + i\eta}       \, , \label{eq:SigInf_II_11} \\
 & \Sigma^{(\infty, 2)\,12}_{\alpha \beta} =   \frac 1 2  \sum_{\gamma \,  \delta }   \int_{C\uparrow}  \frac{d\,\omega}{2\pi i}\,  v_{\alpha  \bar\beta , \gamma \bar\delta} \,  G^{(1) 12}_{\gamma \delta}(\omega)  \nonumber \\
     & \qquad=   \sum_{\substack{\gamma \,  \delta \\ k_1 \, k_2}}  \frac{v_{\alpha \bar\beta , \gamma \delta} }{2}
      \frac{  \hfb{U}^{k_1}_\gamma  Q^{\rm fb}_{k_1 \, k_2}  \hfb{U}^{k_2}_\delta  + \left(  \bar{\hfb{V}}^{k_1}_\delta  Q^{\rm fb}_{k_1  k_2}  \bar{\hfb{V}}^{k_2}_\gamma \right)^*  }{ - (\hfe{k_1} + \hfe{k_2}) + i\eta}     \, , \label{eq:SigInf_II_12} 
\end{align}
\end{subequations}
where ${C\uparrow}$ is a counterclockwise path along the real axis and including the whole positive imaginary plane and $Q^{\rm fb}_{k_1  k_2}$ is defined below in Eq.~\eqref{eq:Qfb_def}.

For the second-order expansion of the propagator two different terms arise, one from iterating two static first-order self-energies
\begin{align}
   \mathbf{G}^{(2a)}_{\alpha \beta}(\omega)  ={}&  \sum_{\gamma \, \delta \, \mu \, \nu} \mathbf{G}^{(0)}_{\alpha \gamma}(\omega) \, \mathbf{\Sigma}^{(\infty) I}_{\gamma \delta} \, \mathbf{G}^{(0)}_{\delta \mu}(\omega)  \, \mathbf{\Sigma}^{(\infty) I}_{\mu \nu} \, \mathbf{G}^{(0)}_{\nu \beta}(\omega)  \, ,
\label{eq:G2a_prop}
\end{align}
and the other including a single second-order self-energy
\begin{align}
    \mathbf{G}^{(2b)}_{\alpha \beta}(\omega)  ={}&  \sum_{\gamma \, \delta} \mathbf{G}^{(0)}_{\alpha \gamma}(\omega) \, \widetilde{\mathbf{\Sigma}}^{(ADC2)}_{\gamma \delta}(\omega) \, \mathbf{G}^{(0)}_{\delta \beta}(\omega) \, .
\label{eq:G2b_prop}
\end{align}
\begin{widetext}
Both Eqs.~\eqref{eq:G2a_prop} and~\eqref{eq:G2b_prop} imply similar frequency integrals combining three poles.
After performing such integrals, one obtains
\begin{subequations}\label{eq:SigInf_III}
\begin{align}
\Sigma^{(\infty, 3a)\,11}_{\alpha \beta} ={}&  \sum_{\gamma \,  \delta }   \int  \frac{d\,\omega}{2\pi i}\,  v_{\alpha \gamma, \beta \delta} \,  G^{(2a) 11}_{\delta \gamma}(\omega)  \nonumber \\
     ={}&   \sum_{\substack{ k_1 \, k_2 \, k_3 \\ \gamma \,  \delta}}  v_{\alpha \gamma, \beta \delta}   \frac 1{[ - (\hfe{k_1} + \hfe{k_2}) + i\eta][- (\hfe{k_1} + \hfe{k_3}) + i\eta]}  
                         \left\{    \hfb{U}^{k_2}_\delta  Q^{\rm fb}_{k_2  k_1} \,  ( \hfb{U}^{k_3}_\gamma  Q^{\rm fb}_{k_3  k_1} \!)^*  -  ( \bar{\hfb{V}}^{k_2}_\delta  Q^{\rm fb}_{k_1  k_2} \!)^* \;    \bar{\hfb{V}}^{k_3}_\gamma  Q^{\rm fb}_{k_1 k_3}      \right.    \nonumber \\
   &\qquad \left. + ( \bar{\hfb{V}}^{k_1}_\delta  Q^{\rm fb}_{k_2  k_1}\!)^* \,   ( \hfb{U}^{k_3}_\gamma  Q^{\rm ff}_{k_3 k_2}\!)^*  +  \hfb{U}^{k_1}_\delta  Q^{\rm fb}_{k_1  k_2}  \,   \bar{\hfb{V}}^{k_3}_\gamma  Q^{\rm ff}_{k_3  k_2}    
                         +           \hfb{U}^{k_2}_\delta  Q^{\rm ff}_{k_2  k_3}  \,   \bar{\hfb{V}}^{k_1}_\gamma  Q^{\rm fb}_{k_3  k_1}   + ( \bar{\hfb{V}}^{k_2}_\delta  Q^{\rm ff}_{k_2  k_3}\!)^* \;   ( \hfb{U}^{k_1}_\gamma  Q^{\rm fb}_{k_1  k_3}\!)^*  \right\}  
     \, , \label{eq:SigInf_IIIa_11}  \\
\Sigma^{(\infty, 3a)\,12}_{\alpha \beta} ={}&   \frac 1 2  \sum_{\gamma \,  \delta }   \int  \frac{d\,\omega}{2\pi i}\,  v_{\alpha  \bar\beta , \gamma \bar\delta} \,  G^{(2a) 12}_{\gamma \delta}(\omega)  \nonumber \\
    ={}&   \sum_{\substack{ k_1 \, k_2 \, k_3 \\ \gamma \,  \delta}}  \frac{v_{\alpha \bar\beta , \gamma \delta} }{2}   \frac 1{[ - (\hfe{k_1} + \hfe{k_2}) + i\eta][- (\hfe{k_1} + \hfe{k_3}) + i\eta]}  
                \left\{              \hfb{U}^{k_2}_\gamma  Q^{\rm fb}_{k_2  k_1}       \,    ( \bar{\hfb{V}}^{k_3}_\delta  Q^{\rm fb}_{k_3  k_1}\!)^*   -  ( \bar{\hfb{V}}^{k_2}_\gamma  Q^{\rm fb}_{k_1  k_2}\!)^*  \;  \hfb{U}^{k_3}_\delta  Q^{\rm fb}_{k_1  k_3}     \right.    \nonumber \\
   &\qquad \left. + ( \bar{\hfb{V}}^{k_1}_\gamma  Q^{\rm fb}_{k_2  k_1}\!)^* \;   ( \bar{\hfb{V}}^{k_3}_\delta  Q^{\rm ff}_{k_3  k_2}\!)^*   +            \hfb{U}^{k_1}_\gamma  Q^{\rm fb}_{k_1  k_2}        \,  \hfb{U}^{k_3}_\delta  Q^{\rm ff}_{k_3  k_2}    
                         +           \hfb{U}^{k_2}_\gamma  Q^{\rm ff}_{k_2  k_3}       \,               \hfb{U}^{k_1}_\delta  Q^{\rm fb}_{k_3  k_1}         +  ( \bar{\hfb{V}}^{k_2}_\gamma  Q^{\rm ff}_{k_2  k_3}\!)^* \,   ( \bar{\hfb{V}}^{k_1}_\delta  Q^{\rm fb}_{k_1  k_3}\!)^*   \right\}  
      \, , \label{eq:SigInf_IIIa_12} 
\end{align}
and
\begin{align}
\Sigma^{(\infty, 3b)\,11}_{\alpha \beta} ={}&  \sum_{\gamma \,  \delta }   \int  \frac{d\,\omega}{2\pi i}\,  v_{\alpha \gamma, \beta \delta} \,  G^{(2b) 11}_{\delta \gamma}(\omega)  \nonumber \\
   ={}&   \sum_{\substack{\gamma \,  \delta \\ r \, k_4 \, k_5}}  v_{\alpha \gamma, \beta \delta}  \,
     \frac {   \hfb{U}^{k_4}_\delta  R^{\rm fb}_{k_4 r}  \,   ( R^{\rm fb}_{k_5 r}  \hfb{U}^{k_5}_\gamma \!)^* -  ( \bar{\hfb{V}}^{k_4}_\delta  R^{\rm fb}_{k_4 r} \!)^* \;    R^{\rm fb}_{k_5 r}  \bar{\hfb{V}}^{k_5}_\gamma  }  {[ - (\hfe{k_1} + \hfe{k_2} + \hfe{k_3} + \hfe{k_4}) + i\eta]\,[- (\hfe{k_1} + \hfe{k_2} + \hfe{k_3} + \hfe{k_5}) + i\eta]}  \nonumber \\
    & +  \sum_{\substack{\gamma \,  \delta \\ r \, k_4 \, k_5}}  v_{\alpha \gamma, \beta \delta}  \,
     \frac {  \hfb{U}^{k_5}_\delta  R^{\rm ff}_{k_5 r}   \,    R^{\rm fb}_{k_4 r}  \bar{\hfb{V}}^{k_4}_\gamma  -    \hfb{U}^{k_4}_\delta  R^{\rm fb}_{k_4 r} \, R^{\rm ff}_{k_5 r}  \bar{\hfb{V}}^{k_5}_\gamma
         +  (  \bar{\hfb{V}}^{k_4}_\delta  R^{\rm fb}_{k_4 r}  \, R^{\rm ff}_{k_5 r}  \hfb{U}^{k_5}_\gamma )^* -  ( \bar{\hfb{V}}^{k_5}_\delta  R^{\rm ff}_{k_5 r} \,  R^{\rm fb}_{k_4 r}  \hfb{U}^{k_4}_\gamma )^* } {[ - (\hfe{k_1} + \hfe{k_2} + \hfe{k_3} + \hfe{k_4}) + i\eta]\,[- (\hfe{k_4} + \hfe{k_5}) + i\eta]}       \, , \label{eq:SigInf_IIIb_11} \\
\Sigma^{(\infty, 3b)\,12}_{\alpha \beta} ={}&   \frac 1 2  \sum_{\gamma \,  \delta }   \int  \frac{d\,\omega}{2\pi i}\,  v_{\alpha  \bar\beta , \gamma \bar\delta} \,  G^{(2b) 12}_{\gamma \delta}(\omega)  \nonumber \\
  ={}&   \sum_{\substack{\gamma \,  \delta \\ r \, k_4 \, k_5}}  \frac{v_{\alpha \bar\beta , \gamma \delta} }{2}  \,
      \frac {   \hfb{U}^{k_4}_\gamma  R^{\rm fb}_{k_4 r} \,   ( R^{\rm fb}_{k_5 r}  \bar{\hfb{V}}^{k_5}_\delta \!)^* -  ( \bar{\hfb{V}}^{k_4}_\gamma  R^{\rm fb}_{k_4 r} \!)^* \; R^{\rm fb}_{k_5 r}  \hfb{U}^{k_5}_\delta   }{[ - (\hfe{k_1} + \hfe{k_2} + \hfe{k_3} + \hfe{k_4}) + i\eta]\, [- (\hfe{k_1} + \hfe{k_2} + \hfe{k_3} + \hfe{k_5}) + i\eta]}    \nonumber \\
    & +     \sum_{\substack{\gamma \,  \delta \\ r \, k_4 \, k_5}}  \frac{v_{\alpha \bar\beta , \gamma \delta} }{2}  \,
      \frac { \hfb{U}^{k_5}_\gamma  R^{\rm ff}_{k_5 r}   \,    R^{\rm fb}_{k_4 r}  \hfb{U}^{k_4}_\delta  -    \hfb{U}^{k_4}_\gamma  R^{\rm fb}_{k_4 r} \, R^{\rm ff}_{k_5 r}  \hfb{U}^{k_5}_\delta
         +  (  \bar{\hfb{V}}^{k_4}_\gamma  R^{\rm fb}_{k_4 r}  \, R^{\rm ff}_{k_5 r}  \bar{\hfb{V}}^{k_5}_\delta )^* -  ( \bar{\hfb{V}}^{k_5}_\gamma  R^{\rm ff}_{k_5 r} \,  R^{\rm fb}_{k_4 r}  \bar{\hfb{V}}^{k_4}_\delta )^* }{[ - (\hfe{k_1} + \hfe{k_2} + \hfe{k_3} + \hfe{k_4}) + i\eta]\,[- (\hfe{k_4} + \hfe{k_5}) + i\eta]}     \, . \label{eq:SigInf_IIIb_12} 
\end{align}
\end{subequations}

In  Eqs.~\eqref{eq:SigInf_II} and~\eqref{eq:SigInf_III} the following tensors of rank two and four in the quasiparticle indices have been employed
\begin{subequations}\label{eq:Qtens_def}
\begin{align}
  Q^{\rm ff}_{k_1 \, k_2}  \equiv{}&   \sum_{\alpha \beta} \; \left[
     (\hfb{U}^{k_1}_\alpha )^*  \, \Sigma^{(I)11}_{\alpha \beta} \, \hfb{U}^{k_2}_\beta   \;-\;    \bar{\hfb{V}}^{k_2}_\alpha  \, \Sigma^{(I)11}_{\alpha \beta} \, (\bar{\hfb{V}}^{k_1}_\beta )^*
 \;+\;  (\hfb{U}^{k_1}_\alpha )^*  \, \Sigma^{(I)12}_{\alpha \beta} \, \hfb{V}^{k_2}_\beta   \;+\;            \hfb{U}^{k_2}_\alpha \, ( \Sigma^{(I)12}_{\alpha \beta} \, \hfb{V}^{k_1}_\beta )^*  \right] \, ,  \label{eq:Qff_def} \\
  Q^{\rm fb}_{k_1 \, k_2}  \equiv{}&    \sum_{\alpha \beta} \; \left[
         (\hfb{U}^{k_1}_\alpha )^*  \, \Sigma^{(I)11}_{\alpha \beta} \, (\bar{\hfb{V}}^{k_2}_\beta)^*   \;-\;              \hfb{U}^{k_2}_\alpha  \, \Sigma^{(I)11}_{\alpha \beta} \, (\bar{\hfb{V}}^{k_1}_\beta )^*
 \;+\;  (\hfb{U}^{k_1}_\alpha )^*  \, \Sigma^{(I)12}_{\alpha \beta} \, (\bar{\hfb{U}}^{k_2}_\beta)^*   \;+\;   ( \bar{\hfb{V}}^{k_2}_\alpha \,  \Sigma^{(I)12}_{\alpha \beta} \,         \hfb{V}^{k_1}_\beta )^*  \right]  \, ,    \label{eq:Qfb_def}
\end{align}
\end{subequations}
and
\begin{subequations}\label{eq:Rtens_def}
\begin{align}
  R^{\rm ff}_{k \, r}  \equiv{}&   \sum_\alpha \; \left[           \mathcal{C}^{(\rm{I})}_{\alpha, r} \,  (\hfb{U}^{k}_\alpha )^* \; + \;    \bar{\mathcal D}^{(\rm{I})}_{r, \alpha}  \, (\bar{\hfb{V}}^{k}_\alpha )^*  \right]  \,  ,   \label{eq:Rff_def}\\
  R^{\rm fb}_{k  \, r}  \equiv{}&   \sum_\alpha \; \left[ (  \bar{\mathcal D}^{(\rm{I})}_{r, \alpha} \,   \hfb{U}^{k}_\alpha )^* \; + \;    (       \mathcal{C}^{(\rm{I})}_{\alpha, r} \,  \bar{\hfb{V}}^{k}_\alpha )^*  \right]  \, ,    \label{eq:Ffb_def}
\end{align}
\end{subequations}
where $ \mathcal{C}^{(\rm{I})}$ aned $ \bar{\mathcal D}^{(\rm{I})}$ are the first-order coupling amplitudes defined in Eqs.~\eqref{eq:ADC2} and the index $r$ encapsulate three quasiparticle excitations as defined in Eq.~\eqref{eq:r_indx}.
\end{widetext}

\section{\label{App:Freq_ints}Frequency integrals}

The ADC coupling amplitudes and denominators discussed in Sec.~\ref{Sec:ADCn} are obtained by direct comparison to the analytic expression of all relevant diagrams~\cite{Barb2017LNP}. The Feynman rules for computing diagrams have been introduced in Ref.~\cite{Soma2011GkI}, where the full calculation of all ADC(2)  diagrams has also been discussed in detail.  The case of third-order diagrams is essentially analogous, involving more complicated frequency integral. However, one must also pay attention in grouping together classes of different diagrams in such a way that important symmetries are preserved, and in particular permutations signs imposed by  the Pauli  principle. 
This last section outlines the computation for the third-order diagrams of Fig.~\ref{fig:ADC3_A} as an example.

\begin{widetext}
Let us start with the computations of diagram~\ref{Fig_Aa} contributing to $\widetilde{\Sigma}^{11}(\omega)$. By applying the Feynman rules from Ref.~\cite{Soma2011GkI}, we have
\begin{align}
  \widetilde{\Sigma}^{11 \, ({\rm Fig.\ref{Fig_Aa}})}_{\alpha \beta}(\omega) ={}& \frac 14 \int \frac{d\,\omega_1}{2\pi i} \frac{d\,\omega_2}{2\pi i} \frac{d\,\omega_3}{2\pi i}
   \,   v_{\alpha \lambda, \mu \nu}   \,  G^{11}_{\mu \mu'}(\omega_2)  G^{11}_{\nu \nu'}(\omega+\omega_1-\omega_2)  \,   v_{\mu' \nu', \gamma' \delta'}  \nonumber \\
    &\qquad \qquad \times   G^{11}_{\gamma' \gamma}(\omega_3)  G^{11}_{\delta' \delta}(\omega+\omega_1-\omega_3)  \, v_{\gamma \delta, \beta \lambda'}  \, G^{11}_{\lambda' \lambda}(\omega_1)   \, ,
\label{eq:DiagA1_1st}
\end{align}
where we use the convention that repeated indices are summed over.
The fist step is to substitute the spectral representation~\eqref{eq:Gkv_props} of the propagators and apply Cauchy's theorem to perform integrals over $\omega_2$ and  $\omega_3$
\begin{align}
   \widetilde{\Sigma}^{11 \, ({\rm Fig.\ref{Fig_Aa}})}_{\alpha \beta}(\omega) ={}&  \frac 14   \sum_{\substack{k_1 k_2 k_3 \\ \; k_4 k_5}}  \int \frac{d\,\omega_1}{2\pi i} 
   \,   v_{\alpha \lambda, \mu \nu}   \,    
    \left\{   \frac{          {\mathcal U}^{k_1}_{\mu}       {\mathcal U}^{k_2}_{\nu}       \; \;  (       {\mathcal U}^{k_1}_{\mu'}        {\mathcal U}^{k_2}_{\nu'})^*   }{\omega+\omega_1 - (\omega_{k_1} + \omega_{k_2}) + i\eta} 
         +    \frac{  (\bar{\mathcal V}^{k_1}_{\mu} \bar{\mathcal V}^{k_2}_{\nu})^*    \;\,    \bar{\mathcal V}^{k_1}_{\mu'}  \bar{\mathcal V}^{k_2}_{\nu'}       }{\omega+\omega_1 + (\omega_{k_1} + \omega_{k_2}) - i\eta}    \right\}
     \,   v_{\mu' \nu', \gamma' \delta'}  \nonumber \\
    &\qquad  \times 
    \left\{   \frac{          {\mathcal U}^{k_4}_{\gamma'}       {\mathcal U}^{k_5}_{\delta'}       \; \;  (       {\mathcal U}^{k_4}_{\gamma}         {\mathcal U}^{k_5}_{\delta})^*   }{\omega+\omega_1 - (\omega_{k_4} + \omega_{k_5}) + i\eta} 
         +    \frac{  (\bar{\mathcal V}^{k_4}_{\gamma'} \bar{\mathcal V}^{k_5}_{\delta'})^*    \;\,    \bar{\mathcal V}^{k_4}_{\gamma}   \bar{\mathcal V}^{k_5}_{\delta}       }{\omega+\omega_1 + (\omega_{k_4} + \omega_{k_5}) - i\eta}    \right\}
       \, v_{\gamma \delta, \beta \lambda'}  \,   \nonumber \\ 
    &\qquad  \times 
      \left\{   \frac{        {\mathcal U}^{k_3}_{\lambda'}    \; \;        {\mathcal U}^{k_3}_\lambda{}^*     }{\omega_1 - \omega_{k_3} + i\eta} +    \frac{  \bar{\mathcal V}^{k_3}_{\lambda'}{}^*        \;\,  \bar{\mathcal V}^{k_3}_\lambda }{\omega_1 + \omega_{k_3} - i\eta}     \right\} \, .
\label{eq:DiagA1_2nd}
\end{align}
The final integral yields six (time ordered) Goldstone contributions. We consider only the three forward going ones
\begin{align}
   \widetilde{\Sigma}^{11 \, ({\rm Fig.\ref{Fig_Aa}})}_{\alpha \beta}(\omega) ={}& 
       \frac{~ v_{\alpha \lambda, \mu \nu}   \, (\bar{\mathcal V}^{k_1}_{\mu} \bar{\mathcal V}^{k_2}_{\nu})^*    \;\,    \bar{\mathcal V}^{k_1}_{\mu'}       \bar{\mathcal V}^{k_2}_{\nu'}   \,   v_{\mu' \nu', \gamma' \delta'} ~}{ - (\omega_{k_1} + \omega_{k_2} +\omega_{k_4} + \omega_{k_5}) +i\eta} \; \frac 1 2 \; 
       \frac{ ~ {\mathcal U}^{k_4}_{\gamma'}       {\mathcal U}^{k_5}_{\delta'}    \bar{\mathcal V}^{k_3}_{\lambda}     \; \;  (       {\mathcal U}^{k_4}_{\gamma}         {\mathcal U}^{k_5}_{\delta}   \bar{\mathcal V}^{k_3}_{\lambda'} )^* ~ }{~ \omega - (\omega_{k_4} +\omega_{k_5} + \omega_{k_3}) +i\eta~} \;\frac 1 2 \;  v_{\gamma \delta, \beta \lambda'} \nonumber \\
&+ v_{\alpha \lambda, \mu \nu}   \; 
      \frac{  {\mathcal U}^{k_1}_{\mu}       {\mathcal U}^{k_2}_{\nu}   \bar{\mathcal V}^{k_3}_{\lambda}      \; \;  (       {\mathcal U}^{k_1}_{\mu'}        {\mathcal U}^{k_2}_{\nu'}   \bar{\mathcal V}^{k_3}_{\lambda'}  )^*  }{~ \omega - (\omega_{k_1} +\omega_{k_2} + \omega_{k_3}) +i\eta~} \; \frac 1 2  \;
      \frac{  ~  v_{\mu' \nu', \gamma' \delta'} \,   (\bar{\mathcal V}^{k_4}_{\gamma'} \bar{\mathcal V}^{k_5}_{\delta'})^*    \;\,    \bar{\mathcal V}^{k_4}_{\gamma}       \bar{\mathcal V}^{k_5}_{\delta}      \, v_{\gamma \delta, \beta \lambda'} ~}{ - (\omega_{k_1} + \omega_{k_2} +\omega_{k_4} + \omega_{k_5}) +i\eta} \; \frac 1 2  \nonumber \\
&+ v_{\alpha \lambda, \mu \nu}   \; \frac 1 2 \:
      \frac{  {\mathcal U}^{k_1}_{\mu}       {\mathcal U}^{k_2}_{\nu}   \bar{\mathcal V}^{k_3}_{\lambda}      \; \;  (       {\mathcal U}^{k_1}_{\mu'}        {\mathcal U}^{k_2}_{\nu'}  )^*  }{~ \omega - (\omega_{k_1} +\omega_{k_2} + \omega_{k_3}) +i\eta ~}   \;  v_{\mu' \nu', \gamma' \delta'} \;    
      \frac{  {\mathcal U}^{k_4}_{\gamma'}       {\mathcal U}^{k_5}_{\delta'}       \; \;  (       {\mathcal U}^{k_4}_{\gamma}         {\mathcal U}^{k_5}_{\delta}   \bar{\mathcal V}^{k_3}_{\lambda'} )^*  }{~ \omega - (\omega_{k_4} +\omega_{k_5} + \omega_{k_3}) +i\eta ~} \, \frac 1 2 \, v_{\gamma \delta, \beta \lambda'}   +  \ldots \nonumber \\
 ={}& \mathcal{J}^{(2a)}_{\alpha, k_4 k_5 k_3}   \frac 1{~\omega - \mathcal{E}^{(\rm{0})}_{k_4 k_5 k_3} + i\eta~} (\mathcal{M}_{\beta, k_4 k_5 k_3})^* 
    +   \mathcal{M}_{\alpha, k_1 k_2 k_3}  \frac 1{~\omega - \mathcal{E}^{(\rm{0})}_{k_1 k_2 k_3}+ i\eta~}  (\mathcal{J}^{(2a)}_{\beta, k_1 k_2 k_3})^*   \nonumber \\
   & +  \mathcal{M}_{\alpha, k_1 k_2 k_3}  \frac 1 {~\omega - \mathcal{E}^{(\rm{0})}_{k_1 k_2 k_3} + i\eta~}  \, \frac 1 2 \, \mathcal{E}^{(pp)}_{k_1 k_2, k_4 k_5}  \frac 1{~\omega - \mathcal{E}^{(\rm{0})}_{k_4 k_5 k_3} + i\eta~}  (\mathcal{M}_{\beta, k_4 k_5 k_3})^*   +  \ldots   \, ,
\label{eq:DiagAc_3rd}
\end{align}
where we have defined
\begin{align}
  \mathcal{M}_{\alpha, k_1 k_2 k_3}  \equiv{}&   \frac 1{\sqrt 2} \, v_{\alpha \lambda, \mu \nu}    \,  {\mathcal U}^{k_1}_{\mu}       {\mathcal U}^{k_2}_{\nu}   \bar{\mathcal V}^{k_3}_{\lambda}   \, , \\ 
  \mathcal{J}^{(2a)}_{\alpha, k_4 k_5 k_3}  \equiv{}&     ~ v_{\alpha \lambda, \mu \nu}   \, \frac 1 {\sqrt 8} \; \frac{ (\bar{\mathcal V}^{k_1}_{\mu} \bar{\mathcal V}^{k_2}_{\nu})^*    \;\,   
     \bar{\mathcal V}^{k_1}_{\mu'}       \bar{\mathcal V}^{k_2}_{\nu'}   \,   v_{\mu' \nu', \gamma' \delta'} \; {\mathcal U}^{k_4}_{\gamma'}       {\mathcal U}^{k_5}_{\delta'}    \bar{\mathcal V}^{k_3}_{\lambda}  }{ - (\omega_{k_1} + \omega_{k_2} +\omega_{k_4} + \omega_{k_5}) +i\eta} \nonumber \\
     ={}&   \frac 1{\sqrt 2} \, \frac { v_{\alpha \lambda, \mu \nu}  } 2  \;  (\bar{\mathcal V}^{k_1}_{\mu} \bar{\mathcal V}^{k_2}_{\nu})^*    \;  t_{k_1 k_2}^{k_4 k_5} \;  \bar{\mathcal V}^{k_3}_{\lambda}  \, ,
\label{eq:MJ2a_def}
\end{align}
and $ \mathcal{E}^{(\rm{0})}$ and $\mathcal{E}^{(pp)}$ are given by Eqs.~\eqref{eq:ADC2_E} and~\eqref{eq:3rd_Epp}.

Matrices $\mathcal{M}$ and $\mathcal{J}$  define the coupling amplitudes between single-particle states and ISCs. However, the sole diagram of Fig.~\ref{Fig_Aa} is not sufficient to guarantee the correct antisymmetrization among quasiparticles. Part of the missing terms are introduced
%by the second diagram in Fig.~\ref{fig:ADC3_A}:
%\begin{align}
%  \widetilde{\Sigma}^{11 \, ({\rm Fig.~\ref{Fig_Ab}})}_{\alpha \beta}(\omega)  ={}&  \frac {-1}2 \int \frac{d\,\omega_1}{2\pi i} \frac{d\,\omega_2}{2\pi i} \frac{d\,\omega_3}{2\pi i}
%   \,   v_{\alpha \bar\lambda, \mu \bar\nu}   \,  G^{11}_{\mu \mu'}(\omega_2)  G^{21}_{\lambda \nu'}(\omega+\omega_1-\omega_2)  \,   v_{\mu' \nu', \gamma' \delta'}  \nonumber \\
%    &\qquad \qquad \times   G^{11}_{\gamma' \gamma}(\omega_3)  G^{11}_{\delta' \delta}(\omega+\omega_1-\omega_3)  \, v_{\gamma \delta, \beta \lambda'}  \, G^{12}_{\lambda' \nu}(\omega_1)    \nonumber \\
% ={}&  2 \mathcal{J}^{(2b)}_{\alpha, k_4 k_5 k_3}   \frac {1}{~\omega - \mathcal{E}^{(\rm{0})}_{k_4 k_5 k_3} + i\eta~} (\mathcal{M}_{\beta, k_4 k_5 k_3})^* 
%    -   \mathcal{M}_{\alpha, k_1 k_3 k_2}  \frac 1{~\omega - \mathcal{E}^{(\rm{0})}_{k_1 k_2 k_3}+ i\eta~}  (2 \mathcal{J}^{(2a)}_{\beta, k_1 k_2 k_3})^*   \nonumber \\
%   & -  \mathcal{M}_{\alpha, k_1 k_3 k_2}  \frac 1 {~\omega - \mathcal{E}^{(\rm{0})}_{k_1 k_2 k_3} + i\eta~}   \mathcal{E}^{(pp)}_{k_1 k_2, k_4 k_5}  \frac 1{~\omega - \mathcal{E}^{(\rm{0})}_{k_4 k_5 k_3} + i\eta~}  (\mathcal{M}_{\beta, k_4 k_5 k_3})^*   +  \ldots   \, ,
%\label{eq:DiagAb}
%\end{align}
%
by the third diagram in Fig.~\ref{fig:ADC3_A}
\begin{align}
  \widetilde{\Sigma}^{11 \, ({\rm Fig.~\ref{Fig_Ac}})}_{\alpha \beta}(\omega)  ={}&  \frac {-1}2 \int \frac{d\,\omega_1}{2\pi i} \frac{d\,\omega_2}{2\pi i} \frac{d\,\omega_3}{2\pi i}
   \,   v_{\alpha \lambda, \mu \nu}   \,  G^{11}_{\mu \mu'}(\omega_2)  G^{11}_{\nu \nu'}(\omega+\omega_1-\omega_2)  \,   v_{\mu' \nu', \gamma' \delta'}  \nonumber \\
    &\qquad \qquad \times   G^{11}_{\gamma' \gamma}(\omega_3)  G^{12}_{\delta' \lambda'}(\omega+\omega_1-\omega_3)  \, v_{\gamma \bar\delta, \beta \bar\lambda'}  \, G^{21}_{\delta \lambda}(\omega_1)    \nonumber \\
 ={}&  - 2 \mathcal{J}^{(2a)}_{\alpha, k_4 k_5 k_3}   \frac {1}{~\omega - \mathcal{E}^{(\rm{0})}_{k_4 k_5 k_3} + i\eta~} (\mathcal{M}_{\beta, k_4 k_3 k_5})^* 
    +   \mathcal{M}_{\alpha, k_1 k_2 k_3}  \frac 1{~\omega - \mathcal{E}^{(\rm{0})}_{k_1 k_2 k_3}+ i\eta~}  (\mathcal{J}^{(2b)}_{\beta, k_1 k_2 k_3})^*   \nonumber \\
   & -  \mathcal{M}_{\alpha, k_1 k_4 k_3}  \frac 1 {~\omega - \mathcal{E}^{(\rm{0})}_{k_1 k_2 k_3} + i\eta~}  \mathcal{E}^{(pp)}_{k_1 k_2, k_4 k_5}  \frac 1{~\omega - \mathcal{E}^{(\rm{0})}_{k_4 k_5 k_3} + i\eta~}  (\mathcal{M}_{\beta, k_4 k_3 k_5})^*   +  \ldots   \, ,
\label{eq:DiagAc}
\end{align}
%
%and
% by the fourth diagram in Fig.~\ref{fig:ADC3_A}:
%\begin{align}
%  \widetilde{\Sigma}^{11 \, ({\rm Fig.~\ref{Fig_Ad}})}_{\alpha \beta}(\omega)  ={}&  - \int \frac{d\,\omega_1}{2\pi i} \frac{d\,\omega_2}{2\pi i} \frac{d\,\omega_3}{2\pi i}
%   \,   v_{\alpha \bar\lambda, \mu \nu}   \,  G^{11}_{\mu \mu'}(\omega_2)  G^{21}_{\lambda \nu'}(\omega+\omega_1-\omega_2)  \,   v_{\mu' \nu', \gamma' \delta'}  \nonumber \\
%    &\qquad \qquad \times   G^{11}_{\gamma' \gamma}(\omega_3)  G^{12}_{\delta' \lambda'}(\omega+\omega_1-\omega_3)  \, v_{\gamma \delta, \beta \bar\lambda'}  \, G^{12}_{\nu \delta}(-\omega_1)    \nonumber \\
% ={}&  - 2 \mathcal{J}^{(b2)}_{\alpha, k_ k_ k_}   \frac {1}{~\omega - \mathcal{E}^{(\rm{0})}_{k_4 k_5 k_3} + i\eta~} (\mathcal{M}_{\beta, k_ k_ k_})^* 
%    -  \mathcal{M}_{\alpha, k_ k_ k_}  \frac 1{~\omega - \mathcal{E}^{(\rm{0})}_{k_1 k_2 k_3}+ i\eta~}  (2 \mathcal{J}^{(2b)}_{\beta, k_ k_ k_})^*   \nonumber \\
%   & + 2 \mathcal{M}_{\alpha, k_ k_ k_}  \frac 1 {~\omega - \mathcal{E}^{(\rm{0})}_{k_1 k_2 k_3} + i\eta~}   \mathcal{E}^{(pp)}_{k_1 k_2, k_4 k_5}  \frac 1{~\omega - \mathcal{E}^{(\rm{0})}_{k_4 k_5 k_3} + i\eta~}  (\mathcal{M}_{\beta, k_ k_ k_})^*   +  \ldots   \, ,
%\label{eq:DiagAd}
%\end{align}
with
\begin{align}
  \mathcal{J}^{(2b)}_{\alpha, k_1 k_2 k_3}  \equiv{}&    \frac 1{\sqrt 2}    v_{\alpha \lambda, \mu \nu} \left(\bar{\mathcal V}^{k_4}_\nu \mathcal{U}^{k_5}_\lambda \right)^{\!*}  t^{k_1 k_2}_{k_4 k_5} \,  \mathcal{U}^{k_3}_\mu   \, .
\label{eq:J2b_def}
\end{align}

The first terms on the right hand side in Eqs.~\eqref{eq:DiagAc_3rd} and~\eqref{eq:DiagAc} become fully antisymmetrized once they are summed together. By using the antisymmetry of $\mathcal{M}_{\alpha, k_1 k_2 k_3}$ and $\mathcal{J}^{(2a)}_{\alpha, k_1 k_2 k_3}$ with respect to the exchange of their first two quasiparticle indices and the independence of $ \mathcal{E}^{(\rm{0})}_{k_1 k_2 k_3}$ under any permutation, one finds
\begin{align}
\mathcal{J}^{(2a)}_{\alpha, k_4 k_5 k_3}   (\mathcal{M}_{\alpha, k_4 k_5 k_3} - 2 \mathcal{M}_{\alpha, k_4 k_3 k_5})^*  ={}&
\mathcal{J}^{(2a)}_{\alpha, k_4 k_5 k_3}   (\mathcal{M}_{\alpha, k_4 k_5 k_3}  +  \mathcal{M}_{\alpha, k_3 k_4 k_5})^*  - \mathcal{J}^{(2a)}_{\alpha, k_5 k_4 k_3}   (- \mathcal{M}_{\alpha, k_4 k_3 k_5})^*  \nonumber \\
={}& \mathcal{J}^{(2a)}_{\alpha, k_4 k_5 k_3}   (\mathcal{M}_{\alpha, k_4 k_5 k_3}  +  \mathcal{M}_{\alpha, k_3 k_4 k_5} + \mathcal{M}_{\alpha, k_5 k_3 k_4})^*   \nonumber \\
={}&   \mathcal{J}^{(2a)}_{\alpha, k_4 k_5 k_3}   (\sqrt{3} \, \mathcal{C}^{(I)}_{\alpha, k_4 k_5 k_3} )^*  \nonumber \\
={}& \frac 1 {\sqrt 3}  [ \mathcal{J}^{(2a)}_{\alpha, k_4 k_5 k_3} +  \mathcal{J}^{(2a)}_{\alpha, k_5 k_3 k_4 } +  \mathcal{J}^{(2a)}_{\alpha, k_3 k_4 k_5} ] \,  ( \mathcal{C}^{(I)}_{\alpha, k_4 k_5 k_3} )^* \nonumber \\
={}&   \mathcal{C}^{(IIa)}_{\alpha, k_4 k_5 k_3}   (\mathcal{C}^{(I)}_{\alpha, k_4 k_5 k_3} )^*  \, ,
%%%\mathcal{P}_{123}
\label{eq:Eg_adc3_JM_to_CC}
\end{align}
where Eq.~\eqref{eq:defM123} was also used.  Similarly, diagrams \ref{Fig_Ab} and~\ref{Fig_Ad} provide the missing contributions needed to antisymmetrize the second  term of Eqs.~\eqref{eq:DiagAc_3rd} and~\eqref{eq:DiagAc}, respectively.
The full antisymmetrization of the terms with double denominators requires all four diagrams of Fig.~\ref{fig:ADC3_A}.  When all contributions are added together, one obtains
\begin{align}
  \widetilde{\Sigma}^{11 \, ({\rm Fig.~\ref{fig:ADC3_A}a+b+c+d})}_{\alpha \beta}(\omega)  ={}&
      [\mathcal{C}^{(IIa)}_{\alpha, k_1 k_2 k_3} + \mathcal{C}^{(IIb)}_{\alpha, k_1 k_2 k_3} ]  \frac 1{~\omega - \mathcal{E}^{(\rm{0})}_{k_1 k_2 k_3} + i\eta~} (\mathcal{C}^{(I)}_{\beta, k_1 k_2 k_3})^*    \nonumber \\
    & +   \mathcal{C}^{(I)}_{\alpha, k_1 k_2 k_3}  \frac 1{~\omega - \mathcal{E}^{(\rm{0})}_{k_1 k_2 k_3}+ i\eta~}  [\mathcal{C}^{(IIa)}_{\beta, k_1 k_2 k_3} + \mathcal{C}^{(IIb)}_{\beta, k_1 k_2 k_3}]^*   \nonumber \\
   & +  \mathcal{C}^{(I)}_{\alpha, k_1 k_2 k_3}  \frac 1 {~\omega - \mathcal{E}^{(\rm{0})}_{k_1 k_2 k_3} + i\eta~}  
     \, \frac 1 2 \,   [ \mathcal{E}^{(pp)}_{k_1 k_2, k_4 k_5} \delta_{k_3 k_6} + \mathcal{E}^{(pp)}_{k_3 k_1, k_6 k_4} \delta_{k_2 k_5} + \mathcal{E}^{(pp)}_{k_2 k_3, k_5 k_6} \delta_{k_1 k_4} ]    \nonumber \\
     & \qquad \qquad \times \frac 1{~\omega - \mathcal{E}^{(\rm{0})}_{k_4 k_5 k_3} + i\eta~}  (\mathcal{C}^{(I)}_{\beta, k_4 k_5 k_3})^*   +  \ldots   \, .
\label{eq:Eg_adc3_Aterms}
\end{align}
\end{widetext}

The perturbative expansion of $\widetilde{\Sigma}^{11}(\omega)$ is obtained by substituting Eqs.~\eqref{eq:CDE_exp} into Eq.~\eqref{eq:Sig_tild11} and then expanding the inverse matrices with respect to the $\mathcal{E}^{(I)}$ appearing in the denominators. By comparing the third-order terms of this expansion to Eq.~\eqref{eq:Eg_adc3_Aterms}, one identifies the amplitudes and denominators of Eqs.~\eqref{eq:3rd_CIIa},~\eqref{eq:3rd_CIIb} and~\eqref{eq:3rd_EIa}.

The above example clarifies how a proper approximation to the self-energy may require to gather contributions of specific time ordering across different Feynman diagrams. The ADC($n$) framework ensures that all Feynman diagrams are included in full up to order $n$, while all other terms beyond this (including non-perturbative resummations) appear for selected time ordering and do not necessarily constitute full Feynman amplitudes.
We note that there exist mixed approximations in the literature (that is, intermediate among different ADC($n$) orders) that can be obtained by suppressing some of the Goldstone distributions. For example, third-order corrections due to $\mathcal{C}^{(II)}$ are known to be important to reproduce correct separation energies of dominant quasiparticle peaks, both in atomic nuclei and molecules. This was the basis of the outer-valence Green’s function (OVGF) method, one of the earliest approximations used by quantum chemists for ionisation potentials and affinities~\cite{Niessen1984PhysRep}.  Conversely, the two-particle--one-hole Tamm-Dancoff approximation extends ADC(2) by including the $\mathcal{E}^{(I)}$ contributions to the energy denominators but neglects the remaining  ADC(3) contributions~\cite{Schirmer19782phTDA,Rijsdijk1996TDA2p1h}. In each of these cases, one needs to add consistently selected time orderings from different diagrams.

\bibliography{Biblio_scgf}% Produces the bibliography via BibTeX.

\end{document}